\documentclass[onecolumn,preprintnumbers,showkeys,showpacs]{article}
\usepackage{amsmath, amsthm, amscd, amsfonts, amssymb, graphicx, hhline}
\usepackage{dcolumn}
\usepackage{bm}
\usepackage{multirow}
\usepackage[]{color}
\usepackage[papersize={210.0mm,297.0mm},dvips]{geometry}
\providecommand{\keywords}[1]
{
  \begin{quote}
  \small Keywords: #1
  \end{quote}
}
\providecommand{\pacs}[1]
{
  \begin{quote}
  \small PACS numbers: #1
  \end{quote}
}

\begin{document}
\title{Three-band extension for the Glashow-Weinberg-Salam model}
\author{Konstantin V. Grigorishin\footnote{\small konst.phys@gmail.com}\\ \small \textit{Bogolyubov Institute for Theoretical Physics of the National Academy}\\ \small \textit{of Sciences of Ukraine, 14-b Metrolohichna str. Kyiv, 03143, Ukraine}}
\date{}
\maketitle
\begin{abstract}
By analogy with the Ginzburg-Landau theory of multi-band superconductors with inner (interband) Josephson couplings we formulate the three-band Glashow-Weinberg-Salam model with weak Josephson couplings between strongly asymmetrical condensates of scalar (Higgs) fields. Unlike usual single-band model, we found three Higgs bosons corresponding to three generations of particles, moreover the heaviest of them corresponds to the already discovered H-boson and decays into fermions of only the third generation through Yukawa interaction. The other two decay into fermions of the first and second generations accordingly, but they are difficult to observe due to very poor conditions for production. We found two sterile ultra-light Leggett bosons, the Bose condensates of which form the dark halos of galaxies and their clusters (i.e so called "dark matter"). The masses of the Leggett bosons are determined by the coefficient of the interband coupling and can be arbitrarily small ($\sim 10^{-20}\mathrm{eV}$) due to non-perturbativeness of the interband coupling. Since propagation of Leggett bosons is not accompanied by current, these bosons are not absorbed by gauge fields unlike the common-mode Goldstone bosons. Three coupled condensates of the scalar fields are related to the existence of three generations of leptons, where each generation interacts with the corresponding condensate getting mass. The interflavour mixing between the generations of active neutrinos and sterile right-handed neutrinos in the three-band system causes the existence of mass states of neutrino without interaction with the Higgs condensates.
\end{abstract}




\footnotesize
\tableofcontents

\normalsize
\section{Introduction}\label{intro}

The Standard Model (SM) is an $SU(3)_{c}\otimes SU(2)_{I}\otimes U(1)_{Y}$ gauge theory. Here, $SU(3)_{c}$ - the symmetry of the strong color interaction of quarks and gluons. The group of the weak isospin $I$ and the weak hypercharge $Y$, $SU(2)_{I}\otimes U(1)_{Y}$, describes electro-weak interaction of quarks and leptons mediated by the corresponding gauge bosons $\vec{A}_{\mu},B_{\mu}$. Due to the coefficient $a<0$ in the potential for the scalar field $a\varphi^{+}\varphi+\frac{b}{2}(\varphi^{+}\varphi)^{2}$, the complex scalar field $\varphi=|\varphi|e^{\mathrm{i}\theta}$ acquires a nonzero vacuum expectation value, which can be suppose as $|\langle0|\varphi|0\rangle|=\sqrt{\frac{|a|}{b}}\equiv\varphi_{0}$, and the $SU(2)_{I}\otimes U(1)_{Y}$ electroweak symmetry is spontaneously broken down to the $U(1)_{Q}$ gauge symmetry of electromagnetism with the electrical charge $Q=I_{z}+\frac{Y}{2}$. Here the Higgs mechanism takes place: the phase $\theta$ is absorbed by the gauge fields, and while three linear combinations of the gauge fields interact with the condensate $\varphi_{0}$ and become massive (i.e $W^{+},W^{-},Z$ bosons), but the photon $\gamma_{\mu}=A_{\mathrm{z}\mu}\sin\alpha+B_{\mu}\cos\alpha$ remains massless: $\frac{g^{2}}{4}\varphi_{0}^{2}(A_{\mathrm{x}\mu}A^{\mu}_{\mathrm{x}}+A_{\mathrm{y}\mu}A^{\mu}_{\mathrm{y}})
+\frac{1}{4}\varphi_{0}^{2}\left(g^{2}A_{\mathrm{z}\mu}A^{\mu}_{\mathrm{z}}+f^{2}B_{\mu}B^{\mu}\right)=
\frac{g^{2}}{2}\varphi_{0}^{2}W_{\mu}W^{\ast\mu}+\frac{\widetilde{g}^{2}}{4}\varphi_{0}^{2}Z_{\mu}Z^{\mu}$ (here $\sin\alpha=0.47$ is the Weinberg angle, $e=1/\sqrt{137}$ is an elementary charge in the Gaussian system of units, $g=\frac{e}{\sin\alpha}$, $f=\frac{e}{\cos\alpha}$, $\widetilde{g}^{2}=g^{2}+f^{2}$). In addition, the Dirac fields $\psi$ (spinor) interact with the condensate by the Yukawa interaction $\chi(\overline{\psi}_{L}\varphi\psi_{R}+\overline{\psi}_{R}\varphi^{+}\psi_{L})$, and, as the result, leptons obtain masses $m_{Di}=\chi_{i}\varphi_{0}$ (where $i=e,\mu,\tau$ - electron, muon, tauon); it is analogously for quarks, however neutrino remains strictly massless, and it is supposed that the right-handed neutrino $\nu_{R}$ and the left-handed antineutrino $\nu_{L}^{C}$ are absent \cite{sad,ryder}, but in various extensions of SM the existence of additional neutrinos with different parameters is allowed, for example, the neutrino minimal standard model ($\nu \mathrm{MSM}$) supposes existence of three sterile right-handed neutrinos $\nu_R$ \cite{asaka}. It should be noted that the lepton mixing and the quark mixing occur in such a way that some elements of the mixing matrices, i.e. the PMNS matrix for neutrino mixing and the CKM matrix for quark mixing, are complex $e^{\pm i\delta_{\mathrm{CP}}}$), which results in the CP violation \cite{buras,branco1,pich,mannel,esteban,giganti}.

\begin{figure}[ht]
\begin{center}
\includegraphics[width=9cm]{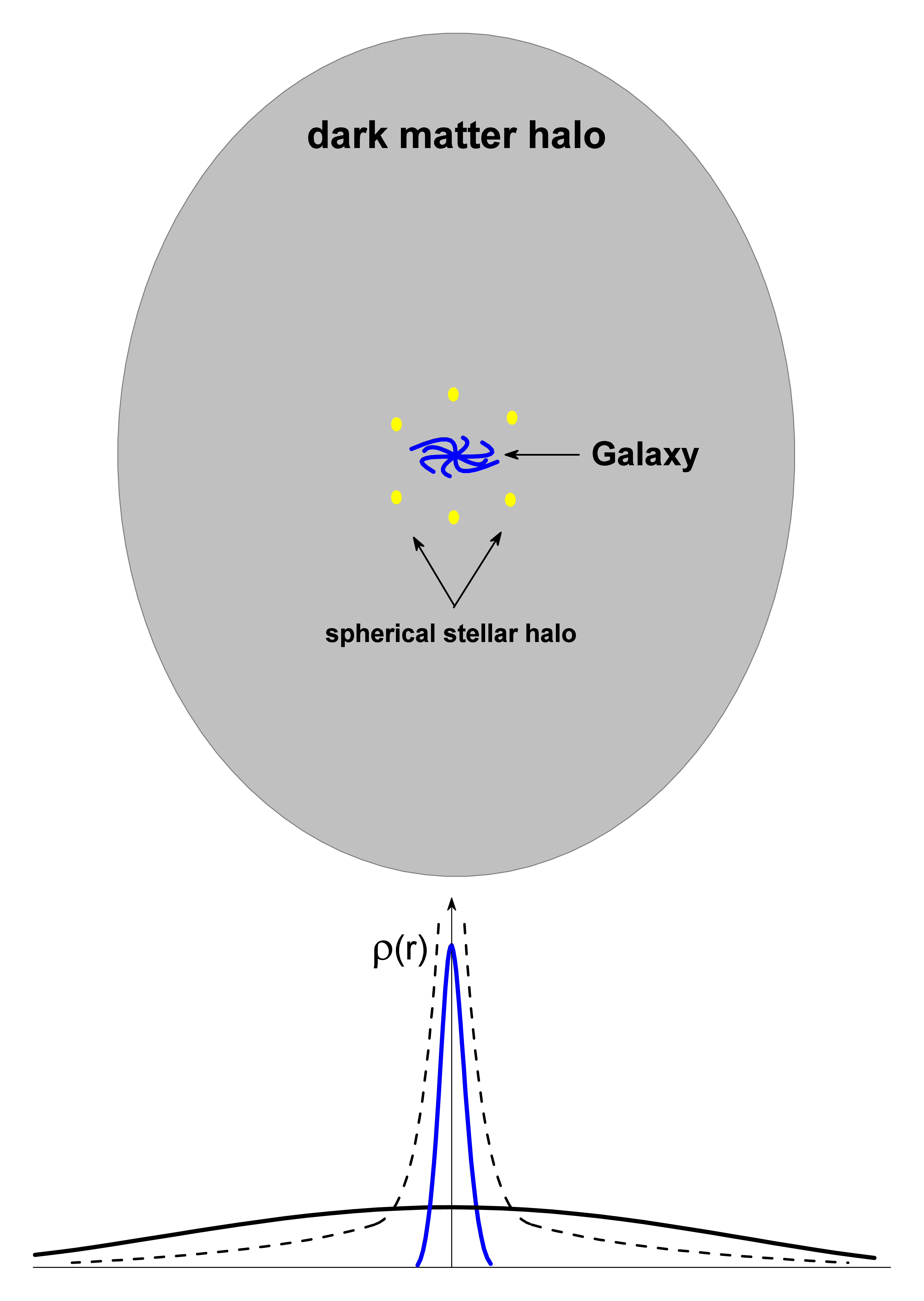}
\end{center}
\caption{Top figure: the region of DM halo compared with size of the galaxy and its stellar halo. Lower figure: corresponding profiles $\rho(r)$ of DM density (dark line) and density of barionic matter (blue line). Dash line is result of numerical simulations for distribution of DM density, where we can see a singularity - "cusp".}
\label{Fig01}
\end{figure}

SM with its minimal Higgs structure successfully describes the nature of fundamental particles. Especially, the Glashow-Weinberg-Salam (GWS) model of the electro-weak interaction provides an extremely successful description of observed electro-weak phenomena. However, SM in its present form is unable to describe number of extremely important phenomena. In a present work we would like to discuss some of them:

1) \textbf{Dark matter}. At present time, it is well known that the total mass-energy of the observable universe consists of $5\%$ ordinary (baryonic, leptonic, photonic) matter, $26\%$ dark matter (DM) and $69\%$ in a form of energy known as the dark energy \cite{plank}. Thus, DM constitutes $81\%$ of the total mass. So, the total mass of Milky Way taking into account DM, is estimated as $M\sim 0.8\ldots1.2\times10^{12}M_{\odot}$ and the radius of the DM halo is estimated as $r_{0}\sim120\mathrm{kpc}$ \cite{batt}. On the contrary, mass of baryonic matter in Milky Way is estimated as $M_{\mathrm{B}}\sim 5\ldots7\times10^{10}M_{\odot}$, and radius of the disk is estimated as $r_{\mathrm{B}}\sim 25\mathrm{kpc}$. Thus, DM constitutes $94\%$ of the total mass of the Milky Way, and region occupied by the relatively dense baryonic matter is very small region in a central part of the DM halo. Thus, Milky Way (in the same way other galaxies and galaxy clusters) is immersed in an almost homogeneous cloud of DM as illustrated in Fig.\ref{Fig01}. Moreover, density perturbations in the baryon-electron-photon plasma before recombination do not grow because of high light pressure; instead, the perturbations produce sound waves that propagate in the plasma. Because DM particles do not interact with photons, that nothing prevents them from forming self-gravitating clusters. After recombination, baryons fall into potential wells formed by DM. Galaxies form in those regions where DM form self-gravitating cluster originally \cite{rubakov1}. Thus, without DM no structures would have been formed, no galaxies, no life.

Thus, we have situation, when SM does not describe $81\%$ of matter in universe. Attempts have been repeatedly made to expand the SM so that it would include particles of DM. Since such particles do not manifest themselves in any way except through gravity (do not absorb, radiate or scatter electromagnetic waves and do not cause any significant nuclear reactions), then these particles must be almost sterile: they do not interact with photon and do not participate in strong interactions, only the weak interactions is allowed. So, it has been proposed as candidates for DM particles, for example, sterile (right-chiral) neutrinos \cite{boyar1,kopp1,kopp2,boyar2} with mass $\sim 1\mathrm{keV}$, neutralinos (as WIMP) with mass $>10^{2}\mathrm{GeV}$ \cite{feng,drees}, axions with mass $\sim 10^{-2}\mathrm{eV}$ \cite{feng,hook}, light scalaron of $f(R)$ gravity with mass $\sim 10^{-3}\mathrm{eV}$ \cite{shtan}, and many other \cite{feng,bertone}. At present moment, no DM candidate particles have been detected.

In order to form potential wells, the DM particles must be nonrelativistic, because relativistic particles travel through gravitational wells instead of being trapped there. On the other hand, according to numerical simulations, a DM halo should tend to produce densities in galactic centers as $\rho\sim r^{\alpha}$ with $\alpha\approx -1$: so called cusp in density profile \cite{blok,popo1,popo2,robles}. At the same time, the observed distributions DM halo is almost flat in the centre of a DM cloud $\rho\sim r^{0}$. For example, distributions of mass in a DM halo profile and in ordinary barionic matter are schematic shown in Fig.\ref{Fig01}. The cuspy halo problem is proposed to solve by heating of the DM gas in central region, as for example, proposed in \cite{dorosh}. Another solution of this problem is, instead to propose of complicated mechanism of heating of the DM gas, to suppose such property of DM particles, which makes impossible formation of a cusp. If DM is composed by some kind of ultra-light bosons ($10^{-24}\lesssim m\lesssim 1\mathrm{eV}$), then such Bose-gas can form Bose-Einstein condensate \cite{robles,bald,hu,lee}. The latest state of development of this hypothesis is presented in the review \cite{ferr}. Due to the uncertainty principle the central cusp is washed out to the flat profile, moreover the formation of small structures (galaxy satellites) is suppressed, many of which are predicted by the cold DM theory. Such a model has different names in the literature, such as Fuzzy Dark Matter (FDM), ultra-light DM, BEC Dark Matter, wave DM, scalar field DM, and others. Estimation of the ultra-light boson masses lies within a wide range. From $\sim 10^{-24}\mathrm{eV}$, which was obtained by comparing the de-Broglie wave-length of DM to the typical size of the DM halo in galaxies ($\sim 100\mathrm{kpc}$) \cite{bald}. If we suppose that the DM halo has some structure: core from BEC of size $\sim 1\mathrm{kpc}$ and Bose gas behaving as the cold DM, then mass $\sim 10^{-22}\mathrm{eV}$ \cite{hu,lee,robles1,matos,ferr} is assumed. At the same time, observations of stellar kinematics is dwarf galaxies give mass $\sim 10^{-22}\ldots 10^{-20}\mathrm{eV}$ \cite{broad,gold,pozo}. Obviously, in these models, the ultra-light bosons are assumed to be noninteracting or to interact very weakly with each other. If we suppose a strong interaction between bosons, then they can form a superfluid Bose liquid (as $\mathrm{HeII}$). In this case the mass of boson can be $\sim 1\mathrm{eV}$ \cite{ferr}. However, obviously, in such a model, in addition to unknown particles, there is also an interaction of unknown nature. Thus, we can see, that hypothesis about FDM adequately describes dark halo, despite some backlash in boson masses. However, nature of the ultra-light weak interacting (or even sterile) bosons remains unknown: these bosons do not fit into the framework of SM. 

2) \textbf{Neutrino masses}. Observation of the neutrino oscillations in vacuum means presence mass of neutrinos \cite{bilen1,bilen2,gersh,bilen3,bilen4,giunti}, but only the differences in the squares of the masses can be measured: $|\Delta m^{2}_{23}|\equiv|m^{2}_{3}-m^{2}_{2}|\approx 2.51\cdot 10^{-3}\mathrm{eV}^{2}$, $|\Delta m^{2}_{12}|\approx 7.41\cdot 10^{-5}\mathrm{eV}^{2}$ \cite{NuFIT,ester}, and the upper limits of the masses $\sqrt{m_{\nu e}^{2}},\sqrt{m_{\nu \mu}^{2}},\sqrt{m_{\nu \tau}^{2}}$ can be experimentally determined from $\beta$-decay of tritium, pion decay, $\tau$-decays into multi-pion final states accordingly \cite{formag,nucc}. Cosmological data (anisotropy of cosmic microwave background radiation, formation of structures, etc.) impose restrictions on masses: $\sum_{\nu}m_{\nu}<0.19\mathrm{eV}$ \cite{lorenz}, $\sum_{\nu}m_{\nu}<0.28\mathrm{eV}$ \cite{thomas}. Formally, we can write the Dirac mass term (Yukawa interaction) for both charged lepton and neutrino:
\begin{equation}\label{1.1}
  \mathcal{U}_{D}=\chi_{l}(\overline{\psi}_{L}\Psi l_{R}+l_{R}\Psi^{+}\psi_{L})+
  \chi_{\nu}(\overline{\psi}_{L}\widetilde{\Psi}\nu_{R}+\overline{\nu}_{R}\widetilde{\Psi}^{+}\psi_{L})
  =m_{Dl}(\overline{l}_{L}l_{R}+\overline{l}_{R}l_{L})+
  m_{D\nu}(\overline{\nu}_{L}\nu_{R}+\overline{\nu}_{R}\nu_{L}),
\end{equation}
where the isospinor $\psi_{L}=\left(\begin{array}{c}
                                                                                \nu_{L} \\
                                                                                l_{L} \\
                                                                              \end{array}\right)$
is a left-handed dublet, $l_{R}$ are $\nu_{R}$ right-handed singlets (here $\nu_{L}$ is a spinor of the active neutrino, $\nu_{R}$ is a spinor of the hypothetical sterile neutrino, $l_{L,R}$ are spinors of charged lepton), $\Psi=\left(\begin{array}{c}
                                                                                0 \\
                                                                                \varphi \\
                                                                              \end{array}\right)$,
$\widetilde{\Psi}=\mathrm{i}\tau_{y}\Psi=\left(\begin{array}{c}
                                                                                \varphi \\
                                                                                0 \\
                                                                              \end{array}\right)$
are isospinors, where $\varphi$ is scalar field with condensate $\langle\varphi\rangle=\varphi_{0}\neq 0$, $\tau_{y}$ is a Pauli matrix. $m_{Dl}=\chi_{l}\varphi_{0}$ and $m_{D\nu}=\chi_{\nu}\varphi_{0}$ are Dirac masses of the charged lepton and the neutrino accordingly. However, the problem is the unnatural difference in the Yukawa constants:
\begin{equation}\label{1.2}
  \chi_{\nu}\sim 10^{-11}\ll \chi_{l}\sim 10^{-6},
\end{equation}
unlike, for example, top and bottom rows of quarks, where their masses are not very different.

In SM the right-handed neutrinos $\nu_{R}$ are absent, hence $ m_{D\nu}=0$. There are several opportunities of extension of SM, where the small neutrino mass appears. So, for example, following review \cite{gersh} it should be noted the Gelmini-Roncadelli model, where it has been proposed an extension of the model with the single scalar field to the scalar doublet, where the additional vacuum condensates appears $\varphi_{1}$ so that $\varphi_{1}\ll\varphi_{0}\sim 250 \mathrm{GeV}$, and neutrino interacts only with the last $\chi\overline{\nu}_{L}^{C}\varphi_{1}\nu_{L}$ (where $C$ is a charge conjugation), hence the neutrino mass can be much less than the electron mass. Or well-known "see-saw" mechanism, where there are two scale of mass $m_{D}\ll m_{M}\sim 10^{14} \mathrm{GeV}$, so that $m_{\nu}\sim\frac{m_{D}^{2}}{m_{M}}\ll m_{D}$ as result of diagonalization of the mass matrix. However, these models assume that the neutrino is a Majorana neutrino which results in the neutrinoless double $\beta$-decay, but this has not been yet observed. Thus, origin of the neutrino mass remains unknown.

3) \textbf{The absence of experimentally detected decays of the Higgs boson into fermions of the second and first generations}.

\begin{figure}[ht]
\begin{center}
\includegraphics[width=9cm]{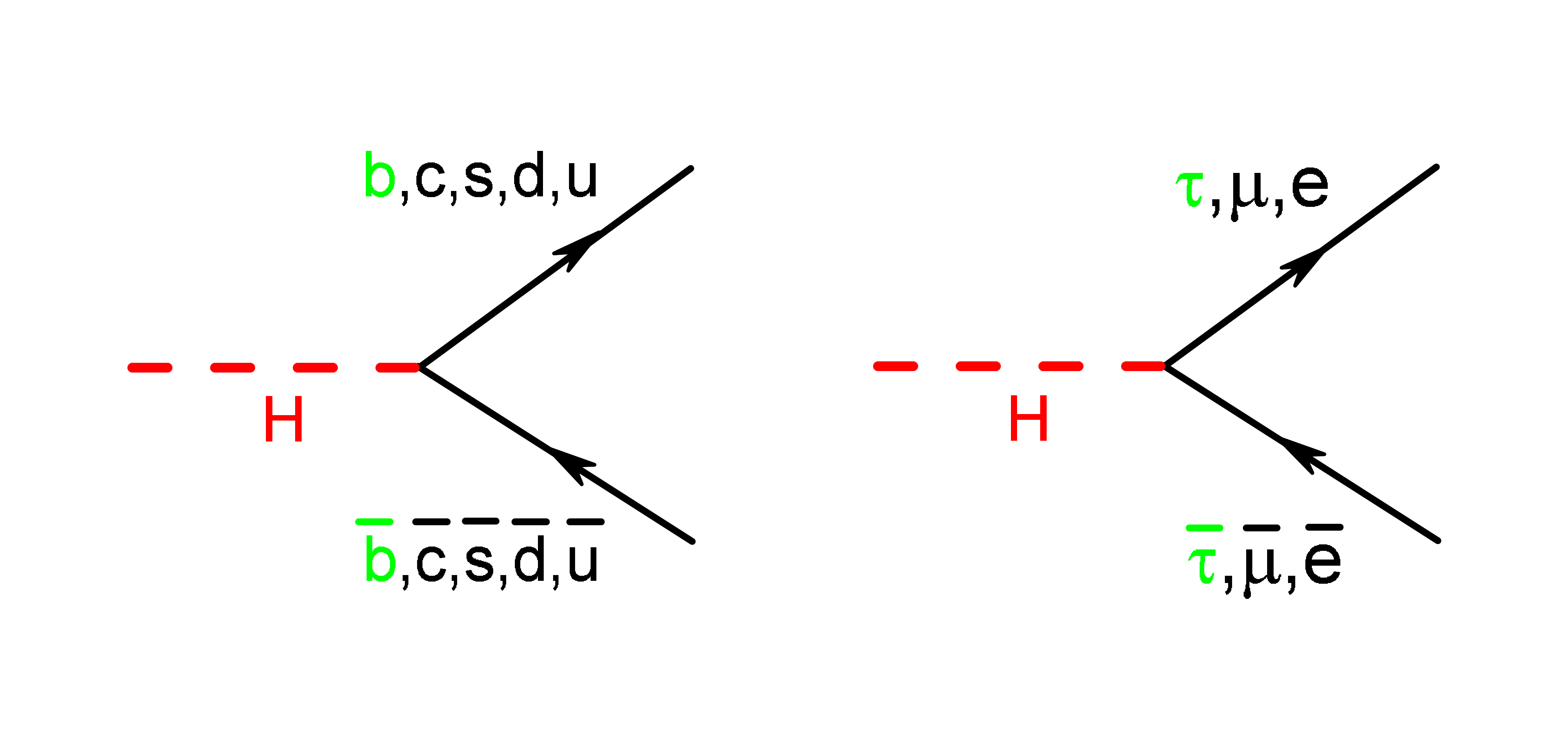}
\end{center}
\caption{Theoretical decays of Higgs boson into quark-antiquark pairs and into lepton-antilepton pairs due to Yukawa coupling. The green font denotes experimentally observed decays with a significance more than 5$\sigma$.}
\label{Fig02}
\end{figure}

There are many types of H-boson decay channels \cite{atlas,cms,atlascms}. So, due to the Yukawa coupling the H-boson can decay into quark-antiquarks pairs (all quarks except $t$-quarks, because $m_{t}>m_{H}$) and into lepton-antilepton pairs, that illustratid in Fig.\ref{Fig02}. According to SM the H-boson should decay as $H\rightarrow b\overline{b}$ with probability $57.5\%$, $H\rightarrow\tau\overline{\tau}$ with probability $6.30\%$, $H\rightarrow c\overline{c}$ with probability $2.90\%$, $H\rightarrow\mu\bar{\mu}$ with probability $\lesssim0.022\%$ \cite{atlascms}. At the same time, there has been no quite reliable experimental evidence found in direct searches by the ATLAS and CMS collaborations \cite{atlas1,cms1} for an H-boson decaying into a charm quark–antiquark pair, a strange quark–antiquark, an electron–positron pair, and a muon–antimuon pair. This fact is usually associated with the small Yukawa constant for the first and second generations of fermions. However, the decay rate into a pair of $c$-quarks is not much less than the decay rate into a pair of $\tau$-lepton (the decay probabilities are $2.9\%$ and $6.4\%$ accordingly). On the other hand, such very rare decays as two-photon decay $H\rightarrow\gamma\gamma$ with probability $\approx0.2\%$ have been detected. Thus, in our opinion, the absence of experimentally detected decays of the H-boson into fermions of the second and first generations can point to new physics, in the sense, that several Higgs fields can exist, so that mass of each generation is caused by corresponding Higgs field.

It should be noted, that the CMS collaboration has reported about the decay $H\rightarrow \mu\overline{\mu}$ with a significance of 3$\sigma$ \cite{cms4}. At the same time, the ATLAS collaboration has reported about decays $H\rightarrow \gamma\mu\overline{\mu},\gamma e\overline{e}$, which occurs through many intermediate channels due to various interactions (via virtual photon, Z,W bosons, quarks) with a significance of 3.2$\sigma$ \cite{atlas4}. Thus, decays $H\rightarrow \mu\overline{\mu}$ and $H\rightarrow \gamma\mu\overline{\mu}$ still need to be securely separated. Unlike $H\rightarrow\gamma\gamma$, it is difficult to discover the signal of $H\to c \overline{c}$, because the background from QCD is several orders of magnitude larger than the signal. Thus, LHC is not well suited for these problems, but multi-TeV lepton-antilepton colliders would be more suitable.


4) \textbf{Why are three generations of fermions needed, the problem of the hierarchy of their masses and lepton oscillations}.

As well known, all fundamental fermions are divided into three generations, that is, three sets of particles with identical interactions but with very different masses (except neutrinos): the first - $u,d$-quarks, $e,\nu_{e}$-leptons (electron and electron neutrino), the second - $c,s$-quarks, $\mu,\nu_{\mu}$-leptons (muon and muon neutrino), the third - $t,b$-quarks, $\tau,\nu_{\tau}$-leptons (tauon and tau neutrino). However, the first generation is sufficient for the substance, why the other two are needed is unclear. So, in Ref.\cite{ibe} the model with two heavy right-handed neutrino is proposed in order to provide generation of baryon asymmetry in earl era of universe and one sterile right-handed neutrino which makes up DM. However this model requires the seesaw mechanism. The origin of the mass hierarchy is unknown at this time. Indeed, for instance, the electron ($m_{e}=0.511\mathrm{MeV}$), the muon ($m_{e}=105.7\mathrm{MeV}$), and the tauon ($m_{e}=1777\mathrm{MeV}$) carry identical gauge quantum numbers, but their masses differ by orders of magnitude (this means that their Yukawa constants $\chi_{e},\chi_{\mu},\chi_{\tau}$ differ by orders of magnitude, because $m_{Di}=\chi_{i}\varphi_{0}$). As stated in the review \cite{troitsky}, an explanation of the hierarchy requires extra spatial dimensions. Moreover, the neutrino oscillations take place with large mixing angles ($\sim\pi/4$), however for charged leptons (electron-muon-tauon) the mixing is absent. It should be noted, that purely quantum-mechanical reasons for the absence of oscillations of charged leptons associated with the processes of their detection were expressed \cite{akhm,kays}: when the production of more than one type of mass-eigenstate charged leptons is kinematically allowed, the charged lepton states are either produced as incoherent mixtures of $e$, $\mu$ and $\tau$, or they lose their coherence over microscopic distances due to large difference in the masses of the basis states $m(\tau)-m(e)\gg m(\nu_{\tau})-m(\nu_{e})$, except at extremely high energies, not accessible to present experiments. However, this does not exclude and others fundamental reasons for the absence of the mixing of charged leptons.


To solve these and other problems of SM a two-Higgs-doublet model (2HDM) as a simple extension of SM is used \cite{branco2,wang,dubinin1,dubinin2,dubinin3}. This model supposes a two-doublet scalar potential:
\begin{eqnarray}\label{1.3}
  V_{\mathrm{2HDM}} &=& m_{11}^{2}\Psi_{1}^{+}\Psi_{1}+m_{22}^{2}\Psi_{2}^{+}\Psi_{2}-m_{12}^{2}(\Psi_{1}^{+}\Psi_{2}+\Psi_{2}^{+}\Psi_{1}) \nonumber\\
   &+&\frac{1}{2}\lambda_{1}(\Psi_{1}^{+}\Psi_{1})^{2}+\frac{1}{2}\lambda_{2}(\Psi_{2}^{+}\Psi_{2})^{2}
   +\lambda_{3}(\Psi_{1}^{+}\Psi_{1})(\Psi_{2}^{+}\Psi_{2})\nonumber\\
   &+&\lambda_{4}(\Psi_{1}^{+}\Psi_{2})(\Psi_{2}^{+}\Psi_{1})
   +\frac{1}{2}\lambda_{5}\left((\Psi_{1}^{+}\Psi_{2})^{2}+(\Psi_{2}^{+}\Psi_{1})^{2}\right)\nonumber\\
   &+&\lambda_{6}(\Psi_{1}^{+}\Psi_{1})(\Psi_{1}^{+}\Psi_{2}+\Psi_{2}^{+}\Psi_{1})
   +\lambda_{7}(\Psi_{2}^{+}\Psi_{2})(\Psi_{1}^{+}\Psi_{2}+\Psi_{2}^{+}\Psi_{1}).
\end{eqnarray}
Here we restrict to the CP-conserving models in which all $\lambda_{i}$ and $m_{ij}^{2}$ are real, at least one of $m_{ii}^{2}<0$, and $\lambda_{1,2}>0$. For illustration and simplicity an exact $Z_{2}$ discrete symmetry can be imposed, i.e $\Psi_{1}\rightarrow -\Psi_{1},\Psi_{2}\rightarrow\Psi_{2}$. Then $m_{12}=0,\lambda_{6,7}=0$. The fields $\Psi_{1,2}$ are $SU(2)$ isospinor:
\begin{equation}\label{1.4}
  \Psi_{1,2}=
  \left(
    \begin{array}{c}
      \phi^{+}_{1,2} \\
      (v_{1,2}+\rho_{1,2}+\mathrm{i}\eta_{1,2})/\sqrt{2} \\
    \end{array}
  \right),\quad\Psi_{1,2}^{+}=
  \left(
    \begin{array}{cc}
      \phi_{1,2}, & (v_{1,2}+\rho_{1,2}-\mathrm{i}\eta_{1,2})/\sqrt{2} \\
    \end{array}
  \right),
\end{equation}
where scalar vacuum condensates $v_{1,2}$ are such, that $\sqrt{v_{1}^{2}+v_{2}^{2}}=246\mathrm{GeV}$, $\langle\rho_{1,2}\rangle=\langle\eta_{1,2}\rangle=\langle\phi_{1,2}\rangle=0$. There are 8 fields:
\begin{equation}\label{1.5}
  \left(
    \begin{array}{c}
      H \\
      h \\
    \end{array}
  \right)=
  \left(
    \begin{array}{cc}
      \cos\alpha & \sin\alpha \\
      \sin\alpha & \cos\alpha \\
    \end{array}
  \right)
  \left(
    \begin{array}{c}
      \rho_{1} \\
      \rho_{2} \\
    \end{array}
  \right)\textmd{ - two neutral scalars (neutral Higgs bosons)}
\end{equation}
\begin{equation}\label{1.6}
  \left(
    \begin{array}{c}
      G^{0} \\
      A \\
    \end{array}
  \right)=
  \left(
    \begin{array}{cc}
      \cos\beta & \sin\beta \\
      \sin\beta & \cos\beta \\
    \end{array}
  \right)
  \left(
    \begin{array}{c}
      \eta_{1} \\
      \eta_{2} \\
    \end{array}
  \right)\textmd{ - two neutral pseudoscalars}
\end{equation}
\begin{equation}\label{1.7}
  \left(
    \begin{array}{c}
      G^{\pm} \\
      H^{\pm} \\
    \end{array}
  \right)=
  \left(
    \begin{array}{cc}
      \cos\beta & \sin\beta \\
      \sin\beta & \cos\beta \\
    \end{array}
  \right)
  \left(
    \begin{array}{c}
      \phi_{1}^{\pm} \\
      \phi_{2}^{\pm} \\
    \end{array}
  \right)\textmd{ - two charged scalars},
\end{equation}
where $G^{0}$ and $G^{\pm}$ are Goldstone bosons which are absorbed as longitudinal components of the $W^{\pm},Z$, $\tan\beta\equiv\frac{v_{2}}{v_{1}}$, $\alpha$ is some angle.

Masses of fermions (quarks and leptons) are result of Yukawa interaction: coupling of left-handed Dirac fields $q_{L}\equiv\left(
                                                                                                      \begin{array}{c}
                                                                                                        u_{L} \\
                                                                                                        d_{L} \\
                                                                                                      \end{array}
                                                                                                    \right), l_{L}\equiv\left(
                                                                                                      \begin{array}{c}
                                                                                                        \nu_{eL} \\
                                                                                                        e_{L} \\
                                                                                                      \end{array}
                                                                                                    \right)$
with right-handed Dirac fields $u_{R},d_{R},e_{R}$ via isospinor fields $\Psi=\Psi_{1},\Psi_{2}$:
\begin{equation}\label{1.8}
  U_{D}=\sqrt{2}\chi_{u}\left(\overline{q}_{L}\widetilde{\Psi}u_{R}+\overline{u}_{R}\widetilde{\Psi}^{+}q_{L}\right)
+\sqrt{2}\chi_{d}\left(\overline{q}_{L}\Psi d_{R}+\overline{d}_{R}\Psi^{+}q_{L}\right)
+\sqrt{2}\chi_{e}\left(\overline{l}_{L}\Psi e_{R}+\overline{e}_{R}\Psi^{+}l_{L}\right),
\end{equation}
where $\psi=\psi^{+}\gamma_{0}$ is Dirac conjugated spinor, $\widetilde{\Psi}=i\tau_{y}\Psi$; $\chi_{u},\chi_{d},\chi_{e}$ are Yukawa constants for $u$-quark, $d$-quark and electron accordingly. Neutrino $\nu_{e}$ remains massless.
Since there are two fields $\Psi_{1},\Psi_{2}$, four options of interaction with fermions ($u,d$ quarks and electron $e$) are possible, which illustrated in Tab.\ref{tab01}. It is analogously for the second $c,s,\mu,\nu_{\mu}$ and third $t,b,\tau,\nu_{\tau}$ generations.

\begin{table}[ht]
\begin{center}
\begin{tabular}{|c|c|c|c|}
  \hline
   & u & d & e  \\
\hline
  Type I & $\Psi_{1}$ & $\Psi_{1}$ & $\Psi_{1}$  \\
\hline
  Type II & $\Psi_{1}$ & $\Psi_{2}$ & $\Psi_{2}$  \\
\hline
  Lepton-specific & $\Psi_{1}$ & $\Psi_{1}$ & $\Psi_{2}$  \\
\hline
  Flipped & $\Psi_{1}$ & $\Psi_{2}$ & $\Psi_{1}$  \\
  \hline
\end{tabular}
\end{center}
\caption{The four independent types of Yukawa interaction for 2HDM scalar doublets.}
  \label{tab01}
\end{table}

Let $m_{11}^{2}<0,m_{22}^{2}>0$, then we should choice the vacuum as $v=v_{1}=246\mathrm{GeV},v_{2}=0$ \cite{wang,keus}, and the expressions for the boson masses take the simple form \cite{wang}:
\begin{equation}\label{1.9}
  m_{h}^{2}=\lambda_{1}v^{2}=-2m_{11}=(126\mathrm{GeV})^{2},\quad m_{H}^{2}=m_{A}^{2}+\lambda_{5}v^{2},\quad
  m_{H^{\pm}}^{2}=m_{22}+\frac{\lambda_{3}}{2}v^{2},\quad m_{A}^{2}=m_{H^{\pm}}^{2}+\frac{\lambda_{4}-\lambda_{5}}{2}v^{2},
\end{equation}
where the $h$-boson is associated with observed Higgs boson. Because of the exact $Z_{2}$ symmetry, the lightest neutral component $H$ or $A$ is stable and may be considered as a DM candidate. If taking $H$ as DM, it requires $\lambda_{5}<0,\lambda_{4}-|\lambda_{5}|<0$. If taking $A$ as DM, it requires $\lambda_{5}>0,\lambda_{4}-\lambda_{5}<0$. However the model requires \cite{wang}:
\begin{equation}\label{1.10}
m_{A}+m_{H}>m_{Z},\quad 2m_{H^{\pm}}>m_{Z},\quad m_{A}+m_{H^{\pm}}>m_{W},\quad m_{H}+m_{H^{\pm}}>m_{W}
\Longrightarrow m_{A},m_{H}\sim10\ldots100\mathrm{GeV}.
\end{equation}
We have seen above that particles that are candidates for the ultra-light DM should have mass $m_{\mathrm{DM}}\sim 10^{-24}\ldots 1\mathrm{eV}$. Obviously, that $H,A$-bosons not suitable for this role.

We can go another way: in Ref.\cite{azev} a model containing two scalar doublets, $\Psi_{1}$ and $\Psi_{2}$, and a real scalar singlet $\Psi_{S}$ with a specific discrete symmetry $\Psi_{1}\rightarrow\Psi_{1},\Psi_{2}\rightarrow -\Psi_{2},\Psi_{S}\rightarrow-\Psi_{S}$ has been constructed:
\begin{eqnarray}\label{1.11}
  V_{\mathrm{2HDM+S}} &=& m_{11}^{2}\Psi_{1}^{+}\Psi_{1}+m_{22}^{2}\Psi_{2}^{+}\Psi_{2}+\frac{1}{2}m_{S}^{2}\Psi_{S}^{2}
+\Psi_{S}(A\Psi_{1}^{+}\Psi_{2}+A^{\ast}\Psi_{2}^{+}\Psi_{1}) \nonumber\\
   &+&\frac{1}{2}\lambda_{1}(\Psi_{1}^{+}\Psi_{1})^{2}+\frac{1}{2}\lambda_{2}(\Psi_{2}^{+}\Psi_{2})^{2}
   +\lambda_{3}(\Psi_{1}^{+}\Psi_{1})(\Psi_{2}^{+}\Psi_{2})\nonumber\\
   &+&\lambda_{4}(\Psi_{1}^{+}\Psi_{2})(\Psi_{2}^{+}\Psi_{1})
   +\frac{1}{2}\lambda_{5}\left((\Psi_{1}^{+}\Psi_{2})^{2}+(\Psi_{2}^{+}\Psi_{1})^{2}\right)\nonumber\\
   &+&\frac{1}{4}\lambda_{6}\Psi_{S}^{4}+\frac{1}{2}\lambda_{7}\Psi_{1}^{+}\Psi_{1}\Psi_{S}^{2}+\frac{1}{2}\lambda_{8}\Psi_{2}^{+}\Psi_{2}\Psi_{S}^{2}.
\end{eqnarray}
All fermion fields are considered as neutral under this symmetry. As such, only the doublet $\Psi_{1}$ couples to fermions. Thus, DM can be attached to 2HDM Lagrangian as excitations of neutral $\Psi_{S}$ field (which does not interact with either fermions or gauge bosons). Thus we can obtain the desired mass of DM by choosing the coefficients $m_{S}^{2},\lambda_{6},\lambda_{7},\lambda_{8}$ accordingly.

As a further generalization the three-Higgs-doublet model (3HDM) can be formulated \cite{keus,darv1}. Maximal symmetry for such model is $U(1)\otimes U(1)$. Corresponding potential $V_{0}$ is invariant under any phase rotation:
\begin{equation}\label{1.13}
  V_{0}=\sum_{i=1}^{3}\left[m_{ii}^{2}\Psi_{i}^{+}\Psi_{i}+\frac{1}{2}\lambda_{ii}(\Psi_{i}^{+}\Psi_{i})^{2}\right]
+\sum_{i=1,i\neq j}^{3}\left[\lambda_{ij}(\Psi_{i}^{+}\Psi_{i})(\Psi_{j}^{+}\Psi_{j})
+\lambda_{ij}'(\Psi_{i}^{+}\Psi_{j})(\Psi_{j}^{+}\Psi_{i})\right].
\end{equation}
This potential gives three massive neutral scalars $H_{1,2,3}$, two massive charged scalars $H_{1,2}^{\pm}$ and one massless charged scalar $H_{3}^{\pm}$, two massive neutral pseudo-scalars $A_{1,2}$ and one massless neutral pseudo-scalar $A_{3}$. In general case the 3HDM potential symmetric under a group $G$ can be written as
\begin{equation}\label{1.14}
  V_{0}=V_{0}+V_{G},
\end{equation}
where $V_{G}$ is a collection of extra terms ensuring the symmetry group $G$, which can be both continuous and discrete symmetries, both Abelian and non-Abelian symmetries. Classifications of symmetric 3HDM potentials and corresponding Higgs and Goldstone particles is presented in a review \cite{keus}. For clarity, in Appendix \ref{symm} we present some invariant potentials under the simplest transformations. Finally, a $n$-Higgs-doublet model ($n$HDM, $n>3$) can be formulated \cite{darv2}, where the number of scalar, pseudoscalar and charged bosons will be even larger. The lightest of the neutral massive bosons ($H$, $A$ or $S$ types) can be candidate for role of DM.

\emph{Despite the fact that in the models $n$HDM or $n$HDM+S ($n\geq 2$) and in many other the particles-candidates for the role of DM appear (the lightest of massive neutral $H$, $A$ or $S$-bosons), these models generate a lot of other particles (which can be number in tens in multiplets). These particles can be both electrically neutral and charged, both massless and massive, and have not yet been detected in collider experiments or in cosmic rays. In addition, even the proposed DM particles are weakly interacting (as WIMPs) also, that is, they are not completely sterile, hence could have been detected too. In the future, with more in-depth research, the discovery of these particles, of course, cannot be ruled out.}

\begin{figure}[ht]
\begin{center}
\includegraphics[width=12cm]{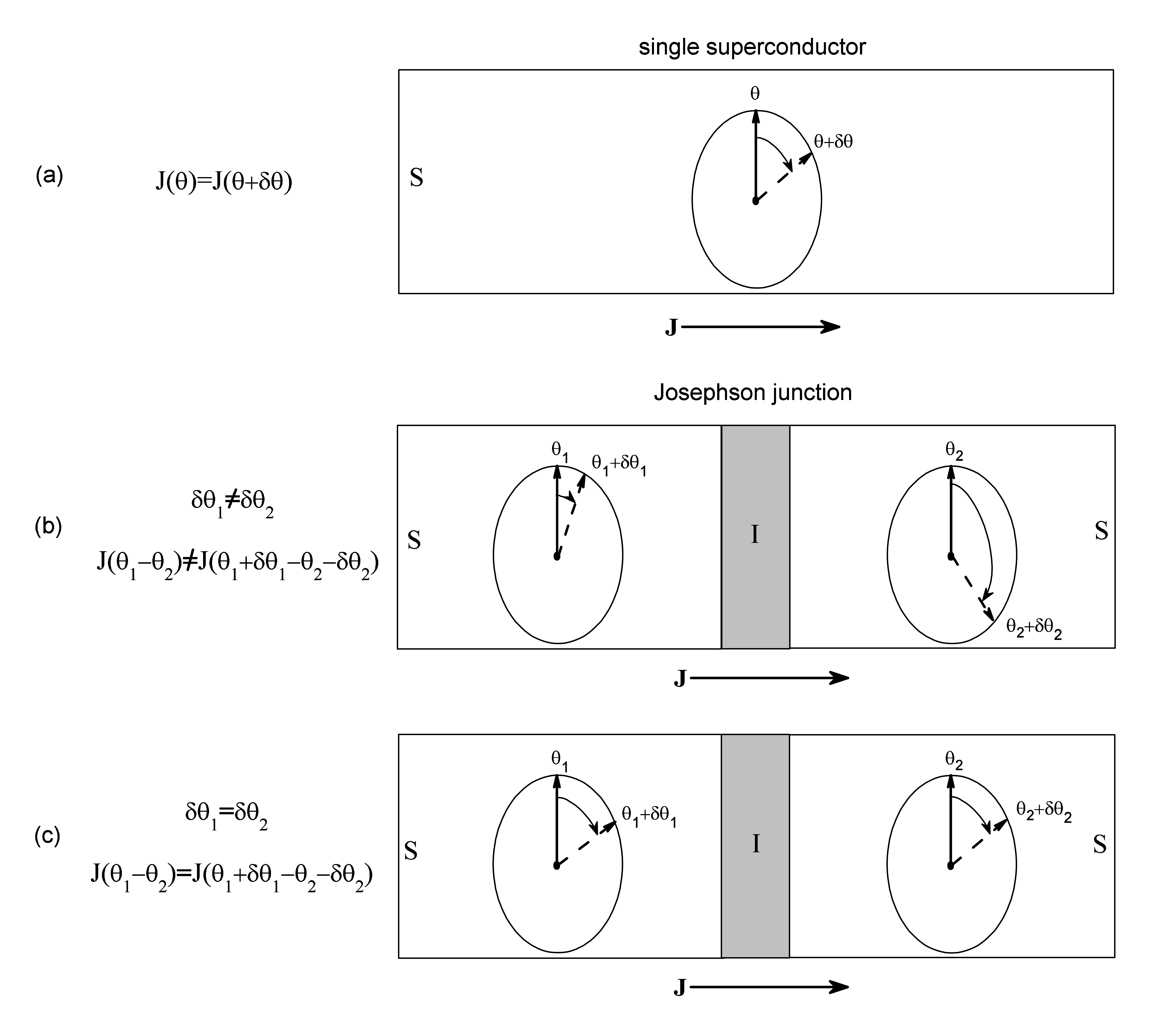}
\end{center}
\caption{(a) - $U(1)$ symmetry of one-piece superconducting sample: any phase rotations do not change current $J$. (b) - independent phase rotations in superconductors separated by a thin insulator with thickness of the order of coherence length (S-I-S Josephson junction) changes the current through the junction. (c) - synchronous phase rotations (so that $\theta_{2}-\theta_{1}=\mathrm{const}$) do not change the current.}
\label{Fig03}
\end{figure}

Historically, GWS theory arose as a field-theoretic, dynamic, relativistic, group (from $U(1)$ symmetry to $SU(2)\otimes U(1)$ symmetry) generalization of the Ginzburg-Landau (GL) theory for superconductors. Attractive forces act between electrons with opposite spins due to the exchange of phonons, overpowering Coulomb repulsion. As a result, electrons bind into effective pairs (so-called Cooper pairs), which at low temperatures condense into the same quantum state (similar to a Bose-Einstein condensate). The resulting coherent state of a collective of Cooper pairs can be described with the many-particles wave function:
\begin{equation}\label{1.15}
  \varphi(\mathbf{r})=|\varphi(\mathbf{r})|e^{\mathrm{i}\theta(\mathbf{r})},
\end{equation}
where both module $|\varphi|$ and phase $\theta$ are functions of spatial coordinates $\mathbf{r}$, moreover the module determines density of superconducting electrons $n_{s}=2|\varphi|^{2}$, and gradient of the phase determines current $\mathbf{J}=\frac{e\hbar}{m}|\varphi|^{2}\nabla\theta$.
Density of free energy is
\begin{equation}\label{1.16}
  \mathcal{F}=\frac{\hbar^{2}}{4m}\left(\nabla-\frac{\mathrm{i}2e}{\hbar c}\mathbf{A}\right)\varphi
    \left(\nabla+\frac{\mathrm{i}2e}{\hbar c}\mathbf{A}\right)\varphi^{+}
    +a\left|\varphi\right|^{2}+\frac{b}{2}\left|\varphi\right|^{4}+\frac{(\nabla\times\mathbf{A})^{2}}{8\pi},
\end{equation}
where $a<0$, $b>0$, $\mathbf{A}$ is a vector-potential of magnetic field, $2m$ and $2e$ are mass and charge of a Cooper pair accordingly. Then the current is
\begin{equation}\label{1.17}
  \mathbf{J}=\frac{e\hbar}{m}\varphi_{0}^{2}\left(\nabla\theta-\frac{2e}{\hbar c}\mathbf{A}\right),
\end{equation}
where
\begin{equation}\label{1.18}
  \varphi_{0}=\sqrt{\frac{-a}{b}}
\end{equation}
is an equilibrium magnitude of module of the field $\varphi$. Free energy (\ref{1.16}) and current (\ref{1.17}) are invariants under the gauge $U(1)$ transformation, i.e when the phase is rotated by a certain angle $\delta\theta$: $\theta\rightarrow\theta+\delta\theta$, which is function of a point $\delta\theta(\mathbf{r})$ in general case:
\begin{eqnarray}
  \mathcal{F}\left(\varphi\rightarrow\varphi e^{\mathrm{i}\delta\theta},
\varphi^{+}\rightarrow\varphi^{+}e^{-\mathrm{i}\delta\theta},\mathbf{A}\rightarrow\mathbf{A}+\frac{\hbar c}{2e}\nabla\delta\theta\right) &=& \mathcal{F}\left(\varphi,\varphi^{+},\mathbf{A}\right)\label{1.19} \\
  \mathbf{J}\left(\theta\rightarrow\theta+\delta\theta,\mathbf{A}\rightarrow\mathbf{A}+\frac{\hbar c}{2e}\nabla\delta\theta\right) &=& \mathbf{J}\left(\theta,\mathbf{A}\right)\label{1.20}
\end{eqnarray}
This means, that any phase rotations do not change either the energy of the system or the current flowing through the superconductor. This symmetry is illustrated schematically in Fig.\ref{Fig03}a. Moreover, equation for magnetic field has the form (in the gauge $\nabla\cdot\mathbf{A}=0$):
\begin{equation}\label{1.21}
  \nabla\times\frac{\partial\mathcal{F}}{\partial(\nabla\times\mathbf{A})}-\frac{\partial\mathcal{F}}{\partial\mathbf{A}}=0
\Rightarrow\Delta\mathbf{A}=\frac{8\pi e^{2}\varphi_{0}^{2}}{mc^{2}}\mathbf{A}\equiv\frac{1}{\lambda^{2}}\mathbf{A}
\propto m_{A}^{2}\mathbf{A},
\end{equation}
where value reciprocal of the magnetic penetration depth $\lambda$ plays role of mass of a photon $m_{A}$. Dynamics generalization of GL theory has been done in Ref.\cite{grig1}, where has been demonstrated, that Higgs mass in such system is
\begin{equation}\label{1.22}
  m_{H}=\sqrt{2}\kappa m_{A}\propto\frac{1}{\xi},
\end{equation}
where $\kappa\equiv\lambda/\xi$ is a GL parameter, $\xi$ is a coherence length. Then for type-I superconductors $m_{H}<m_{A}$, and for type-II superconductors $m_{H}>m_{A}$.

Now, let us cut our superconductor into two parts and space them far apart. We obtain two independent condensates:
\begin{equation}\label{1.23}
  \varphi_{1}=|\varphi_{1}|e^{\mathrm{i}\theta_{1}},\quad\varphi_{2}=|\varphi_{2}|e^{\mathrm{i}\theta_{2}}.
\end{equation}
Then, let us bring them closer to a distance on the order of the coherence length $\xi\sim 1/m_{H}$. The remaining slit can be filled, for example, with some insulator as demonstrated in Fig.\ref{Fig03}b,c. A Copper pair from the bank 1 with condensate $\varphi_{1}$ can tunnel to the bank 2 with condensate $\varphi_{2}$, which described with nondiagonal matrix elements \cite{schmidt}:
\begin{equation}\label{1.24}
  H_{12}=\int\varphi_{1}^{+}\widehat{H}\varphi_{2}\mathrm{d}V,\quad H_{21}=\int\varphi_{2}^{+}\widehat{H}\varphi_{1}\mathrm{d}V
  ,\quad K\equiv |H_{12}|=|H_{21}|.
\end{equation}
The value $K$ is determined by properties of the junction. Such device is called as Josephson junction, and matrix elements (\ref{1.24}) are  Josephson coupling. Then current through the junction is
\begin{equation}\label{1.25}
  J=\frac{4K\varphi_{0}^{2}}{\hbar}\sin(\theta_{2}-\theta_{1}).
\end{equation}
It is not difficult to see, that Josephson coupling breaks the $U(1)$ gauge invariance, because the current (\ref{1.25}) depends on the phase differences $\theta_{2}-\theta_{1}$. Thus, if we rotate phases $\theta_{1}$ and $\theta_{2}$ in each bank independently, then the current $J$ changes. In order to keep the current constant we must rotate the phases synchronously, i.e so that $\theta_{2}-\theta_{1}=\mathrm{const}$.

The Josephson junction can be realised in the momentum space also: if in some material two conduction bands take place (for example, in magnesium diboride  $\mathrm{MgB}_{2}$, nonmagnetic borocarbides $\mathrm{LuNi_{2}B_{2}C}$, $\mathrm{YNi_{2}B_{2}C}$ and some oxypnictide compounds), then
in each band the condensate of Cooper pairs can exist: $\varphi_{1}$ and $\varphi_{2}$ accordingly. In a bulk isotropic s-wave superconductor the GL free energy functional can be written as
\cite{grig0,asker7,asker2,yerin1,grig2,grig3}:
\begin{eqnarray}\label{1.26}
    F&=&\int d^{3}r[\frac{\hbar^{2}}{4m_{1}}\left|\nabla\varphi_{1}\right|^{2}+\frac{\hbar^{2}}{4m_{2}}\left|\nabla\varphi_{2}\right|^{2}\nonumber\\
    &+&a_{1}\left|\varphi_{1}\right|^{2}+a_{2}\left|\varphi_{2}\right|^{2}+\frac{b_{1}}{2}\left|\varphi_{1}\right|^{4}+\frac{b_{2}}{2}\left|\varphi_{2}\right|^{4}
    +\epsilon\left(\varphi_{1}^{+}\varphi_{2}+\varphi_{1}\varphi_{2}^{+}\right)],
\end{eqnarray}
where $m_{1,2}$ denote the effective mass of carriers in the corresponding band, the coefficients $a_{1,2}$ are given as $a_{i}=\gamma_{i}(T-T_{ci})$ where $\gamma_{i}$ are some constants, the coefficients $b_{1,2}$ are independent of temperature, the quantity $\epsilon$ describes interband mixing of the two condensate: proximity effect or internal Josephson effect. If we switch off the interband interaction $\epsilon=0$, then we will have two independent superconductors with different critical temperatures $T_{c1}$ and $T_{c2}$, because the intraband interactions can be different. Thus, a two-band superconductor is understood as two single-band superconductors with the corresponding condensates of Cooper pairs $\varphi_{1}$ and $\varphi_{2}$ (so that densities of superconducting electrons are $n_{\mathrm{s}1}=2|\varphi_{1}|^{2}$ and $n_{\mathrm{s}2}=2|\varphi_{2}|^{2}$ accordingly), but these two condensates are coupled by the internal proximity effect $\epsilon\left(\varphi_{1}^{+}\varphi_{2}+\varphi_{1}\varphi_{2}^{+}\right)$.

\begin{figure}[ht]
\begin{center}
\includegraphics[width=9cm]{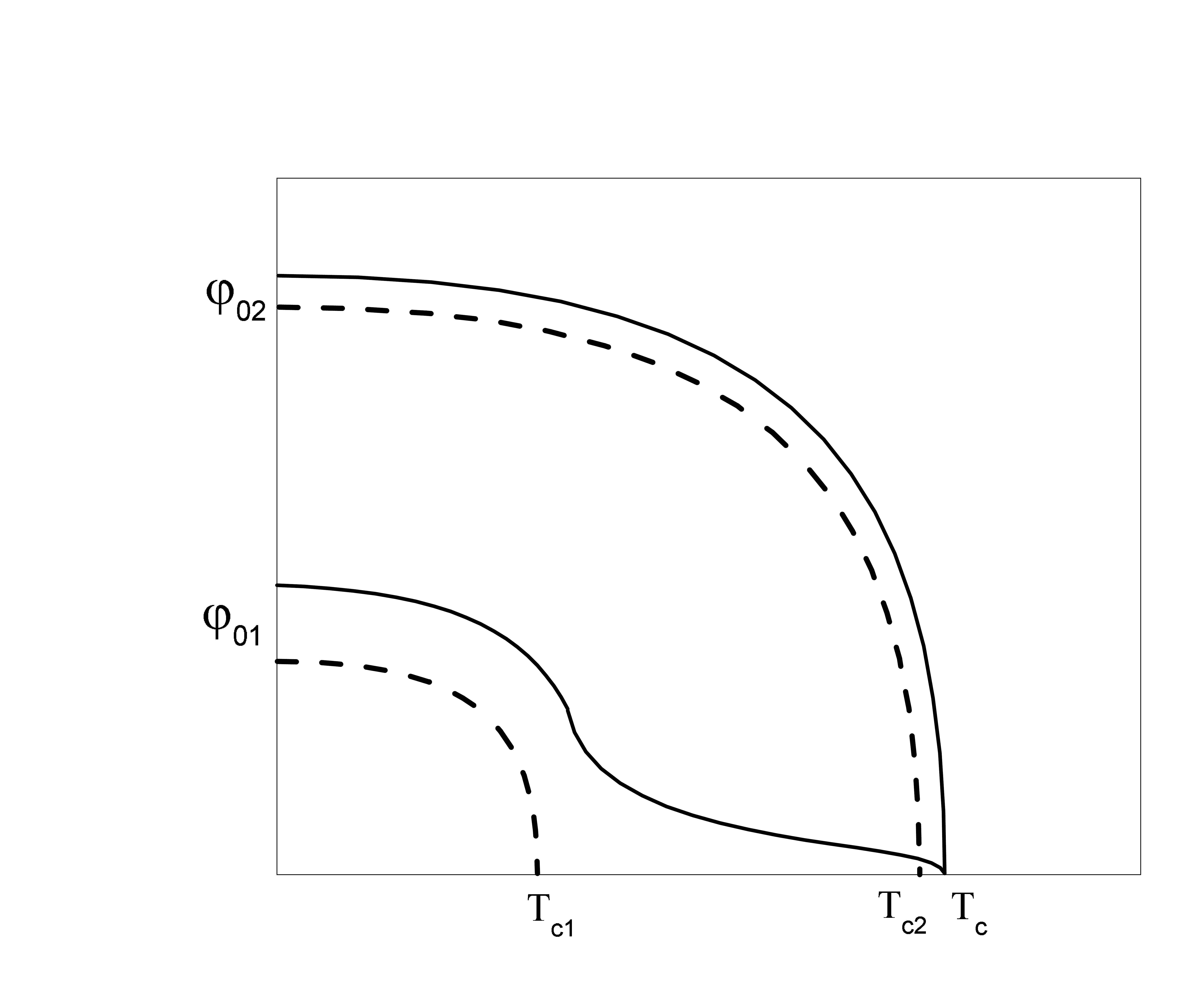}
\end{center}
\caption{The amplitudes of the condensates $\varphi_{01}(T)$ and $\varphi_{02}(T)$ as solutions of Eq.(\ref{1.27}), if the interband coupling is absent, i.e. $\epsilon=0$ (dash lines), and if the interband coupling is weak, i.e. $\epsilon\neq 0$, $|\epsilon|\ll |a_{1}(0)|$ (solid lines). The applying of the weak interband coupling drags the smaller parameter $\varphi_{01}$ up to new critical temperature $T_{c}\gg T_{c1}$. The effect on the larger parameter $\varphi_{02}$ is not so significant.}
\label{Fig04}
\end{figure}

Minimization of the free energy functional with respect to the amplitudes of condensates, if $\nabla\varphi_{1,2}=0$, gives
\begin{equation}\label{1.27}
\left\{\begin{array}{c}
  a_{1}\varphi_{1}+\epsilon\varphi_{2}+b_{1}\varphi_{1}^{3}=0 \\
  a_{2}\varphi_{2}+\epsilon\varphi_{1}+b_{2}\varphi_{2}^{3}=0 \\
\end{array}\right\},
\end{equation}
where the equilibrium values $\varphi_{1,2}$ are assumed to be real (i.e. the phases $\theta_{1,2}$ are $0$ or $\pi$) in absence of current and magnetic field. Near the critical temperature $T_{c}$ we have $\varphi_{1,2}^{3}\rightarrow 0$, hence, we can find the critical temperature equating to zero the determinant of the linearized system (\ref{1.27}):
\begin{equation}\label{1.28}
a_{1}a_{2}-\epsilon^{2}=\gamma_{1}\gamma_{2}(T_{c}-T_{c1})(T_{c}-T_{c2})-\epsilon^{2}=0.
\end{equation}
Solving this equation, we find $T_{c}>T_{c1},T_{c2}$, moreover, the solution does not depend on the sign of $\epsilon$. The sign determines the equilibrium phase difference of the condensates $|\varphi_{1}|e^{i\theta_{1}}$ and $|\varphi_{2}|e^{i\theta_{2}}$:
\begin{equation}\label{1.29}
    \begin{array}{cc}
      \cos(\theta_{1}-\theta_{2})=1 & \mathrm{if}\quad\epsilon<0  \\
      \cos(\theta_{1}-\theta_{2})=-1 & \mathrm{if}\quad\epsilon>0 \\
    \end{array},
\end{equation}
that follows from Eq.(\ref{1.27}). The case $\epsilon<0$ corresponds to an attractive interband interaction (for example, in $\mathrm{MgB}_{2}$, where $s^{++}$ wave symmetry occurs), the case $\epsilon>0$ corresponds to a repulsive interband interaction (for example, in iron-based superconductors, where $s^{+-}$ wave symmetry occurs) \cite{asker7}. The solutions of Eq.(\ref{1.27}) $\varphi_{01},\varphi_{02}$ are illustrated in Fig.\ref{Fig04} for the case of strongly asymmetrical bands $T_{c1}\ll T_{c2}$. We can see, that the effect of interband coupling $\epsilon\neq 0$, even if the coupling is weak $|\epsilon|\ll |a_{1}(0)|$, is \emph{nonperturbative}: applying of the interband coupling drags the smaller parameter $\varphi_{01}$ up to the new critical temperature $T_{c}\gg T_{c1}$. At the same time, the effect on the larger parameter $\varphi_{2}$ is not so significant - applying of the interband coupling only slightly increases the critical temperature compared with $T_{c2}$: $T_{c}\gtrsim T_{c2}$.

In the module-phase representation (\ref{1.23}) the interband mixing takes the form:
\begin{equation}\label{1.30}
  \epsilon\left(\varphi_{1}^{+}\varphi_{2}+\varphi_{1}\varphi_{2}^{+}\right)=
2\epsilon|\varphi_{1}||\varphi_{2}|\cos(\theta_{1}-\theta_{2}).
\end{equation}
Thus, the Josephson term describes interference between the Cooper pairs condensates $\varphi_{1}$ and $\varphi_{2}$. As in the Josephson junction,  the Josephson term breaks the $U(1)$ gauge invariance, because this term depends on the phase differences $\theta_{1}-\theta_{2}$. In Ref.\cite{grig2} it has been investigated normal oscillations of internal degrees of freedom (Higgs mode and Goldstone mode) of two-band superconductors using the dynamical generalization of GL theory, which has been formulated in Ref.\cite{grig1}. It is demonstrated, that, due to the internal proximity effect, the Goldstone modes from each band transform into normal oscillations for all bands: common mode oscillations with acoustic spectrum, which are absorbed by the gauge field because propagation of these collective excitations is accompanied by current; and anti-phase oscillations with an energy gap in spectrum (mass) determined with the interband coupling $m_{L}\sim\sqrt{|\epsilon|}$, which can be associated with the Leggett mode. Propagation of the Leggett mode is not accompanied by current, hence this mode "survives". Analogously, for three-band superconductors \cite{grig3}, it has been demonstrated, that the Goldstone modes from each band transform to normal oscillations for all bands: common mode oscillations with acoustic spectrum, which are absorbed by the gauge field, and two massive modes for anti-phase oscillations which are analogous to the Leggett mode and determined with the coefficients of interband coupling $\epsilon_{12},\epsilon_{13},\epsilon_{23}$.

The free energy functional $F=\int d^{3}r\mathcal{F}$ can be written in general $n$-band system, where potential has the form:
\begin{equation}\label{1.31}
  V=V_{0}+\sum_{i<k}^{n}\epsilon_{ik}\left(\varphi_{i}^{+}\varphi_{k}+\varphi_{i}\varphi_{k}^{+}\right),
\end{equation}
where the potential
\begin{equation}\label{1.32}
  V_{0}=\sum_{i=1}^{n}a_{i}\left|\varphi_{i}\right|^{2}+\frac{b_{i}}{2}\left|\varphi_{i}\right|^{4}
\end{equation}
is a sum of independent potentials of each condensate.  Potential $V_{0}$ is invariant under any phase rotation. Since the condensates in three-band system are coupled by the Josephson terms $\epsilon_{ik}\left(\varphi_{i}^{+}\varphi_{k}+\varphi_{i}\varphi_{k}^{+}\right)=\epsilon_{ik}|\varphi_{i}||\varphi_{k}|\cos(\theta_{i}-\theta_{k})$, the spontaneously broken $U(1)$ symmetry of the ground state is shared throughout the system: the presence of the condensate $\langle\varphi_{i}\rangle\neq 0$ in some band induces the condensation in other bands $\langle\varphi_{k}\rangle\neq 0$, that is the internal proximity effect takes place. At the same time, global gauge symmetry $U(1)^{n-1}$ of the potential $V_{0}$ with $n>1$ is broken down by the Josephson terms \cite{keus}, because these terms depend on the phase differences $\theta_{i}-\theta_{k}$.  In the $n$-band case, we have $n-1$
phase-difference modes (Leggett) modes. These modes acquire masses because the phase differences are fixed near minimums of the  potential $V$. In Ref.\cite{hase} the total rule has been formulated: in the $n$-band system the global symmetry $U(1)^{n-1}$ is broken down by the Josephson terms to $U(1)^{n-3}$ symmetry. Thus, in $n>3$-band system $n-3$ massless Leggett modes must be present. Ultimately, the system with potential (\ref{1.31}) is invariant under synchronic $U(1)$ gauge transformation, i.e. when each scalar field is turned by the same phase $\theta$: $\varphi_{k}\rightarrow\varphi_{k}e^{i\theta}$. Hence, as demonstrated for two- and three-band superconductors in Ref.\cite{grig2,grig3}, the common mode phase oscillations are absorbed by the gauge field, however oscillations of the phase differences $\theta_{i}-\theta_{k}$ occur.

Proceeding from aforesaid, we can use the analogy with multi-band superconductors to formulate the appropriate extension of SM, formalizing the superconducting order parameter $\varphi$ as a scalar field. Such model allows obtain particle-candidates for role of DM - analog of Leggett modes, because 1) masses of these bosons can be arbitrarily small due to nonperturbativness of interband coupling $m_{L}\sim\sqrt{|\epsilon|}$, 2) Since propagation of the Leggett mode is not accompanied by current, then they can be "sterile" in the field theory. However, symmetry of the GL free energy is $U(1)_{Q}$, but symmetry of GWS Lagrangian is $SU(2)_{I}\otimes U(1)_{Y}$. Accordingly, instead the scalar field $\varphi$ we have isospinor $\Psi$ similar to Eq.(\ref{1.4}). Hence, we must try to represent the interband coupling $\epsilon\left(\Psi_{1}^{+}\Psi_{2}+\Psi_{1}\Psi_{2}^{+}\right)$ in the form of interference between the fields $\Psi_{1}$ and $\Psi_{2}$, similar to Eq.(\ref{1.30}). Then, we can suppose the coefficients $\lambda_{n>2}=0$ in  Lagrangians (\ref{1.3},\ref{1.11}) or the coefficients $\lambda_{i\neq j}=0$ in Lagrangian (\ref{1.13}). This approach relieves us of large number of other particles (for example, charged Higgs bosons $H^{\pm}$) which could be easily detected experimentally. However, the sense of formulation of the model different from SM is not so much in solving the DM problem, but in solving a whole complex of problems. Thus, except the DM problem, we propose nature of oscillations and masses of neutrinos, leaving them as Dirac fermions. At the same time we demonstrate, why oscillations of charged leptons (electron-muon-tauon) are absent and why masses of such leptons differ by orders and why are three fermion generations are needed. The model proposes three neutral H-bosons, that explains the absence of experimentally detected decays of the already discovered H-boson into fermions of the second and first generations, but these two additional H-bosons very weekly interact with gauge and Dirac fields, that makes their detecting difficultly, but still possible, that can be an experimental test.

Our paper is organized by the following way. In Sect.\ref{spontanU1} we formulate model with three scalar fields (bands) with spontaneous breaking of $U(1)$ gauge symmetry in each and with the Josephson couplings between them. In such system we obtain both Higgs and Goldstone modes and introduce concept of band states and flavour states of the scalar fields. In Sect.\ref{spontanU1gauge} the Higgs effect on abelian (electromagnetic) field in the three-band system is considered. In Sect.\ref{dirac} we connect the three-bandness with three generations of fermions, and we consider the band states and flavour states of Dirac fields. In Sect.\ref{spontanSU2} and Sect.\ref{spontanSU2U1} we consider the three-band system with spontaneous breaking of $SU(2)_{I}$ and $SU(2)_{I}\otimes U(1)_{Y}$ gauge symmetries accordingly, and with the Josephson couplings between bands. The Higgs effect on both abelian and Yang–Mills gauge fields is considered. In Sect.\ref{mixing} the lepton mixing  is described and mechanism of origin of neutrino "masses" is proposed. In Sect.\ref{particle} we summarize results of the three-band GWS model as the systematic of elementary particles, where the particles, that make up DM, are present. Moreover we propose additional two neutral H-bosons also, estimate their masses and analyzes their productions and decays. Mechanism of the fermions mass hierarchy is proposed. In Sect.\ref{darkmatter} we estimate masses of L-bosons as DM particles and demonstrate, that such ultra-light bosons solve the central cusp problem. In Sect.\ref{higgsTc} we consider the masses of H-bosons at critical temperature.

\section{Spontaneous breaking of $U(1)$ gauge symmetry in the three-band system with the Josephson couplings}\label{spontanU1}
\subsection{The three-band Lagrangian with the Josephson terms}

Let we have three complex scalar fields, which are equivalent to two real scalar fields each: modulus $\left|\varphi(x)\right|$ and phase $\theta(x)$ (the modulus-phase representation):
\begin{equation}\label{2.1}
    \varphi_{1}(x)=\left|\varphi_{1}(x)\right|e^{\mathrm{i}\theta_{1}(x)}, \quad\varphi_{2}(x)=\left|\varphi_{2}(x)\right|e^{\mathrm{i}\theta_{2}(x)},
    \quad\varphi_{3}(x)=\left|\varphi_{2}(x)\right|e^{\mathrm{i}\theta_{3}(x)}.
\end{equation}
Here $x\equiv(t,\textbf{r})$, and we will use the system of units, where $c=\hbar=1$. This fields should minimize some action $S$ in the Minkowski space:
\begin{equation}\label{2.2}
    S=\int\mathcal{L}(\varphi_{1},\varphi_{2},\varphi_{3},\varphi^{+}_{1},\varphi^{+}_{2},\varphi^{+}_{3})\mathrm{d}^{4}x,
\end{equation}
where the Lagrangian $\mathcal{L}$ is a sum of three gauge-invariant Lagrangians (ordinary single-band Lagrangians) and Josephson terms (the interband two-by-two coupling of the scalar fields $\varphi_{i}\varphi_{j}^{+}+\varphi_{i}^{+}\varphi_{j}$):
\begin{eqnarray}\label{2.3}
    \mathcal{L}&=&\partial_{\mu}\varphi_{1}\partial^{\mu}\varphi_{1}^{+}+\partial_{\mu}\varphi_{2}\partial^{\mu}\varphi_{2}^{+}
    +\partial_{\mu}\varphi_{3}\partial^{\mu}\varphi_{3}^{+}\nonumber\\
    &-&a_{1}\left|\varphi_{1}\right|^{2}-a_{2}\left|\varphi_{2}\right|^{2}-a_{3}\left|\varphi_{3}\right|^{2}
    -\frac{b_{1}}{2}\left|\varphi_{1}\right|^{4}-\frac{b_{2}}{2}\left|\varphi_{2}\right|^{4}-\frac{b_{3}}{2}\left|\varphi_{3}\right|^{4}
    \nonumber\\
    &-&\epsilon\left(\varphi_{1}^{+}\varphi_{2}+\varphi_{1}\varphi_{2}^{+}\right)-\epsilon\left(\varphi_{1}^{+}\varphi_{3}+\varphi_{1}\varphi_{3}^{+}\right)
    -\epsilon\left(\varphi_{2}^{+}\varphi_{3}+\varphi_{2}\varphi_{3}^{+}\right),
\end{eqnarray}
where $\partial_{\mu}\equiv\frac{\partial}{\partial x^{\mu}}\equiv\left(\frac{\partial}{\partial t},\nabla\right),
\quad\partial^{\mu}\equiv\frac{\partial}{\partial x_{\mu}}\equiv\left(\frac{\partial}{\partial t},-\nabla\right)$ are covariant and contravariant differential operators accordingly.  The coefficients $a_{1,2,3}<0$ and the coefficients $b_{1,2,3}>0$ belong to the corresponding band. The case $\epsilon<0$ corresponds to attractive interband interaction, the case $\epsilon>0$ corresponds to repulsive interband interaction. If we switch off the interband interaction $\epsilon=0$, then we will have three independent scalar field $\varphi_{i}$. It should be noted, that the considered model is similar to 3HDM \cite{keus}, but without any specific symmetry in the sense of Appendix \ref{symm}, except symmetry under the synchronic $U(1)$-transformation:
\begin{equation}\label{2.3a}
 \mathcal{L}\left(\varphi_{1}\rightarrow\varphi_{1}e^{\mathrm{i}\delta\theta}
,\varphi_{2}\rightarrow\varphi_{2}e^{\mathrm{i}\delta\theta},\varphi_{3}\rightarrow\varphi_{3}e^{\mathrm{i}\delta\theta}\right)
 = \mathcal{L}\left(\varphi_{1},\varphi_{2},\varphi_{3}\right),
\end{equation}
i.e all phases $\theta_{1},\theta_{2},\theta_{3}$ must be rotated equally, so that $\theta_{2}-\theta_{1}=\mathrm{const},\theta_{3}-\theta_{1}=\mathrm{const},\theta_{3}-\theta_{1}=\mathrm{const}$.

Lagrange equations for functional (\ref{2.2}) are
\begin{eqnarray}\label{2.4}
    &&\partial^{\mu}\partial_{\mu}\varphi_{1}+a_{1}\varphi_{1}+\epsilon\varphi_{2}+\epsilon\varphi_{3}+b_{1}\left|\varphi_{1}\right|^{2}\varphi_{1}=0\nonumber\\
    &&\partial^{\mu}\partial_{\mu}\varphi_{2}+a_{2}\varphi_{2}+\epsilon\varphi_{1}+\epsilon\varphi_{3}+b_{2}\left|\varphi_{2}\right|^{2}\varphi_{2}=0\\
    &&\partial^{\mu}\partial_{\mu}\varphi_{3}+a_{3}\varphi_{3}+\epsilon\varphi_{1}+\epsilon\varphi_{2}+b_{3}\left|\varphi_{3}\right|^{2}\varphi_{3}=0,\nonumber
\end{eqnarray}
where $\partial^{\mu}\partial_{\mu}=\partial_{\mu}\partial^{\mu}=\frac{\partial^{2}}{\partial t^{2}}-\Delta$. The current for such Lagrangian is
\begin{equation}\label{2.5}
  J^{\mu}=\sum_{j=1}^{3}\frac{\partial\mathcal{L}}{\partial(\partial_{\mu}\varphi_{j})}(-\mathrm{i}\varphi_{j})
  +\frac{\partial\mathcal{L}}{\partial(\partial_{\mu}\varphi_{j}^{+})}(\mathrm{i}\varphi_{j}^{+})
  =\mathrm{i}\sum_{j=1}^{3}\left(\varphi_{j}^{+}\partial^{\mu}\varphi_{j}-\varphi_{j}\partial^{\mu}\varphi_{j}^{+}\right)
=-2\sum_{j=1}^{3}|\varphi_{j}|^{2}\partial^{\mu}\theta_{j},
\end{equation}
where we have used the modulus-phase representation (\ref{2.1}). Using equations of motion (\ref{2.4}) it can be shown that $\partial_{\mu}J^{\mu}=0$.

\begin{figure}[ht]
\begin{center}
\includegraphics[width=9cm]{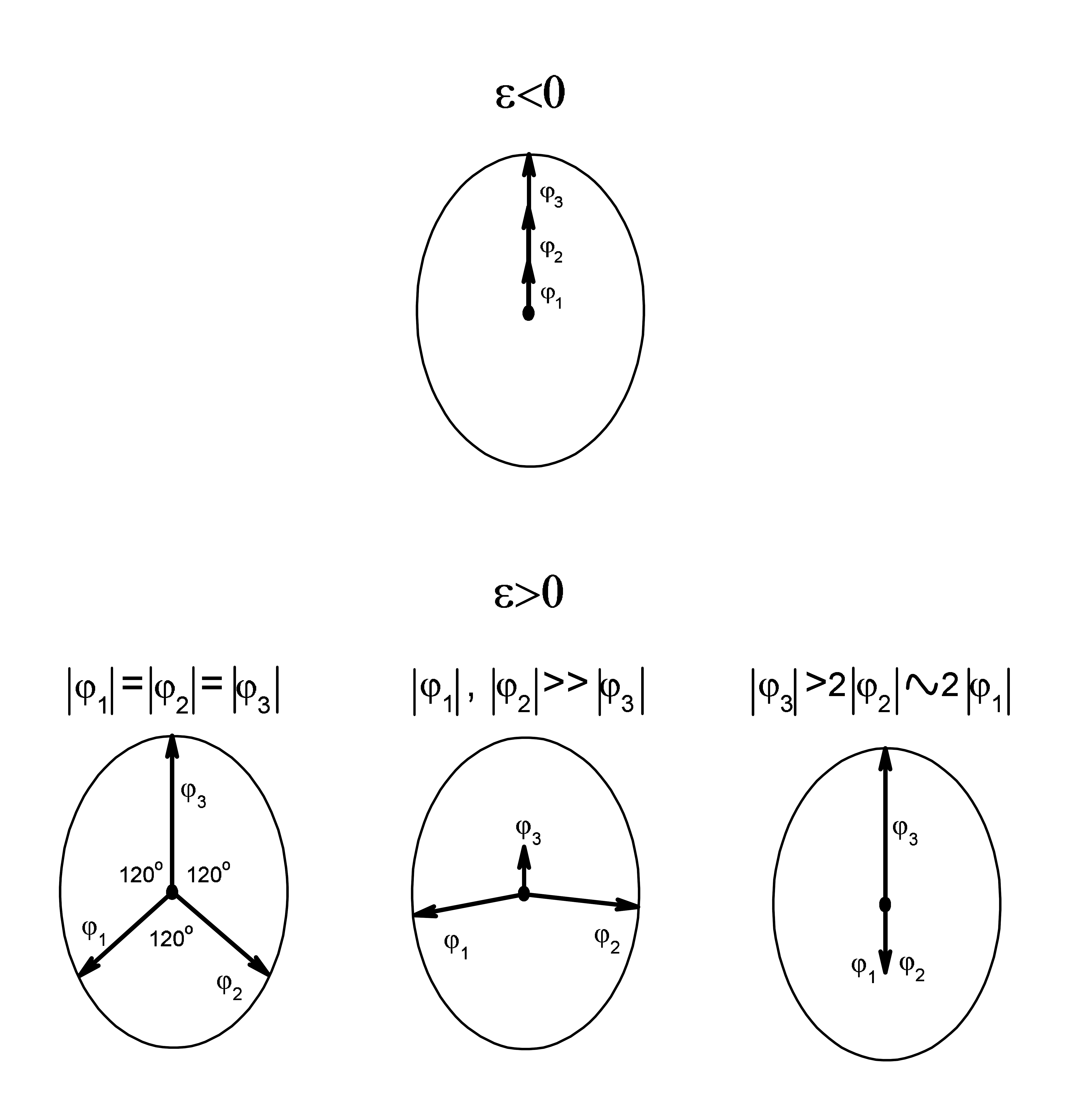}
\end{center}
\caption{The possible configurations of mutual arrangement of the scalar fields $\varphi_{1},\varphi_{2},\varphi_{3}$ corresponding to some limit cases as solutions of Eq.(\ref{2.9})}
\label{Fig1}
\end{figure}

Let us consider stationary and spatially homogeneous case, i.e. $\partial_{t}\varphi=0,\nabla\varphi=0$. Then from Eqs.(\ref{2.4}) we obtain:
\begin{equation}\label{2.7}
\left\{\begin{array}{c}
  a_{1}\varphi_{1}+\epsilon\varphi_{2}+\epsilon\varphi_{3}+b_{1}|\varphi_{1}|^{2}\varphi_{1}=0 \\
  a_{2}\varphi_{2}+\epsilon\varphi_{1}+\epsilon\varphi_{3}+b_{2}|\varphi_{2}|^{2}\varphi_{2}=0 \\
  a_{3}\varphi_{3}+\epsilon\varphi_{1}+\epsilon\varphi_{2}+b_{3}|\varphi_{3}|^{2}\varphi_{3}=0 \\
\end{array}\right\},
\end{equation}
which can be rewritten in the form:
\begin{equation}\label{2.8}
\left\{\begin{array}{c}
  a_{1}|\varphi_{1}|+\epsilon|\varphi_{2}|e^{\mathrm{i}(\theta_{2}-\theta_{1})}+\epsilon|\varphi_{3}|e^{\mathrm{i}(\theta_{3}-\theta_{1})}+b_{1}|\varphi_{1}|^{3}=0 \\
  a_{2}|\varphi_{2}|+\epsilon|\varphi_{1}|e^{\mathrm{i}(\theta_{1}-\theta_{2})}+\epsilon|\varphi_{3}|e^{\mathrm{i}(\theta_{3}-\theta_{2})}+b_{2}|\varphi_{2}|^{3}=0 \\
  a_{3}|\varphi_{3}|+\epsilon|\varphi_{1}|e^{\mathrm{i}(\theta_{1}-\theta_{3})}+\epsilon|\varphi_{2}|e^{\mathrm{i}(\theta_{2}-\theta_{3})}+b_{3}|\varphi_{3}|^{3}=0 \\
\end{array}\right\},
\end{equation}
or in an expanded form:
\begin{equation}\label{2.9}
\left\{\begin{array}{c}
  a_{1}|\varphi_{1}|+\epsilon|\varphi_{2}|\cos(\theta_{2}-\theta_{1})+\epsilon|\varphi_{3}|\cos(\theta_{3}-\theta_{1})+b_{1}|\varphi_{1}|^{3}=0 \\
  a_{2}|\varphi_{2}|+\epsilon|\varphi_{1}|\cos(\theta_{1}-\theta_{2})+\epsilon|\varphi_{3}|\cos(\theta_{3}-\theta_{2})+b_{2}|\varphi_{2}|^{3}=0 \\
  a_{3}|\varphi_{3}|+\epsilon|\varphi_{1}|\cos(\theta_{1}-\theta_{3})+\epsilon|\varphi_{2}|\cos(\theta_{2}-\theta_{3})+b_{3}|\varphi_{3}|^{3}=0 \\
  |\varphi_{2}|\sin(\theta_{2}-\theta_{1})+|\varphi_{3}|\sin(\theta_{3}-\theta_{1})=0 \\
  |\varphi_{1}|\sin(\theta_{1}-\theta_{2})+|\varphi_{3}|\sin(\theta_{3}-\theta_{2})=0 \\
  |\varphi_{1}|\sin(\theta_{1}-\theta_{3})+|\varphi_{2}|\sin(\theta_{2}-\theta_{3})=0\\
\end{array}\right\}.
\end{equation}
In a case $\epsilon>0$ for absolutely symmetrical bands $a_{1}=a_{2}=a_{3}$, $b_{1}=b_{2}=b_{3}$ we obtain $\cos(\theta_{j}-\theta_{k})=-\frac{1}{2}$. In a case $\epsilon<0$ we obtain $\cos(\theta_{j}-\theta_{k})=0$ for any bands. Possible configurations corresponding to some limit cases are illustrated in Fig.\ref{Fig1}. As an approximation in the case of weak coupling $\epsilon\ll |a_{1}|,|a_{2}|,|a_{3}|$, we can assume $|\varphi_{i}|=\sqrt{\frac{|a_{i}|}{b_{i}}}$ and then substitute them in Eq.(\ref{2.9}) to find the angles $\theta_{i}-\theta_{k}$.

Substituting representation (\ref{2.1}) in Lagrangian (\ref{2.3}) we obtain:
\begin{eqnarray}\label{2.10}
    \mathcal{L}&=&
    \partial_{\mu}|\varphi_{1}|\partial^{\mu}|\varphi_{1}|
    +\partial_{\mu}|\varphi_{2}|\partial^{\mu}|\varphi_{2}|
    +\partial_{\mu}|\varphi_{3}|\partial^{\mu}|\varphi_{3}|\nonumber\\
    &+&|\varphi_{1}|^{2}\partial_{\mu}\theta_{1}\partial^{\mu}\theta_{1}
    +|\varphi_{2}|^{2}\partial_{\mu}\theta_{2}\partial^{\mu}\theta_{2}
    +|\varphi_{3}|^{2}\partial_{\mu}\theta_{3}\partial^{\mu}\theta_{3}\nonumber\\
    &-&a_{1}\left|\varphi_{1}\right|^{2}-\frac{b_{1}}{2}\left|\varphi_{1}\right|^{4}
    -a_{2}\left|\varphi_{2}\right|^{2}-\frac{b_{2}}{2}\left|\varphi_{2}\right|^{4}
    -a_{3}\left|\varphi_{3}\right|^{2}-\frac{b_{3}}{2}\left|\varphi_{3}\right|^{4}\nonumber\\
    &-&2\epsilon|\varphi_{1}||\varphi_{2}|\cos\theta_{12}-2\epsilon|\varphi_{1}||\varphi_{3}|\cos\theta_{13}-2\epsilon|\varphi_{2}||\varphi_{3}|\cos\theta_{23}.
\end{eqnarray}
Let us consider small variations of the modules from their equilibrium values: $|\varphi_{1,2,3}|=\varphi_{01,02,03}+\phi_{1,2,3}$, where $|\phi_{1,2,3}|\ll\varphi_{01,02,03}$. Then, $|\varphi|^{2}\approx\varphi_{0}^{2}+2\varphi_{0}\phi+\phi^{2}$, $|\varphi|^{4}\approx\varphi_{0}^{4}+4\varphi_{0}^{3}\phi+6\varphi_{0}^{2}\phi^{2}$, $|\varphi_{1}||\varphi_{2}|\approx\varphi_{01}\varphi_{02}+\varphi_{01}\phi_{2}+\varphi_{02}\phi_{1}+\phi_{1}\phi_{2}$, and Lagrangian (\ref{2.10}) takes the form:
\begin{eqnarray}\label{2.11}
    \mathcal{L}&=&
    \partial_{\mu}|\phi_{1}|\partial^{\mu}|\phi_{1}|+\partial_{\mu}|\phi_{2}|\partial^{\mu}|\phi_{2}| +\partial_{\mu}|\phi_{3}|\partial^{\mu}|\phi_{3}|
    +\varphi_{01}^{2}\partial_{\mu}\theta_{1}\partial^{\mu}\theta_{1}
    +\varphi_{02}^{2}\partial_{\mu}\theta_{2}\partial^{\mu}\theta_{2}
    +\varphi_{03}^{2}\partial_{\mu}\theta_{3}\partial^{\mu}\theta_{3}\nonumber\\
    &-&\phi_{1}^{2}\left(a_{1}+3b_{1}\varphi_{01}^{2}\right)-\phi_{2}^{2}\left(a_{2}+3b_{2}\varphi_{02}^{2}\right)
    -\phi_{3}^{2}\left(a_{3}+3b_{3}\varphi_{03}^{2}\right)
    -2\epsilon\phi_{1}\phi_{2}\cos\theta_{12}-2\epsilon\phi_{1}\phi_{3}\cos\theta_{13}-2\epsilon\phi_{2}\phi_{3}\cos\theta_{23}\nonumber\\
    &-&2\phi_{1}\left(\epsilon\varphi_{02}\cos\theta_{12}+\epsilon\varphi_{03}\cos\theta_{13}+a_{1}\varphi_{01}+b_{1}\varphi_{01}^{3}\right)
    -2\phi_{2}\left(\epsilon\varphi_{01}\cos\theta_{12}+\epsilon\varphi_{03}\cos\theta_{23}+a_{2}\varphi_{02}+b_{2}\varphi_{02}^{3}\right)\nonumber\\
    &-&2\phi_{3}\left(\epsilon\varphi_{01}\cos\theta_{13}+\epsilon\varphi_{02}\cos\theta_{23}+a_{3}\varphi_{03}+b_{3}\varphi_{03}^{3}\right)\nonumber\\
    &-&2\epsilon\varphi_{01}\varphi_{02}\cos\theta_{12}-2\epsilon\varphi_{01}\varphi_{03}\cos\theta_{13}-2\epsilon\varphi_{02}\varphi_{03}\cos\theta_{23}\nonumber\\
    &-&a_{1}\varphi_{01}^{2}-\frac{b_{1}}{2}\varphi_{01}^{4}-a_{2}\varphi_{02}^{2}-\frac{b_{2}}{2}\varphi_{02}^{4}
    -a_{3}\varphi_{03}^{2}-\frac{b_{3}}{2}\varphi_{03}^{4}.
\end{eqnarray}
We can consider small variations of the phase differences from their equilibrium values:  $\cos\theta_{ik}=\cos(\theta_{ik}-\theta_{ik}^{0}+\theta_{ik}^{0})
=\cos(\theta_{ik}-\theta_{ik}^{0})\cos\theta_{ik}^{0}-\sin(\theta_{ik}-\theta_{ik}^{0})\sin\theta_{ik}^{0}
\approx\left(1-\frac{(\theta_{ik}-\theta_{ik}^{0})^{2}}{2}\right)\cos\theta_{ik}^{0}-(\theta_{ik}-\theta_{ik}^{0})\sin\theta_{ik}^{0}$. Then the potential energy in Lagrangian (\ref{2.11}) takes the form:
\begin{eqnarray}\label{2.12a2}
\mathcal{U}&\approx&\mathcal{U}_{\phi}+\mathcal{U}_{\theta}+\mathcal{U}_{\phi\theta}
+a_{1}\varphi_{01}^{2}+\frac{b_{1}}{2}\varphi_{01}^{4}+a_{2}\varphi_{02}^{2}+\frac{b_{2}}{2}\varphi_{02}^{4}
+a_{3}\varphi_{03}^{2}+\frac{b_{3}}{2}\varphi_{03}^{4}\nonumber\\
&+&2\epsilon\cos\theta_{12}^{0}\varphi_{01}\varphi_{02}+2\epsilon\cos\theta_{13}^{0}\varphi_{01}\varphi_{03}
+2\epsilon\cos\theta_{23}^{0}\varphi_{02}\varphi_{03},
\end{eqnarray}
where the last nine terms determine global potential (as the "mexican hat"), $\mathcal{U}_{\phi}$ determines a potential for the module excitations $\phi_{1,2,3}$:
\begin{eqnarray}\label{2.12a3}
\mathcal{U}_{\phi}&=&\phi_{1}^{2}\left(a_{1}+3b_{1}\varphi_{01}^{2}\right)+\phi_{2}^{2}\left(a_{2}+3b_{2}\varphi_{02}^{2}\right)
+\phi_{3}^{2}\left(a_{2}+3b_{3}\varphi_{03}^{2}\right)\nonumber\\
&+&\phi_{1}\phi_{2}2\epsilon\cos\theta_{12}^{0}  +\phi_{1}\phi_{3}2\epsilon\cos\theta_{13}^{0}+\phi_{2}\phi_{3}2\epsilon\cos\theta_{23}^{0}\nonumber\\
&+&2\phi_{1}\left(\epsilon\cos\theta_{12}^{0}\varphi_{02}+\epsilon\cos\theta_{13}^{0}\varphi_{03}+a_{1}\varphi_{01}+b_{1}\varphi_{01}^{3}\right)\nonumber\\
&+&2\phi_{2}\left(\epsilon\cos\theta_{12}^{0}\varphi_{01}+\epsilon\cos\theta_{23}^{0}\varphi_{03}+a_{2}\varphi_{02}+b_{2}\varphi_{02}^{3}\right)\nonumber\\
&+&2\phi_{3}\left(\epsilon\cos\theta_{13}^{0}\varphi_{01}+\epsilon\cos\theta_{23}^{0}\varphi_{02}+a_{3}\varphi_{03}+b_{3}\varphi_{03}^{3}\right).
\end{eqnarray}
The terms at $\phi_{1,2,3}$ have to be zero, then
\begin{equation}\label{2.12a4}
\left\{\begin{array}{c}
  \epsilon\cos\theta_{12}^{0}\varphi_{02}+\epsilon\cos\theta_{13}^{0}\varphi_{03}+a_{1}\varphi_{01}+b_{1}\varphi_{01}^{3}=0 \\
  \epsilon\cos\theta_{12}^{0}\varphi_{01}+\epsilon\cos\theta_{23}^{0}\varphi_{03}+a_{2}\varphi_{02}+b_{2}\varphi_{02}^{3}=0 \\
  \epsilon\cos\theta_{13}^{0}\varphi_{01}+\epsilon\cos\theta_{23}^{0}\varphi_{02}+a_{3}\varphi_{03}+b_{3}\varphi_{03}^{3}=0 \\
\end{array}\right\},
\end{equation}
that corresponds to the first three equations in Eq.(\ref{2.9}). $\mathcal{U}_{\theta}$ determines a potential for the phase excitations $\theta_{1,2,3}$:
\begin{eqnarray}\label{2.12a5}
\mathcal{U}_{\theta}&=&-2\epsilon\varphi_{01}\varphi_{02}\frac{(\theta_{12}-\theta_{12}^{0})^{2}}{2}\cos\theta_{12}^{0}
    -2\epsilon\varphi_{01}\varphi_{03}\frac{(\theta_{13}-\theta_{13}^{0})^{2}}{2}\cos\theta_{13}^{0}
    -2\epsilon\varphi_{02}\varphi_{03}\frac{(\theta_{23}-\theta_{23}^{0})^{2}}{2}\cos\theta_{23}^{0}\nonumber\\
    &-&2\epsilon\varphi_{01}\varphi_{02}(\theta_{12}-\theta_{12}^{0})\sin\theta_{12}^{0}
    -2\epsilon\varphi_{01}\varphi_{03}(\theta_{13}-\theta_{13}^{0})\sin\theta_{13}^{0}
    -2\epsilon\varphi_{02}\varphi_{03}(\theta_{23}-\theta_{23}^{0})\sin\theta_{23}^{0},
\end{eqnarray}
In order for the linear terms $(\theta_{ij}-\theta_{ij}^{0})$ not to affect the equations of motion, the following condition must be satisfied:
\begin{eqnarray}\label{2.12a6}
\left\{\begin{array}{c}
  \varphi_{02}\sin\theta_{12}^{0}+\varphi_{03}\sin\theta_{13}^{0}=0 \\
  \varphi_{01}\sin\theta_{12}^{0}+\varphi_{03}\sin\theta_{32}^{0}=0 \\
  \varphi_{01}\sin\theta_{13}^{0}+\varphi_{02}\sin\theta_{23}^{0}=0\\
\end{array}\right\},
\end{eqnarray}
that corresponds to the second three equations in Eq.(\ref{2.9}). $\mathcal{U}_{\phi\theta}$ determines interaction between the module excitations and the phase excitations:
\begin{eqnarray}\label{2.12a7}
\mathcal{U}_{\phi\theta}=&-&\phi_{1}\phi_{2}\epsilon
\left((\theta_{12}-\theta_{12}^{0})^{2}\cos\theta_{12}^{0}+2(\theta_{12}-\theta_{12}^{0})\sin\theta_{12}^{0}\right)\nonumber\\ &-&\phi_{1}\phi_{3}\epsilon
\left((\theta_{13}-\theta_{13}^{0})^{2}\cos\theta_{13}^{0}+2(\theta_{13}-\theta_{13}^{0})\sin\theta_{13}^{0}\right)\nonumber\\
&-&\phi_{2}\phi_{3}\epsilon
\left((\theta_{23}-\theta_{23}^{0})^{2}\cos\theta_{23}^{0}+2(\theta_{23}-\theta_{23}^{0})\sin\theta_{23}^{0}\right)\nonumber\\
&-&\phi_{1}\epsilon\left((\theta_{12}-\theta_{12}^{0})^{2}\cos\theta_{12}^{0}\varphi_{02}
+(\theta_{13}-\theta_{13}^{0})^{2}\cos\theta_{13}^{0}\varphi_{03}\right)\nonumber\\
&-&\phi_{2}\epsilon\left((\theta_{12}-\theta_{12}^{0})^{2}\cos\theta_{12}^{0}\varphi_{01}
+(\theta_{23}-\theta_{23}^{0})^{2}\cos\theta_{23}^{0}\varphi_{03}\right)\nonumber\\
&-&\phi_{3}\epsilon\left((\theta_{13}-\theta_{13}^{0})^{2}\cos\theta_{13}^{0}\varphi_{01}
+(\theta_{23}-\theta_{23}^{0})^{2}\cos\theta_{23}^{0}\varphi_{02}\right)\nonumber\\
&-&2\phi_{1}\epsilon\left((\theta_{12}-\theta_{12}^{0})\sin\theta_{12}^{0}\varphi_{02}
+(\theta_{13}-\theta_{13}^{0})\sin\theta_{13}^{0}\varphi_{03}\right)\nonumber\\
&-&2\phi_{2}\epsilon\left((\theta_{12}-\theta_{12}^{0})\sin\theta_{12}^{0}\varphi_{01}
+(\theta_{23}-\theta_{23}^{0})\sin\theta_{23}^{0}\varphi_{03}\right)\nonumber\\
&-&2\phi_{3}\epsilon\left((\theta_{13}-\theta_{13}^{0})\sin\theta_{13}^{0}\varphi_{01}
+(\theta_{23}-\theta_{23}^{0})\sin\theta_{23}^{0}\varphi_{02}\right).
\end{eqnarray}
We can see, that the first six terms are of the third $\phi_{i}\phi_{k}(\theta_{ik}-\theta_{ik}^{0}),\phi_{i}(\theta_{ik}-\theta_{ik}^{0})^{2}$ and the forth $\phi_{i}\phi_{k}(\theta_{ik}-\theta_{ik}^{0})^{2}$ order, hence they can be neglected. At the same time, the last three terms are of the second order $\phi_{i}(\theta_{ik}-\theta_{ik}^{0})$. In the case $\epsilon<0$ we have all $\theta_{ik}^{0}=0$, that is $\sin\theta_{ik}^{0}=0$, hence the oscillations of modules and of phases are not hybridized in this case. Additionally, if $\theta_{ik}-\theta_{ik}^{0}=0$, that takes place for the common mode oscillations (the Goldstone mode with an acoustic spectrum), therefore in this case, the hybridization is also absent. Thus, the Leggett modes and the Higgs modes are hybridized in the case $\epsilon>0$ only, that is the phase-amplitude modes can take place. However, as it will be demonstrated in Sect.\ref{dirac}, only case $\epsilon<0$ has a physical sense, hence, we will consider the normal oscillations without the phase-amplitude hybridization further.

\subsection{Goldstone modes}

Let us consider movement of the phases $\theta_{1,2,3}$. Corresponding Lagrange equations for Lagrangian (\ref{2.11}) are:
\begin{eqnarray}\label{2.13}
  &&\varphi_{01}^{2}\partial_{\mu}\partial^{\mu}\theta_{1}
  -\varphi_{01}\varphi_{02}\epsilon\sin(\theta_{1}-\theta_{2})-\varphi_{01}\varphi_{03}\epsilon\sin(\theta_{1}-\theta_{3})=0\nonumber\\
  &&\varphi_{02}^{2}\partial_{\mu}\partial^{\mu}\theta_{2}
  +\varphi_{01}\varphi_{02}\epsilon\sin(\theta_{1}-\theta_{2})-\varphi_{02}\varphi_{03}\epsilon\sin(\theta_{2}-\theta_{3})=0\\
  &&\varphi_{03}^{2}\partial_{\mu}\partial^{\mu}\theta_{3}
  +\varphi_{01}\varphi_{03}\epsilon\sin(\theta_{1}-\theta_{2})+\varphi_{02}\varphi_{03}\epsilon\sin(\theta_{1}-\theta_{3})=0.\nonumber
\end{eqnarray}
The phases can be written in the form of harmonic oscillations:
\begin{equation}\label{2.14}
    \begin{array}{c}
      \theta_{1}=\theta_{1}^{0}+Ae^{\mathrm{i}(\mathbf{qr}-\omega t)}\equiv\theta_{1}^{0}+Ae^{-\mathrm{i}q_{\mu}x^{\mu}} \\
      \theta_{2}=\theta_{2}^{0}+Be^{\mathrm{i}(\mathbf{qr}-\omega t)}\equiv\theta_{2}^{0}+Be^{-\mathrm{i}q_{\mu}x^{\mu}} \\
      \theta_{3}=\theta_{3}^{0}+Ce^{\mathrm{i}(\mathbf{qr}-\omega t)}\equiv\theta_{3}^{0}+Ce^{-\mathrm{i}q_{\mu}x^{\mu}} \\
    \end{array},
\end{equation}
where $q_{\mu}=\left(\omega,-\mathbf{q}\right)$, $x^{\mu}=\left(t,\mathbf{r}\right)$, $\theta_{1,2,3}^{0}$ are equilibrium phases. Eq.(\ref{2.13}) can be linearised supposing $\cos\theta_{ik}\approx\cos\theta_{ik}^{0}$, $\sin\theta_{ik}=\sin(\theta_{ik}-\theta_{ik}^{0}+\theta_{ik}^{0})
\approx(\theta_{ik}-\theta_{ik}^{0})\cos\theta_{ik}^{0}+\sin\theta_{ik}^{0}$ and using Eq.(\ref{2.12a6}):
\begin{eqnarray}\label{2.15}
  &&\varphi_{01}^{2}\partial_{\mu}\partial^{\mu}\theta_{1}
  -\varphi_{01}\varphi_{02}\epsilon\cos\theta_{12}^{0}(\theta_{12}-\theta_{12}^{0})
  -\varphi_{01}\varphi_{03}\epsilon\cos\theta_{13}^{0}(\theta_{13}-\theta_{13}^{0})=0\nonumber\\
  &&\varphi_{02}^{2}\partial_{\mu}\partial^{\mu}\theta_{2}
  +\varphi_{01}\varphi_{02}\epsilon\cos\theta_{12}^{0}(\theta_{12}-\theta_{12}^{0})
  -\varphi_{02}\varphi_{03}\epsilon\cos\theta_{23}^{0}(\theta_{23}-\theta_{23}^{0})=0\\
  &&\varphi_{03}^{2}\partial_{\mu}\partial^{\mu}\theta_{3}
  +\varphi_{01}\varphi_{03}\epsilon\cos\theta_{13}^{0}(\theta_{13}-\theta_{13}^{0})
  +\varphi_{02}\varphi_{03}\epsilon\cos\theta_{23}^{0}(\theta_{23}-\theta_{23}^{0})=0\nonumber
\end{eqnarray}
Substituting Eq.(\ref{2.14}) in Eq.(\ref{2.15}) we obtain equations for the amplitudes $A,B,C$:
\begin{eqnarray}\label{2.16}
&&A\left(-\frac{\varphi_{02}}{\varphi_{01}}\epsilon\cos\theta_{12}^{0}-
\frac{\varphi_{03}}{\varphi_{01}}\epsilon\cos\theta_{13}^{0}-q_{\mu}q^{\mu}\right)
  +B\frac{\varphi_{02}}{\varphi_{01}}\epsilon\cos\theta_{12}^{0}+C\frac{\varphi_{03}}{\varphi_{01}}\epsilon\cos\theta_{13}^{0}=0\nonumber\\
&&A\frac{\varphi_{01}}{\varphi_{02}}\epsilon\cos\theta_{12}^{0}
+B\left(-\frac{\varphi_{01}}{\varphi_{02}}\epsilon\cos\theta_{12}^{0}-\frac{\varphi_{03}}{\varphi_{02}}\epsilon\cos\theta_{23}^{0}-q_{\mu}q^{\mu}\right)
      +C\frac{\varphi_{03}}{\varphi_{02}}\epsilon\cos\theta_{23}^{0}=0\\
&&A\frac{\varphi_{01}}{\varphi_{03}}\epsilon\cos\theta_{13}^{0}+B\frac{\varphi_{02}}{\varphi_{03}}\epsilon\cos\theta_{23}^{0}
  +C\left(-\frac{\varphi_{01}}{\varphi_{03}}\epsilon\cos\theta_{13}^{0}-\frac{\varphi_{02}}{\varphi_{03}}\epsilon\cos\theta_{23}^{0}-q_{\mu}q^{\mu}\right)=0.\nonumber
\end{eqnarray}
Setting the determinant of the system (\ref{2.16}) equal to zero, we find a dispersion equation:
\begin{equation}\label{2.17}
\left(q_{\mu}q^{\mu}\right)^{3}+\left(q_{\mu}q^{\mu}\right)^{2}b+\left(q_{\mu}q^{\mu}\right)c=0,
\end{equation}
where
\begin{eqnarray}\label{2.18}
&&b=\epsilon\left[\left(\frac{\varphi_{01}}{\varphi_{03}}\cos\theta_{13}+\frac{\varphi_{02}}{\varphi_{03}}\cos\theta_{23}\right)+
\left(\frac{\varphi_{01}}{\varphi_{02}}\cos\theta_{12}+\frac{\varphi_{03}}{\varphi_{02}}\cos\theta_{23}\right)+
\left(\frac{\varphi_{02}}{\varphi_{01}}\cos\theta_{12}+\frac{\varphi_{03}}{\varphi_{01}}\cos\theta_{13}\right)\right]\nonumber\\
&&c=\epsilon^{2}[\left(\frac{\varphi_{01}}{\varphi_{02}}\cos\theta_{12}+\frac{\varphi_{03}}{\varphi_{02}}\cos\theta_{23}\right)
\left(\frac{\varphi_{01}}{\varphi_{03}}\cos\theta_{13}+\frac{\varphi_{02}}{\varphi_{03}}\cos\theta_{23}\right)\nonumber\\
&&+\left(\frac{\varphi_{02}}{\varphi_{01}}\cos\theta_{12}+\frac{\varphi_{03}}{\varphi_{01}}\cos\theta_{13}\right)
\left(\frac{\varphi_{01}}{\varphi_{03}}\cos\theta_{13}+\frac{\varphi_{02}}{\varphi_{03}}\cos\theta_{23}\right)\nonumber\\
&&+\left(\frac{\varphi_{02}}{\varphi_{01}}\cos\theta_{12}+\frac{\varphi_{03}}{\varphi_{01}}\cos\theta_{13}\right)
\left(\frac{\varphi_{01}}{\varphi_{02}}\cos\theta_{12}+\frac{\varphi_{03}}{\varphi_{02}}\cos\theta_{23}\right)]
\end{eqnarray}
From Eq.(\ref{2.17}) we can see, that one of dispersion relations is
\begin{equation}\label{2.19}
    q_{\mu}q^{\mu}=0\Rightarrow\omega^{2}=q^{2},
\end{equation}
wherein $A=B=C$, thus this mode is common mode oscillations as the Goldstone mode in single-band GWS model. There are other oscillation modes with such spectrums, that
\begin{eqnarray}
&&m_{L1}^{2}=q_{\mu}q^{\mu}=\frac{1}{2}(-b-\sqrt{b^{2}-4c})\label{2.20a}\\
&&m_{L2}^{2}=q_{\mu}q^{\mu}=\frac{1}{2}(-b+\sqrt{b^{2}-4c}),\label{2.20b}
\end{eqnarray}
i.e. two massive modes, wherein
\begin{equation}\label{2.20}
    A\varphi_{01}^{2}+B\varphi_{02}^{2}+C\varphi_{03}^{2}=0.
\end{equation}
These modes are analogous to Leggett modes in multi-band superconductors \cite{grig1,grig2,grig3}. It should be noted, that if we suppose $\epsilon=0$, then $b=c=0$ and dispersion equation will be $(q_{\mu}q^{\mu})^{3}=0$, that is we obtain independent common mode oscillations in each band. From Eqs.(\ref{2.18},\ref{2.20a},\ref{2.20b}) we can see, that the squared masses of the L-bosons are proportional to the interband coupling $m_{L1,2}^{2}\sim|\epsilon|$.

\begin{figure}[ht]
\begin{center}
\includegraphics[width=9cm]{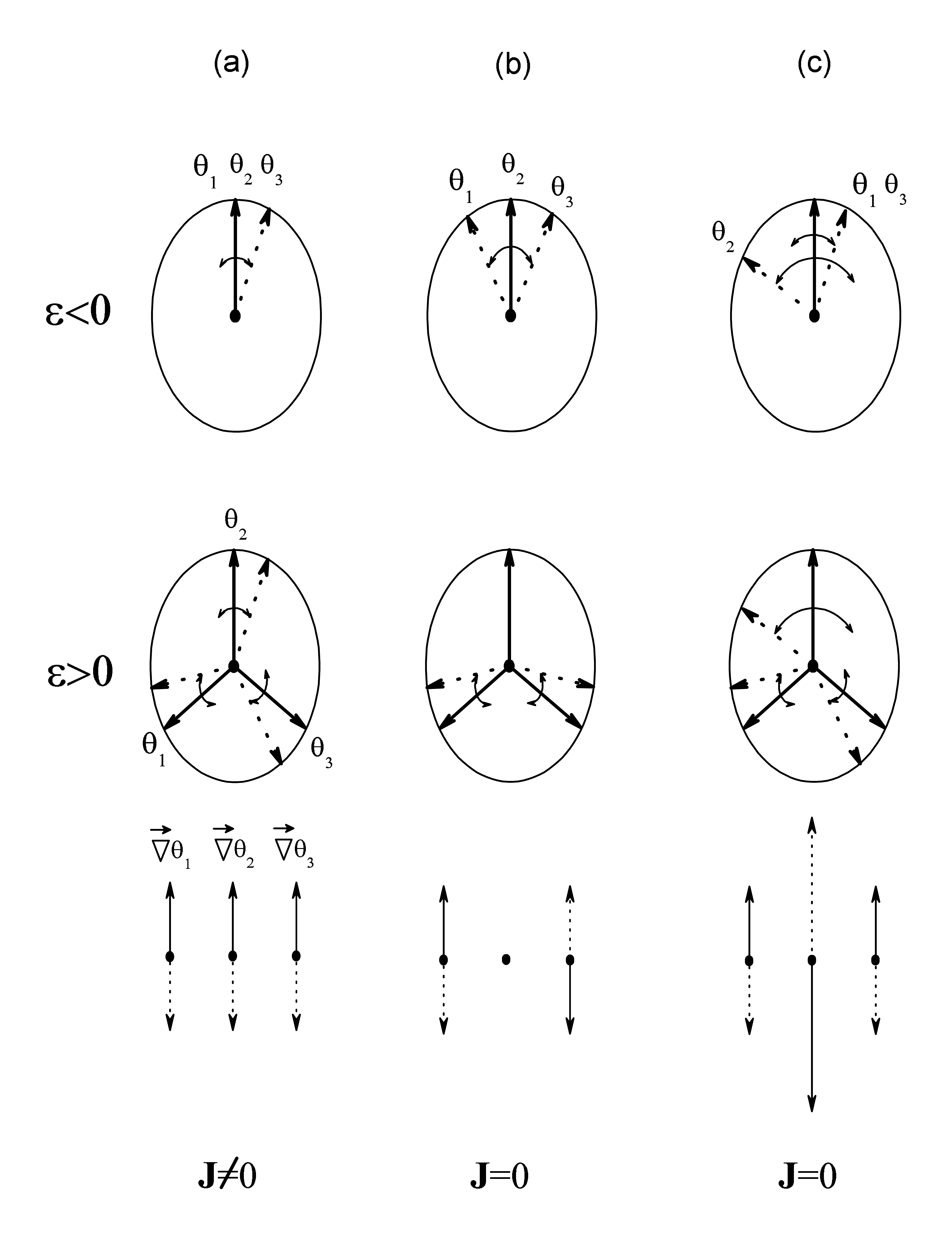}
\end{center}
\caption{Normal oscillations of the phases $\theta_{1},\theta_{2},\theta_{3}$ in symmetrical three-band system $\varphi_{01}=\varphi_{02}=\varphi_{03}$, with attractive interband interactions $\epsilon<0$ and repulsive interband interactions $\epsilon>0$. (a) - common phase oscillations with acoustic spectrum (\ref{2.19}), which accompanied by nonzero current $\mathbf{J}=\varphi_{01}^{2}\nabla\theta_{1}+\varphi_{02}^{2}\nabla\theta_{2}+\varphi_{03}^{2}\nabla\theta_{3}\neq 0$. (b,c) - anti-phase oscillations with massive spectrum (\ref{2.20a},\ref{2.20b}), which are not accompanied by current, i.e. $\mathbf{J}=0$.}
\label{Fig2}
\end{figure}

For example, let us consider a symmetrical three-band system, i.e. $\varphi_{01}=\varphi_{02}=\varphi_{03}$. Then masses of both L-bosons are equal ($b^{2}=4c$):
\begin{eqnarray}\label{2.21}
    &&m_{L1}=m_{L2}=\sqrt{\frac{3}{2}\epsilon},
    \mathrm{ when }\epsilon>0\Rightarrow\cos\theta_{12}=\cos\theta_{13}=\cos\theta_{23}=-\frac{1}{2}\nonumber\\
    \\
    &&m_{L1}=m_{L2}=\sqrt{3|\epsilon|},
    \mathrm{ when }\epsilon<0\Rightarrow\cos\theta_{12}=\cos\theta_{13}=\cos\theta_{23}=1\nonumber
\end{eqnarray}
Amplitudes of the modes (\ref{2.20a},\ref{2.20b}) relate as $A=-C, B=0$ and $A=C, B=-(A+C)$ accordingly. These three Goldstone modes (the acoustic mode (\ref{2.19}) and the Leggett modes (\ref{2.20a},\ref{2.20b})) are shown in Fig.(\ref{Fig2}). If we have the case of strongly asymmetrical bands $\varphi_{01}\ll\varphi_{02}\ll\varphi_{03}$, then the masses of L-bosons are:
\begin{eqnarray}\label{2.21a}
    &&m_{L1}^{2}\sim\min\left\{-\frac{\varphi_{03}}{\varphi_{01}}\epsilon\cos\theta_{13}
    ,-\frac{\varphi_{03}}{\varphi_{02}}\epsilon\cos\theta_{23},-\frac{\varphi_{02}}{\varphi_{01}}\epsilon\cos\theta_{12}\right\}\nonumber\\
    \\
    &&m_{L2}^{2}\sim\max\left\{-\frac{\varphi_{03}}{\varphi_{01}}\epsilon\cos\theta_{13}
    ,-\frac{\varphi_{03}}{\varphi_{02}}\epsilon\cos\theta_{23},-\frac{\varphi_{02}}{\varphi_{01}}\epsilon\cos\theta_{12}\right\},\nonumber
\end{eqnarray}
where we suppose, that all $-\epsilon\cos\theta_{ij}>0$.

The phase oscillations (\ref{2.14}) are accompanied with the current (\ref{2.5}):
\begin{equation}\label{2.22}
  J^{\mu} = 2\mathrm{i}q^{\mu}e^{-\mathrm{i}q_{\mu}x^{\mu}}\left(A\varphi_{01}^{2}+B\varphi_{02}^{2}+C\varphi_{03}^{2}\right)\Rightarrow
  \left[\begin{array}{cc}
    2\mathrm{i}Aq^{\mu}e^{-\mathrm{i}q_{\mu}x^{\mu}}\left(\varphi_{01}^{2}+\varphi_{02}^{2}+\varphi_{03}^{2}\right) &  \textrm{for the acoustic mode} \\
    &\\
    0 &  \textrm{for the Leggett modes}
  \end{array}\right],
\end{equation}
where we have used Eq.(\ref{2.20}). Thus, due to the internal proximity effect, the Goldstone modes from each band transform into common mode oscillations, where $\nabla\theta_{1}=\nabla\theta_{2}=\nabla\theta_{3}$, with the acoustic spectrum - Eq.(\ref{2.19}), and the oscillations of the relative phases $\theta_{i}-\theta_{j}$ between condensates with the energy gap in spectrum determined by the interband coupling $\epsilon$ - Eqs.(\ref{2.20a},\ref{2.20b},\ref{2.21}), which can be identified as a Leggett mode by analogy with multi-band superconductors. Propagation of the acoustic Goldstone mode is accompanied with the current $J^{\mu}\neq 0$, propagation of the Leggett modes (the massive Goldstone modes) are not accompanied with the current $J^{\mu}=0$. If we turn off the interband coupling $\epsilon=0$, then we will have an ordinary Goldstone mode with acoustic spectrum for each band. \emph{Transformation of Goldstone modes from each band into one common mode for all bands and two Leggett modes takes place even at the infinitely small coefficient $\epsilon$: $|\epsilon|\ll |a_{1,2,3}(0)|$. Thus, the effect of interband coupling is nonperturbative.}

\subsection{Higgs modes}\label{higgs}

\begin{figure}[ht]
\begin{center}
\includegraphics[width=9cm]{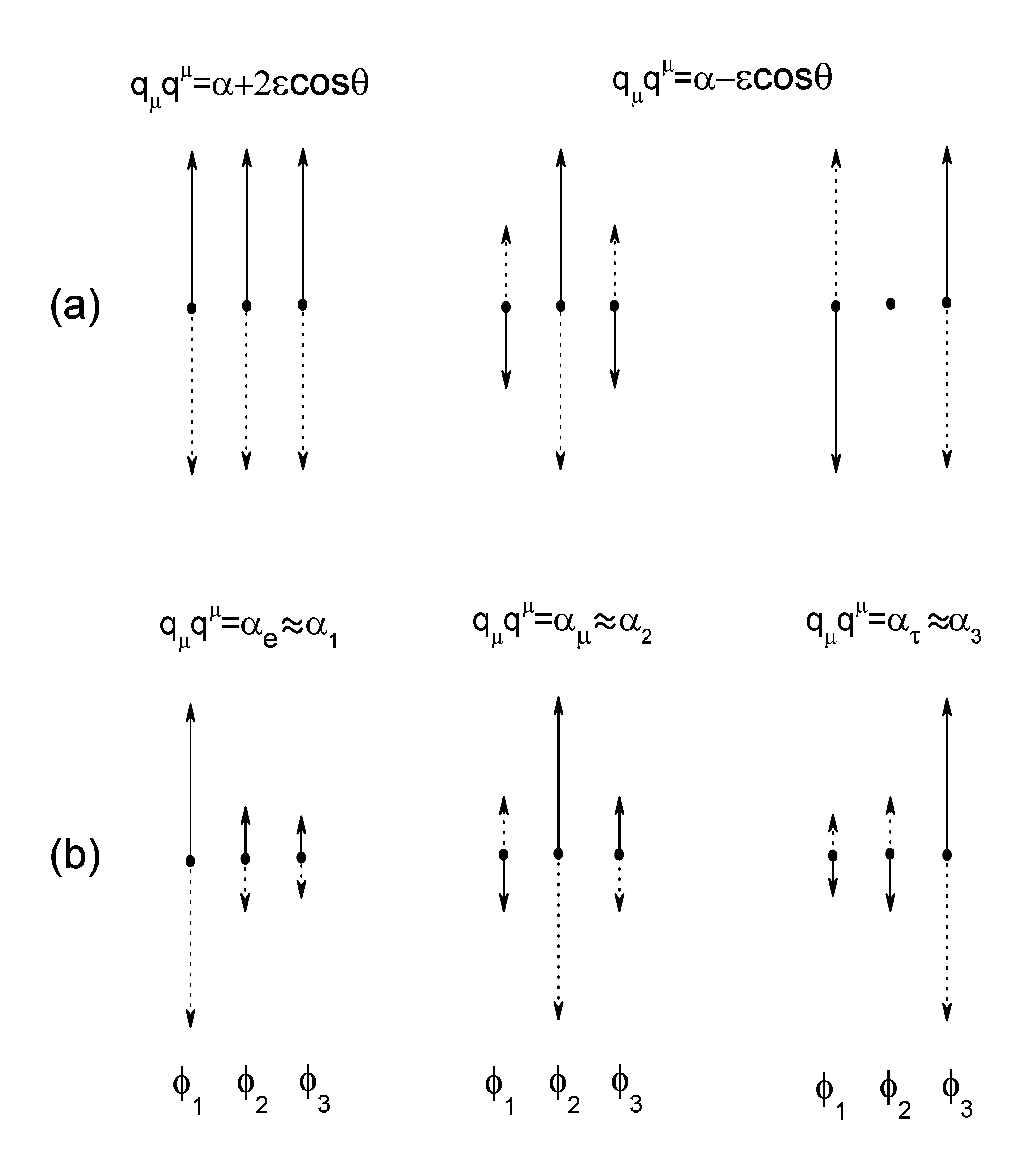}
\end{center}
\caption{Normal oscillations of the small variations of the modules of the scalar fields $\phi_{1},\phi_{2},\phi_{3}$ in a symmetrical case  $\alpha_{1}=\alpha_{2}=\alpha_{3}\equiv\alpha$, $\cos\theta_{12}=\cos\theta_{13}=\cos\theta_{23}\equiv\cos\theta$ (a), and in a case of strongly asymmetrical bands $\alpha_{1}<\alpha_{2}<\alpha_{3}$ (b).}
\label{Fig3}
\end{figure}

Let us consider movement of the modules $|\varphi_{1,2,3}(t,\mathbf{r})|\approx\varphi_{01,02,03}+\phi_{1,2,3}(t,\mathbf{r})$. Corresponding Lagrange equations for Lagrangian (\ref{2.11}) with accounting Eq.(\ref{2.12a4}) are:
\begin{eqnarray}\label{2.23}
  &&\partial_{\mu}\partial^{\mu}\phi_{1}
  +\alpha_{1}\phi_{1}+\epsilon\cos\theta_{12}\phi_{2}+\epsilon\cos\theta_{13}\phi_{3} = 0\nonumber\\
 &&\partial_{\mu}\partial^{\mu}\phi_{2}
  +\alpha_{2}\phi_{2}+\epsilon\cos\theta_{12}\phi_{1}+\epsilon\cos\theta_{23}\phi_{3} = 0\\
  &&\partial_{\mu}\partial^{\mu}\phi_{3}
  +\alpha_{3}\phi_{3}+\epsilon\cos\theta_{13}\phi_{1}+\epsilon\cos\theta_{23}\phi_{2} = 0,\nonumber
\end{eqnarray}
where we have introduced the following notes:
\begin{equation}\label{2.24}
  \alpha_{1}\equiv a_{1}+3b_{1}\varphi_{01}^{2},\quad \alpha_{2}\equiv a_{2}+3b_{2}\varphi_{02}^{2},\quad\alpha_{3}\equiv a_{3}+3b_{3}\varphi_{03}^{2}.
\end{equation}
Then,  in the case of weak coupling $|\epsilon|\ll |a_{1}|,|a_{2}|,|a_{3}|$, where corresponding amplitudes of the condensates can be assumed as $\varphi_{0i}=\sqrt{\frac{|a_{i}|}{b_{i}}}$, we have
\begin{equation}\label{2.25}
\alpha_{i}=-2a_{i}=2|a_{i}|.
\end{equation}
The fields $\phi_{1,2,3}$ can be written in a form of harmonic oscillations: $\phi_{1}=Ae^{-\mathrm{i}q_{\mu}x^{\mu}}$, $\phi_{2}=Be^{-\mathrm{i}q_{\mu}x^{\mu}}$, $\phi_{3}=Ce^{-\mathrm{i}q_{\mu}x^{\mu}}$, where $q_{\mu}x^{\mu}=\omega t-\mathbf{qr}$. Substituting them in Eq.(\ref{2.23}) we obtain equations for the amplitudes $A,B,C$:
\begin{eqnarray}\label{2.26}
   \begin{array}{c}
   A\left(\alpha_{1}-q_{\mu}q^{\mu}\right)+B\epsilon\cos\theta_{12}+C\epsilon\cos\theta_{13}=0 \\
   A\epsilon\cos\theta_{12}+B\left(\alpha_{2}-q_{\mu}q^{\mu}\right)+C\epsilon\cos\theta_{23}=0 \\
   A\epsilon\cos\theta_{13}+B\epsilon\cos\theta_{23}+C\left(\alpha_{3}-q_{\mu}q^{\mu}\right)=0
\end{array}
\end{eqnarray}
Setting the determinant of the system (\ref{2.26}) equal to zero, we find the dispersion equation:
\begin{equation}\label{2.27}
\left(q_{\mu}q^{\mu}\right)^{3}+\left(q_{\mu}q^{\mu}\right)^{2}b+\left(q_{\mu}q^{\mu}\right)c+d=0,
\end{equation}
where
\begin{eqnarray}\label{2.28}
  b &=& -\alpha_{1}-\alpha_{2}-\alpha_{3}\nonumber\\
  c &=& \alpha_{1}\alpha_{2}+\alpha_{1}\alpha_{3}+\alpha_{2}\alpha_{3}
  -\epsilon^{2}(\cos^{2}\theta_{12}+\cos^{2}\theta_{13}+\cos^{2}\theta_{23})\\
  d &=& -\alpha_{1}\alpha_{2}\alpha_{3}-2\epsilon^{3}\cos\theta_{12}\cos\theta_{13}\cos\theta_{23}
  +\epsilon^{2}(\alpha_{1}\cos^{2}\theta_{23}+\alpha_{2}\cos^{2}\theta_{13}+\alpha_{3}\cos^{2}\theta_{12}).\nonumber
\end{eqnarray}
In real physical cases $|\varepsilon|<|a_{1,2,3}|$, hence $b<0,c>0,d<0$. This cubic equation has three real positive roots $q_{\mu}q^{\mu}=m_{H}^{2}$ (squared masses of H-bosons). In a symmetrical case  $\alpha_{1}=\alpha_{2}=\alpha_{3}\equiv\alpha$, $\cos\theta_{12}=\cos\theta_{13}=\cos\theta_{23}\equiv\cos\theta$ we obtain:
\begin{equation}\label{2.29}
  m_{H}^{2}=\alpha+2\epsilon\cos\theta,\quad\alpha-\epsilon\cos\theta, \quad\alpha-\epsilon\cos\theta.
\end{equation}
It should be noted, that these three frequencies are normal modes, but not frequencies of oscillations of each band separately. Amplitudes of these modes relate as, for example, $A=B=C$; $A=C, B=-(A+C)$ and $A=-C, B=0$ accordingly. We can see, that in the case of weak interband coupling $|\epsilon|\ll|a_{1,2,3}|$ masses of H-bosons are almost equal $m_{H}\approx\alpha=\sqrt{2|a|}$. These three Higgs modes are shown in Fig.\ref{Fig3}a.

Let us consider the case of weakly coupled $|\epsilon|\ll\alpha_{1,2,3}$ and strongly asymmetrical bands, because, as we will see below, exactly this case corresponds to the real physical situation. Let us suppose:
\begin{equation}\label{2.29a}
  \varphi_{01}\ll\varphi_{02}\ll\varphi_{03},\quad\alpha_{1}<\alpha_{2}<\alpha_{3},
\end{equation}
where we assume, that small changes in the Higgs mass $m_{H}=\sqrt{\alpha}$ correspond to large changes in the amplitude of the condensate $\varphi_{0}$. Similar behaviour takes place in superconductor: $|a|\propto\mathcal{N}$, where $\mathcal{N}$ is the density of electron states on Fermi surface, then in the case of weak electron-phonon coupling we have $\varphi_{0}\sim\Omega\exp(-1/g\mathcal{N})$. In the asymmetrical case, we can obtain masses of H-bosons $m_{He},m_{H\mu},m_{H\tau}$, i.e frequencies of each normal mode:
\begin{eqnarray}\label{2.30}
 m_{He}^{2}&\approx&\alpha_{1}-\frac{\epsilon^{2}\cos^{2}\theta_{12}}{\alpha_{2}-\alpha_{1}}
 -\frac{\epsilon^{2}\cos^{2}\theta_{13}}{\alpha_{3}-\alpha_{1}}\nonumber\\
 m_{H\mu}^{2}&\approx&\alpha_{2}-\frac{\epsilon^{2}\cos^{2}\theta_{12}}{\alpha_{1}-\alpha_{2}}
 -\frac{\epsilon^{2}\cos^{2}\theta_{23}}{\alpha_{3}-\alpha_{2}}\\
 m_{H\tau}^{2}&\approx&\alpha_{3}-\frac{\epsilon^{2}\cos^{2}\theta_{13}}{\alpha_{1}-\alpha_{3}}
 -\frac{\epsilon^{2}\cos^{2}\theta_{23}}{\alpha_{2}-\alpha_{3}},\nonumber
\end{eqnarray}
and the relations between the amplitudes of these modes:
\begin{eqnarray}\label{2.31}
  q_{\mu}q^{\mu}=\alpha_{1} &\Rightarrow& B=-A\frac{\epsilon\cos\theta_{12}}{\alpha_{2}-\alpha_{1}},\quad C=-A\frac{\epsilon\cos\theta_{13}}{\alpha_{3}-\alpha_{1}} \nonumber\\
  q_{\mu}q^{\mu}=\alpha_{2} &\Rightarrow& A=-B\frac{\epsilon\cos\theta_{12}}{\alpha_{1}-\alpha_{2}},\quad C=-B\frac{\epsilon\cos\theta_{23}}{\alpha_{3}-\alpha_{2}} \\
  q_{\mu}q^{\mu}=\alpha_{2} &\Rightarrow& A=-C\frac{\epsilon\cos\theta_{13}}{\alpha_{1}-\alpha_{3}},\quad B=-C\frac{\epsilon\cos\theta_{23}}{\alpha_{2}-\alpha_{3}}. \nonumber
\end{eqnarray}
We have written the index "$e$" for the lightest boson, the index "$\tau$" for the heaviest boson, and the index "$\mu$" for the boson of medium mass. These three Higgs modes are shown in Fig.(\ref{Fig3}b) for the case, where $\epsilon\cos\theta_{ij}<0$ (as the rule).

Due to the weakness of interband coupling $|\epsilon|\ll\alpha_{1,2,3}$ we can write the following effective diagonalization of the potential energy in the sense, that each \emph{normal} mode $\phi_{e},\phi_{\mu},\phi_{\tau}$ is oscillations of corresponding \emph{effective} band
\begin{equation}\label{2.32a}
|\varphi_{e}|\approx\varphi_{0e}+\phi_{e}(t,\mathbf{r}),\quad
|\varphi_{\mu}|\approx\varphi_{0\mu}+\phi_{\mu}(t,\mathbf{r}),\quad|\varphi_{\tau}|\approx\varphi_{0\tau}+\phi_{\tau}(t,\mathbf{r})
\end{equation}
so, that these effective bands are not coupled:
\begin{eqnarray}\label{2.32}
    \mathcal{U}&=&
    a_{1}\left|\varphi_{1}\right|^{2}+a_{2}\left|\varphi_{2}\right|^{2}+a_{3}\left|\varphi_{3}\right|^{2}
    +\frac{b_{1}}{2}\left|\varphi_{1}\right|^{4}+\frac{b_{2}}{2}\left|\varphi_{2}\right|^{4}+\frac{b_{3}}{2}\left|\varphi_{3}\right|^{4}\nonumber\\
    &+&2\epsilon\cos\theta_{12}^{0}|\varphi_{1}||\varphi_{2}|
    +2\epsilon\cos\theta_{13}^{0}|\varphi_{1}||\varphi_{3}|+2\epsilon\cos\theta_{23}^{0}|\varphi_{2}||\varphi_{3}|\nonumber\\
    &\approx&a_{e}\left|\varphi_{e}\right|^{2}+a_{\mu}\left|\varphi_{\mu}\right|^{2}+a_{\tau}\left|\varphi_{\tau}\right|^{2}
    +\frac{b_{1}}{2}\left|\varphi_{e}\right|^{4}+\frac{b_{2}}{2}\left|\varphi_{\mu}\right|^{4}+\frac{b_{3}}{2}\left|\varphi_{\tau}\right|^{4}
    +\mathcal{O}\left(\frac{|\epsilon|}{m_{Hi}^{2}-m_{Hj}^{2}}\right),
\end{eqnarray}
where the strong band asymmetry (\ref{2.29a}) is supposed, and we have noted:
\begin{equation}\label{2.33}
  m_{He}=\sqrt{-2a_{e}},\quad m_{H\mu}=\sqrt{-2a_{\mu}},\quad m_{H\tau}=\sqrt{-2a_{\tau}},
\end{equation}
that can be named as \emph{the flavour masses} (i.e. eigen frequencies - the masses of H-bosons), and
\begin{equation}\label{2.34}
  m_{H1}=\sqrt{-2a_{1}},\quad m_{H2}=\sqrt{-2a_{2}},\quad m_{H2}=\sqrt{-2a_{2}},
\end{equation}
that can be named as \emph{the band masses} (i.e. frequencies if there was no the interband coupling $\epsilon=0$). Accordingly, the states
$\left(\begin{array}{ccc}\varphi_{1}, & \varphi_{2}, & \varphi_{3} \\ \end{array}\right)$
with equalibrium condensate amplitudes
\begin{equation}\label{2.35}
  \varphi_{01}=\sqrt{\frac{|a_{1}|}{b_{1}}},\quad \varphi_{02}=\sqrt{\frac{|a_{2}|}{b_{2}}},\quad \varphi_{03}=\sqrt{\frac{|a_{3}|}{b_{3}}}
\end{equation}
can be named as \emph{the band states}, and
the states $\left(\begin{array}{ccc}\varphi_{e}, & \varphi_{\mu}, & \varphi_{\tau} \\ \end{array}\right)$
with equalibrium condensate amplitudes
\begin{equation}\label{2.36}
  \varphi_{0e}\approx\sqrt{\frac{|a_{e}|}{b_{1}}},\quad \varphi_{0\mu}\approx\sqrt{\frac{|a_{\mu}|}{b_{2}}},\quad \varphi_{0\tau}\approx\sqrt{\frac{|a_{\tau}|}{b_{3}}}
\end{equation}
can be named as \emph{the flavour states}, i.e they give normal oscillations of the multi-band system. For strongly asymmetrical bands with weak interband coupling the band masses and flavour masses are almost equal: $m_{He}\approx m_{H1},m_{H\mu}\approx m_{H2},m_{H\tau}\approx m_{H3}$. Moreover, the equilibrium amplitudes of the condensates of band states and flavour states are almost equal also: $\varphi_{0e}\approx \varphi_{01},\varphi_{0\mu}\approx \varphi_{02},\varphi_{0\tau}\approx \varphi_{03}$. Indeed, we could see, that due to the strong band asymmetry (\ref{2.29a}) and the weak interband coupling $|\epsilon|\ll\alpha_{1,2,3}$, each collective mode in Eq.(\ref{2.30}) is approximately oscillations of some single band according to the following correspondence $\phi_{e}\approx\phi_{1},\phi_{\mu}\approx\phi_{2},\phi_{\tau}\approx\phi_{3}$ - Fig.\ref{Fig3}b.

Thus, the above transition from the coupled scalar fields $\phi_{1},\phi_{1},\phi_{1}$ to the normal oscillations $\phi_{e},\phi_{\mu},\phi_{\tau}$ with frequencies $m_{He},m_{H\mu},m_{H\tau}$ (the masses of H-bosons) - Eq.(\ref{2.30}) can be considered as diagonalization of the "potential" energy (\ref{2.12a3}):
\begin{eqnarray}\label{2.37}
    \mathcal{U}_{\phi}&=&
    \phi_{1}^{2}\alpha_{1}+\phi_{2}^{2}\alpha_{2}+\phi_{3}^{2}\alpha_{3}
+\phi_{1}\phi_{2}2\epsilon\cos\theta_{12}^{0}  +\phi_{1}\phi_{3}2\epsilon\cos\theta_{13}^{0}+\phi_{2}\phi_{3}2\epsilon\cos\theta_{23}^{0}
    =\alpha_{e}\left|\phi_{e}\right|^{2}+\alpha_{\mu}\left|\phi_{\mu}\right|^{2}+\alpha_{\tau}\left|\phi_{\tau}\right|^{2}\nonumber\\
    &=&\left(
        \begin{array}{ccc}
          \phi_{1}, & \phi_{2}, & \phi_{3} \\
        \end{array}
      \right)
      \left(
        \begin{array}{ccc}
          \alpha_{1} & \epsilon\cos\theta_{12}^{0} & \epsilon\cos\theta_{13}^{0} \\
          \epsilon\cos\theta_{12}^{0} & \alpha_{2} & \epsilon\cos\theta_{23}^{0} \\
          \epsilon\cos\theta_{13}^{0} & \epsilon\cos\theta_{23}^{0} & \alpha_{3} \\
        \end{array}
      \right)
      \left(
        \begin{array}{c}
          \phi_{1} \\
          \phi_{2} \\
          \phi_{3} \\
        \end{array}
      \right)=
      \left(
        \begin{array}{ccc}
          \phi_{e}, & \phi_{\mu}, & \phi_{\tau} \\
        \end{array}
      \right)
      \left(
        \begin{array}{ccc}
          \alpha_{e} & 0 & 0 \\
          0 & \alpha_{\mu} & 0 \\
          0 & 0 & \alpha_{\tau} \\
        \end{array}
      \right)
      \left(
        \begin{array}{c}
          \phi_{e} \\
          \phi_{\mu} \\
          \phi_{\tau} \\
        \end{array}
      \right)\nonumber\\
      &\equiv&\langle\phi_{123}|M_{123}|\phi_{123}\rangle=\langle\phi_{e\mu\tau}|M_{e\mu\tau}|\phi_{e\mu\tau}\rangle .
\end{eqnarray}
Obviously, that $\alpha_{e},\alpha_{\mu},\alpha_{\tau}$ are eigen-values of the matrix $M_{123}$: $M_{e\mu\tau}=\mathrm{diag}(M_{123})$, in addition, the band H-bosons and the flavour H-bosons are connected by the unitary transformation: $|\varphi_{e\mu\tau}\rangle=U|\varphi_{123}\rangle$, $|\varphi_{123}\rangle=U^{T}|\varphi_{e\mu\tau}\rangle$, where $U$ is an unitary matrix $U^{-1}=U^{T}$, which can be written via the mixing angles $\alpha_{12},\alpha_{13},\alpha_{23}$:
\begin{eqnarray}
  U&=&\left(
    \begin{array}{ccc}
      1 & 0 & 0 \\
      0 & c_{23} & s_{23} \\
      0 & -s_{23} & c_{23} \\
    \end{array}
  \right)
  \left(
    \begin{array}{ccc}
      c_{13} & 0 & s_{13} \\
      0 & 1 & 0 \\
      -s_{13} & 0 & c_{13} \\
    \end{array}
  \right)
  \left(
    \begin{array}{ccc}
      c_{12} & s_{12} & 0 \\
      -s_{12} & c_{12} & 0 \\
      0 & 0 & 1 \\
    \end{array}
  \right)\label{2.36a}\\
  U^{T}&=&\left(
    \begin{array}{ccc}
      c_{12} & -s_{12} & 0 \\
      s_{12} & c_{12} & 0 \\
      0 & 0 & 1 \\
    \end{array}
  \right)
  \left(
    \begin{array}{ccc}
      c_{13} & 0 & -s_{13} \\
      0 & 1 & 0 \\
      s_{13} & 0 & c_{13} \\
    \end{array}
  \right)
  \left(
    \begin{array}{ccc}
      1 & 0 & 0 \\
      0 & c_{23} & -s_{23} \\
      0 & s_{23} & c_{23} \\
    \end{array}
  \right),\label{2.36b}
\end{eqnarray}
where $c_{ik}=\cos\alpha_{ik},s_{ik}=\sin\alpha_{ik}$. Then, we obtain equation for the mixing angles $\alpha_{ik}$:
\begin{equation}\label{2.38}
 M_{e\mu\tau}=UM_{123}U^{T}\quad\mathrm{or}\quad M_{123}=U^{T}M_{e\mu\tau}U.
\end{equation}
Supposing the independent mixing for each pair of bands $1\leftrightarrow 2,1\leftrightarrow 3,2\leftrightarrow 3$, we obtain (for example, for $1\leftrightarrow 2$, besides, $a_{1,2}<0,|\epsilon|\ll|a_{1}|<|a_{2}|$):
\begin{eqnarray}\label{2.39}
  &&\tan2\alpha_{12}=\frac{2\epsilon\cos\theta_{12}^{0}}{\alpha_{1}-\alpha_{2}},
  \quad\sin2\alpha_{12}=\frac{2\epsilon\cos\theta_{12}^{0}}{\alpha_{e}-\alpha_{\mu}}\\
  &&\begin{array}{c}
  (\alpha_{e}-\alpha_{\mu})^{2}=(\alpha_{1}-\alpha_{2})^{2}+4\epsilon^{2}\cos^{2}\theta_{12}^{0} \\
  \alpha_{e}+\alpha_{\mu}=\alpha_{1}+\alpha_{2}
\end{array}\Rightarrow
\alpha_{e}\approx \alpha_{1}+\frac{\epsilon^{2}\cos^{2}\theta_{12}^{0}}{\alpha_{1}-\alpha_{2}},\quad
\alpha_{\mu}\approx \alpha_{2}-\frac{\epsilon^{2}\cos^{2}\theta_{12}^{0}}{\alpha_{1}-\alpha_{2}},\nonumber
\end{eqnarray}
which is approximately consistent with Eq.(\ref{2.30}). In the case of weak interband coupling $|\epsilon|\ll\alpha_{1},\alpha_{2},\alpha_{3}$ and asymmetrical bands $\alpha_{1}<\alpha_{2}<\alpha_{3}$ we have, that the mixing angles are very small $|\tan\alpha_{ik}|\ll 1$. This means, that the flavor states almost coincide with the band states (as we can see in Fig.\ref{Fig3}b). Let us estimate the mixing angle $\alpha_{ik}$. In Sect.\ref{darkmatter} it will be demonstrated, that $\epsilon\sim 10^{-40}\mathrm{eV}^{2}$. Since  $\alpha_{i}=m_{Hi}^{2}$, then $\alpha_{2}-\alpha_{1}=m_{H2}^{2}-m_{H1}^{2}\sim m_{H}^{2}\sim10^{4}\mathrm{GeV}^{2}$. Hence
\begin{equation}\label{2.40}
  \alpha_{ik}\sim 10^{-62}.
\end{equation}
Thus, oscillations of H-bosons (unlike the neutrino oscillations) are negligible. On the contrary, in the symmetrical case (\ref{2.29}) we have
\begin{equation}\label{2.41}
  \alpha_{ik}=\frac{\pi}{4},\quad a_{e,\mu,\tau}-a\sim\epsilon.
\end{equation}
Thus, in the symmetrical case each flavor state is the completely mixing of the all band states (as we can see in Fig.\ref{Fig3}a).

\section{Higgs effect for abelian gauge field}\label{spontanU1gauge}

Let us consider interaction of the scalar fields $\varphi_{1,2,3}$, spontaneously breaking the gauge $U(1)$ symmetry, with the gauge field $A_{\mu}$ in its simplest abelian (Maxwell) form. Corresponding gauge invariant Lagrangian has the form:
\begin{eqnarray}\label{3.1}
    \mathcal{L}&=&(\partial_{\mu}+\mathrm{i}eA_{\mu})\varphi_{1}(\partial^{\mu}-\mathrm{i}eA^{\mu})\varphi_{1}^{+}
    +(\partial_{\mu}+\mathrm{i}eA_{\mu})\varphi_{2}(\partial^{\mu}-\mathrm{i}eA^{\mu})\varphi_{2}^{+}
    +(\partial_{\mu}+\mathrm{i}eA_{\mu})\varphi_{3}(\partial^{\mu}-\mathrm{i}eA^{\mu})\varphi_{3}^{+}\nonumber\\
    &-&a_{1}\left|\varphi_{1}\right|^{2}-a_{2}\left|\varphi_{2}\right|^{2}-a_{3}\left|\varphi_{3}\right|^{2}
    -\frac{b_{1}}{2}\left|\varphi_{1}\right|^{4}-\frac{b_{2}}{2}\left|\varphi_{2}\right|^{4}-\frac{b_{3}}{2}\left|\varphi_{3}\right|^{4}
    \nonumber\\
    &-&\epsilon\left(\varphi_{1}^{+}\varphi_{2}+\varphi_{1}\varphi_{2}^{+}\right)-\epsilon\left(\varphi_{1}^{+}\varphi_{3}+\varphi_{1}\varphi_{3}^{+}\right)
    -\epsilon\left(\varphi_{2}^{+}\varphi_{3}+\varphi_{2}\varphi_{3}^{+}\right)-\frac{1}{16\pi}F_{\mu\nu}F^{\mu\nu},
\end{eqnarray}
where $A_{\mu}=(\varphi,-\mathbf{A}),A^{\mu}=(\varphi,\mathbf{A})$ are covariant and contravariant potential of electro-magnetic field, $F_{\mu\nu}=\partial_{\mu}A_{\nu}-\partial_{\nu}A_{\mu}$ is Faraday tensor. Corresponding Lagrange equation
\begin{equation}\label{3.2}
  \partial_{\nu}\frac{\partial\mathcal{L}}{\partial(\partial_{\nu}A_{\mu})}-\frac{\partial\mathcal{L}}{\partial A_{\mu}}=0
\end{equation}
and Maxwell equation $\partial_{\nu}F^{\mu\nu}=-4\pi J^{\mu}$ give current:
\begin{eqnarray}\label{3.3}
  J^{\mu}=-2e\left[|\varphi_{1}|^{2}\left(\partial^{\mu}\theta_{1}+eA^{\mu}\right)+
  |\varphi_{2}|^{2}\left(\partial^{\mu}\theta_{2}+eA^{\mu}\right)+|\varphi_{3}|^{2}\left(\partial^{\mu}\theta_{3}+eA^{\mu}\right)\right].
  \end{eqnarray}
The potential can be transformed as
\begin{equation}\label{3.4}
    A_{\mu}'=A_{\mu}+\frac{1}{e}\left(\alpha\partial_{\mu}\theta_{1}
    +\beta\partial_{\mu}\theta_{2}+\gamma\partial_{\mu}\theta_{3}\right),
\end{equation}
where
\begin{equation}\label{3.5}
\alpha=\frac{|\varphi_{1}|^{2}}{|\varphi_{1}|^{2}+|\varphi_{2}|^{2}+|\varphi_{3}|^{2}},\quad
\beta=\frac{|\varphi_{2}|^{2}}{|\varphi_{1}|^{2}+|\varphi_{2}|^{2}+|\varphi_{3}|^{2}},\quad
\gamma=\frac{|\varphi_{3}|^{2}}{|\varphi_{1}|^{2}+|\varphi_{2}|^{2}+|\varphi_{3}|^{2}},
\end{equation}
so that
\begin{equation}\label{3.6}
    \alpha+\beta+\gamma=1,\quad|\varphi_{2}|^{2}|\varphi_{3}|^{2}\alpha=|\varphi_{1}|^{2}|\varphi_{3}|^{2}\beta=|\varphi_{1}|^{2}|\varphi_{2}|^{2}\gamma.
\end{equation}
Then Eq.(\ref{3.3}) is reduced to "London law":
\begin{equation}\label{3.7}
  J^{\mu}=-2e^{2}\left(|\varphi_{1}|^{2}+|\varphi_{2}|^{2}+|\varphi_{3}|^{2}\right)A^{\mu}\equiv-\frac{1}{4\pi\lambda^{2}}A^{\mu},
\end{equation}
where
\begin{equation}\label{3.8}
  \lambda=\frac{1}{\sqrt{8\pi e^{2}\left(|\varphi_{1}|^{2}+|\varphi_{2}|^{2}+|\varphi_{3}|^{2}\right)}}
\end{equation}
is the "penetration depth" - the length of interaction mediated by the gauge bosons $A_{\mu}$. Thus, screening of el.-mag. field by the scalar fields $\varphi_{1,2,3}$, spontaneously breaking the gauge $U(1)$ symmetry, is analogous to the response of single-band system, but with contribution from each band $|\varphi_{i}|^{2}$. It should be noted, that in Eqs.(\ref{3.5}-\ref{3.8}) the field modules $|\varphi_{1}|^{2},|\varphi_{2}|^{2},|\varphi_{3}|^{2}$ should be replaced with their equilibrium values $\varphi_{01}^{2},\varphi_{02}^{2},\varphi_{03}^{2}$ accordingly.

The modulus-phase representations (\ref{2.1}) can be considered as the local gauge $U(1)$ transformations. 
Then, the covariant derivative is transformed by the follows:
\begin{equation}\label{3.9}
(\partial_{\mu}+\mathrm{i}eA_{\mu})\varphi_{j}=
  e^{\mathrm{i}\theta_{j}}(\partial_{\mu}+\mathrm{i}\partial_{\mu}\theta_{j}+\mathrm{i}eA_{\mu})|\varphi_{j}|.
\end{equation}
Applying the transformation (\ref{3.4}) we can transform Lagrangian (\ref{3.1}) to the following form:
\begin{eqnarray}\label{3.11}
    \mathcal{L}&=&\partial_{\mu}|\varphi_{1}|\partial^{\mu}|\varphi_{1}|+\partial_{\mu}|\varphi_{2}|\partial^{\mu}|\varphi_{2}|+
    \partial_{\mu}|\varphi_{3}|\partial^{\mu}|\varphi_{3}|+
    e^{2}\left(|\varphi_{1}|^{2}+|\varphi_{2}|^{2}+|\varphi_{3}|^{2}\right)A_{\mu}A^{\mu}\nonumber\\
    &-&2\epsilon|\varphi_{1}||\varphi_{2}|\cos(\theta_{1}-\theta_{2})-2\epsilon|\varphi_{1}||\varphi_{3}|\cos(\theta_{1}-\theta_{3})
    -2\epsilon|\varphi_{2}||\varphi_{3}|\cos(\theta_{2}-\theta_{3})\nonumber\\
    &+&\left(|\varphi_{1}|^{2}\beta^{2}+|\varphi_{2}|^{2}\alpha^{2}\right)
    \partial_{\mu}\left(\theta_{1}-\theta_{2}\right)\partial^{\mu}\left(\theta_{1}-\theta_{2}\right)\nonumber\\
    &+&\left(|\varphi_{1}|^{2}\gamma^{2}+|\varphi_{3}|^{2}\alpha^{2}\right)
    \partial_{\mu}\left(\theta_{1}-\theta_{3}\right)\partial^{\mu}\left(\theta_{1}-\theta_{3}\right)\nonumber\\
    &+&\left(|\varphi_{2}|^{2}\gamma^{2}+|\varphi_{3}|^{2}\beta^{2}\right)
    \partial_{\mu}\left(\theta_{2}-\theta_{3}\right)\partial^{\mu}\left(\theta_{2}-\theta_{3}\right)\nonumber\\
    &-&|\varphi_{1}|^{2}2\gamma\beta\partial_{\mu}\left(\theta_{1}-\theta_{2}\right)\partial^{\mu}\left(\theta_{1}-\theta_{3}\right)
    -|\varphi_{2}|^{2}2\alpha\gamma\partial_{\mu}\left(\theta_{1}-\theta_{2}\right)\partial^{\mu}\left(\theta_{2}-\theta_{3}\right)
    -|\varphi_{3}|^{2}2\alpha\beta\partial_{\mu}\left(\theta_{1}-\theta_{3}\right)\partial^{\mu}\left(\theta_{2}-\theta_{3}\right)\nonumber\\
    &-&a_{1}\left|\varphi_{1}\right|^{2}-a_{2}\left|\varphi_{2}\right|^{2}-a_{3}\left|\varphi_{3}\right|^{2}
    -\frac{b_{1}}{2}\left|\varphi_{1}\right|^{4}-\frac{b_{2}}{2}\left|\varphi_{2}\right|^{4}-\frac{b_{3}}{2}\left|\varphi_{3}\right|^{4}-\frac{1}{16\pi}F_{\mu\nu}F^{\mu\nu}.
\end{eqnarray}
We can see, that the phases $\theta_{1},\theta_ {2},\theta_ {3}$ have been excluded from the Lagrangian individually leaving only their differences: $\theta_{1}-\theta_{2},\theta_{1}-\theta_{3},\theta_{2}-\theta_{3}$. Thus, the gauge field $A_{\mu}$ absorbs the Goldstone boson (i.e. the common mode oscillations, where $\nabla\theta_{1}=\nabla\theta_{2}=\nabla\theta_{3}$) with acoustic spectrum (\ref{2.19}). At the same time, the L-bosons (i.e. the oscillations of the phases differences $\theta_{i}-\theta_{j}$) with massive spectrums (\ref{2.20a},\ref{2.20b}) "survive". This "survival" can be explained as follows. Each phase oscillation $\theta_{i}$ is absorbed by the gauge field, but such mutual oscillations of $\theta_{i}$ and $\theta_{k}$ exist, that the gauge fields from each oscillation cancel each other out due to interference, so that the Leggett modes "survive". The phase differences are not normal coordinates, because, firstly, they are not independent: we can suppose, for example, $\theta_{2}-\theta_{3}=\theta_{1}-\theta_{3}-(\theta_{1}-\theta_{2})$; secondly, we can see from Eq.(\ref{3.11}), that there are off-diagonal kinetic terms, as $\partial_{\mu}\left(\theta_{1}-\theta_{2}\right)\partial^{\mu}\left(\theta_{1}-\theta_{3}\right)$. Thus, diagonalising Lagrangian (\ref{3.11}), noticing that $\theta_{23}=\theta_{13}-(\theta_{12})$, we can obtain the Leggett modes (\ref{2.20a},\ref{2.20b}) again.

Substituting the calibrated Lagrangian (\ref{3.11}) in the Eq.(\ref{3.2}) we obtain the equation for the field $A_{\mu}$:
\begin{equation}\label{3.12}
   \partial_{\nu}F^{\nu\mu}+\frac{1}{\lambda^{2}}A^{\mu}=0,
\end{equation}
where
\begin{equation}\label{3.13}
   \frac{1}{\lambda^{2}}=8\pi e^{2}\left(\varphi_{01}^{2}+\varphi_{02}^{2}+\varphi_{03}^{2}\right)\equiv m_{A}^{2}
\end{equation}
is the squared mass of the gauge boson $A^{\mu}$, which is the squared reciprocal "penetration depth" (\ref{3.8}) in the London law (\ref{3.7}). The scalar field $\varphi$ can be written in a dimensionless form: $\varphi=\varphi_{0}\widetilde{\varphi}$, where $\varphi_{0}=\sqrt{\frac{-a}{b}}$ is the equilibrium value. Then the Lagrangian takes the form:
\begin{equation}\label{3.14}
  \mathcal{L}=\partial_{\mu}\varphi\partial^{\mu}\varphi^{+}-a\left|\varphi\right|^{2}-\frac{b}{2}\left|\varphi\right|^{4}=
  \frac{a^{2}}{b}\left[\xi^{2}\partial_{\mu}\widetilde{\varphi}\partial^{\mu}\widetilde{\varphi}^{+}
  -\left|\widetilde{\varphi}\right|^{2}-\frac{1}{2}\left|\widetilde{\varphi}\right|^{4}\right],
\end{equation}
where the length $\xi\equiv\frac{1}{\sqrt{|a|}}$ determines the spatial scale of variations of the scalar field $\varphi$ - the "coherence length". From the other hand, we could see that the mass of H-boson is $m_{H}=\sqrt{2|a|}$. Then we have:
\begin{equation}\label{3.15}
  m_{H}=\frac{\sqrt{2}}{\xi}.
\end{equation}
It is noteworthy that the mass of the Higgs boson and the mass of the gauge boson are related as
\begin{equation}\label{3.16}
    \frac{m_{H}}{m_{A}}=\sqrt{2}\kappa,
\end{equation}
where $\kappa=\lambda/\xi$ is the Ginzburg-Landau parameter. Accordingly, the three-band system is characterised with the three coherence lengths $\xi_{1}=\frac{\sqrt{2}}{m_{H1}}$, $\xi_{2}=\frac{\sqrt{2}}{m_{H2}}$, $\xi_{3}=\frac{\sqrt{2}}{m_{H3}}$, hence with three Ginzburg-Landau parameters $\kappa_{1}\equiv\frac{\lambda}{\xi_{1}}=\frac{m_{H1}}{\sqrt{2}m_{A}}$, $\kappa_{2}\equiv\frac{\lambda}{\xi_{2}}=\frac{m_{H2}}{\sqrt{2}m_{A}}$, $\kappa_{3}\equiv\frac{\lambda}{\xi_{3}}=\frac{m_{H3}}{\sqrt{2}m_{A}}$.

Let us consider the term of interaction of modulus of the scalar fields $|\varphi_{1}|,|\varphi_{2}|,|\varphi_{3}|$ with the gauge field $A_{\mu}$ in Lagrangian (\ref{3.11}). Using the small deviations $|\phi_{i}|\ll\varphi_{0i}$ from the corresponding equilibrium values: $|\varphi_{i}|^{2}\approx\varphi_{0i}^{2}+2\varphi_{0i}\phi_{i}(t,\mathbf{r})$, we obtain
\begin{eqnarray}\label{3.17}
    \mathcal{U}_{\varphi A}&=&e^{2}\left(|\varphi_{1}|^{2}+|\varphi_{2}|^{2}+|\varphi_{3}|^{2}\right)A_{\mu}A^{\mu}\nonumber\\
    &\approx& e^{2}\left(\varphi_{01}^{2}+\varphi_{02}^{2}+\varphi_{03}^{2}\right)A_{\mu}A^{\mu}
    +e^{2}\left(2\varphi_{01}\phi_{1}(t,\mathbf{r})+2\varphi_{02}\phi_{2}(t,\mathbf{r})+2\varphi_{03}\phi_{3}(t,\mathbf{r})\right)A_{\mu}A^{\mu}\nonumber\\
    &\equiv&\frac{m_{A}^{2}}{8\pi}A_{\mu}A^{\mu}
    +e^{2}\left(2\varphi_{01}\phi_{1}(t,\mathbf{r})+2\varphi_{02}\phi_{2}(t,\mathbf{r})+2\varphi_{03}\phi_{3}(t,\mathbf{r})\right)A_{\mu}A^{\mu}\nonumber\\
    &\approx&\frac{m_{A}^{2}}{8\pi}A_{\mu}A^{\mu}
    +e^{2}\left(2\varphi_{01}\phi_{e}(t,\mathbf{r})+2\varphi_{02}\phi_{\mu}(t,\mathbf{r})+2\varphi_{03}\phi_{\tau}(t,\mathbf{r})\right)A_{\mu}A^{\mu},
\end{eqnarray}
where we have used Eq.(\ref{3.13}) and we have taken advantage by the strong band asymmetry and the weakness of interband coupling discussed in Subsect.\ref{higgs}, where we could see that each collective mode is approximately oscillations of some single band according to the following correspondence $\phi_{e}\approx\phi_{1},\phi_{\mu}\approx\phi_{2},\phi_{\tau}\approx\phi_{3}$. As will be demonstrated below $\varphi_{01}:\varphi_{02}:\varphi_{03}=m_{e}:m_{\mu}:m_{\tau}=0.00028:0.059:1$. Thus, \emph{the gauge boson} $A_{\mu}$ \emph{interacts with} $\tau$\emph{-Higgs boson} $\phi_{\tau}$ \emph{predominantly, at the same time the interaction with} $\mu,e$\emph{-Higgs bosons} $\phi_{\mu},\phi_{e}$ \emph{is very weak}.

\section{The band states and the flavour states of Dirac fields.}\label{dirac}

We can consider three Dirac spinor fields $\psi_{1},\psi_{2},\psi_{3}$ as we have considered three scalar fields (\ref{2.1}). The fields are massless, but each field interacts with the corresponding scalar field (i.e in own band). Then, the Lagrangian will have the form:
\begin{eqnarray}\label{4.1}
    \mathcal{L}&=&\mathrm{i}\overline{\psi}_{L1}\gamma^{\mu}\overset{\leftrightarrow}{\partial}_{\mu}\psi_{L1}
    +\mathrm{i}\overline{\psi}_{R1}\gamma^{\mu}\overset{\leftrightarrow}{\partial}_{\mu}\psi_{R1}
    -\chi(\overline{\psi}_{L1}\varphi_{1}\psi_{R1}+\overline{\psi}_{R1}\varphi_{1}^{+}\psi_{L1})\nonumber\\
    &+&\mathrm{i}\overline{\psi}_{L2}\gamma^{\mu}\overset{\leftrightarrow}{\partial}_{\mu}\psi_{L2}
    +\mathrm{i}\overline{\psi}_{R2}\gamma^{\mu}\overset{\leftrightarrow}{\partial}_{\mu}\psi_{R2}
    -\chi(\overline{\psi}_{L2}\varphi_{2}\psi_{R2}+\overline{\psi}_{R2}\varphi_{2}^{+}\psi_{L2})\\
    &+&\mathrm{i}\overline{\psi}_{L3}\gamma^{\mu}\overset{\leftrightarrow}{\partial}_{\mu}\psi_{L3}
    +\mathrm{i}\overline{\psi}_{R3}\gamma^{\mu}\overset{\leftrightarrow}{\partial}_{\mu}\psi_{R3}
    -\chi(\overline{\psi}_{L3}\varphi_{3}\psi_{R3}+\overline{\psi}_{R3}\varphi_{3}^{+}\psi_{L3}),\nonumber
\end{eqnarray}
where $\gamma^{\mu}$ are Dirac matrices, $\overline{\psi}\gamma^{\mu}\overset{\leftrightarrow}{\partial}_{\mu}\psi
\equiv\frac{1}{2}[\overline{\psi}\gamma^{\mu}(\partial_{\mu}\psi)-(\partial_{\mu}\overline{\psi})\gamma^{\mu}\psi]$ is a differential operator, $\overline{\psi}=\psi^{+}\gamma^{0}$ is the Dirac conjugated bispinor; $\psi_{R}=\frac{1}{2}(1+\gamma^{5})\psi$ and $\psi_{L}=\frac{1}{2}(1-\gamma^{5})\psi$ are the right-handed and the left-handed fields accordingly, so that $\psi=\psi_{L}+\psi_{R}$; $\chi$ is the dimensionless coupling constant between the corresponding Dirac field $\psi_{j}$ and scalar field $\varphi_{j}$ (Yukawa constant). Thus, by analogy with the Higgs modes, we will call the states $\psi_{1},\psi_{2},\psi_{3}$ as \emph{the band states}.

Due to presence of condensate of the scalar field $\langle 0|\varphi|0\rangle=\varphi_{0}e^{i\theta^{0}}$ the Dirac fermion takes mass by follows. Let us consider a single-band case, then the term of interaction of the scalar field $\varphi$ with the Dirac field $\psi$  has the following form:
\begin{eqnarray}\label{4.2}
    \mathcal{U}_{D}=\chi(\overline{\psi}_{L}\varphi\psi_{R}+\overline{\psi}_{R}\varphi^{+}\psi_{L})
    =\chi|\varphi|(\overline{\psi}_{L}\psi_{R}+\overline{\psi}_{R}\psi_{L})\cos\theta
    +\mathrm{i}\chi|\varphi|(\overline{\psi}_{L}\psi_{R}-\overline{\psi}_{R}\psi_{L})\sin\theta.
\end{eqnarray}
Here, $\overline{\psi}_{L}\psi_{R}+\overline{\psi}_{R}\psi_{L}$ is a \emph{scalar}, but $\overline{\psi}_{L}\psi_{R}-\overline{\psi}_{R}\psi_{L}$ is a \emph{pseudoscalar}. Hence, in order to obtain the Dirac mass of a fermion we must choose the vacuum so, that $\theta^{0}=0$, that is $m_{D}=\chi\varphi_{0}\cos\theta^{0}=\chi\varphi_{0}$. It should be noted, that in the single-band system this choice of phase is not principal, because due the $U(1)$-symmetry the phase $\theta$ can be always set as $\theta=0$. Then, the Dirac term takes the form:
\begin{eqnarray}\label{4.3}
    \mathcal{U}_{D}
    =\chi|\varphi|(\overline{\psi}_{R}\psi_{L}+\overline{\psi}_{R}\psi_{L})= m_{D}(\overline{\psi}_{R}\psi_{L}+\overline{\psi}_{R}\psi_{L})+\chi\phi(\overline{\psi}_{R}\psi_{L}+\overline{\psi}_{R}\psi_{L}).
\end{eqnarray}
Thus, the initially massless fermion obtains the mass $m_{D}=\chi\varphi_{0}$, due to interaction with the condensate $\varphi_{0}=\sqrt{\frac{-a}{b}}$ of the scalar field $\varphi$. The coupling $\chi\phi$ is interaction of the Dirac field $\psi$ with small variations of the modulus of the scalar field from its equilibrium value: $|\varphi|=\varphi_{0}+\phi$, where $|\phi|\ll\varphi_{0}$, i.e. interaction with the H-boson.

However, in the three-band system (multi-band system) there are many scalar fields: $\left|\varphi_{1}\right|e^{i\theta_{1}}$, $\left|\varphi_{2}\right|e^{i\theta_{2}}$, $\left|\varphi_{3}\right|e^{i\theta_{3}}$, where the equilibrium phase differences $\theta^{0}_{12},\theta^{0}_{13},\theta^{0}_{23}$ are determined with Eq.(\ref{2.9}). In the case of repulsive interband coupling $\epsilon>0$ we can have different phases: $\theta^{0}_{1}\neq\theta^{0}_{2}\neq\theta^{0}_{3}$ - Fig.\ref{Fig1}, for example $\theta^{0}_{12}=\theta^{0}_{23}=\frac{2\pi}{3},\theta^{0}_{13}=\frac{4\pi}{3}$ for symmetrical bands $\varphi_{01}=\varphi_{02}=\varphi_{03}$. This means, that even if we set $\theta^{0}_{1}=0$, then other phases will be $\theta^{0}_{2}=\theta^{0}_{12}\neq 0$, $\theta^{0}_{3}=\theta^{0}_{13}\neq 0$. Hence, the coupling terms (\ref{4.1}) cannot be reduced to the Dirac mass term (\ref{4.3}) due to the pseudoscalar contribution. On the contrary, in the case of attractive interband coupling $\epsilon<0$ we have the same phases: $\theta^{0}_{1}=\theta^{0}_{2}=\theta^{0}_{3}$ - Fig.\ref{Fig1}. This means, that we should suppose $\theta^{0}_{1}=\theta^{0}_{2}=\theta^{0}_{3}=0$, then the coupling terms in Eq.(\ref{4.1}) can be reduced to the Dirac mass term (\ref{4.3}):
\begin{eqnarray}\label{4.4}
    \mathcal{L}&=&\mathrm{i}\overline{\psi}_{L1}\gamma^{\mu}\overset{\leftrightarrow}{\partial}_{\mu}\psi_{L1}
    +\mathrm{i}\overline{\psi}_{R1}\gamma^{\mu}\overset{\leftrightarrow}{\partial}_{\mu}\psi_{R1}
    -\chi|\varphi_{1}|(\overline{\psi}_{L1}\psi_{R1}+\overline{\psi}_{R1}\psi_{L1})\nonumber\\
    &+&\mathrm{i}\overline{\psi}_{L2}\gamma^{\mu}\overset{\leftrightarrow}{\partial}_{\mu}\psi_{L2}
    +\mathrm{i}\overline{\psi}_{R2}\gamma^{\mu}\overset{\leftrightarrow}{\partial}_{\mu}\psi_{R2}
    -\chi|\varphi_{2}|(\overline{\psi}_{L2}\psi_{R2}+\overline{\psi}_{R2}\psi_{L2})\\
    &+&\mathrm{i}\overline{\psi}_{L3}\gamma^{\mu}\overset{\leftrightarrow}{\partial}_{\mu}\psi_{L3}
    +\mathrm{i}\overline{\psi}_{R3}\gamma^{\mu}\overset{\leftrightarrow}{\partial}_{\mu}\psi_{R3}
    -\chi|\varphi_{3}|(\overline{\psi}_{L3}\psi_{R3}+\overline{\psi}_{R3}\psi_{L3}).\nonumber
\end{eqnarray}
Therefore, the masses of the Dirac fields $\psi_{1},\psi_{2},\psi_{3}$ are determined by coupling with the \emph{equilibrium} module of corresponding scalar fields $\varphi_{01},\varphi_{02},\varphi_{03}$:
\begin{equation}\label{4.5}
  m_{D1}=\chi\varphi_{01},\quad m_{D2}=\chi\varphi_{02},\quad m_{D3}=\chi\varphi_{03}.
\end{equation}
Thus, only attractive interband coupling
\begin{equation}\label{4.6}
  \epsilon<0
\end{equation}
has a physical sense, unlike multi-band superconductivity, where the analog of interaction of Dirac spinors and scalar field (the superconducting order parameter) is absent, hence any interband couplings $\epsilon_{ik}$ are allowed \cite{grig2,grig3}. It should be noted, from Fig.\ref{Fig1} we can see that at $\varphi_{01,02}\ll\varphi_{03}$ and $\epsilon>0$  we can suppose $\theta^{0}_{3}=0,\theta^{0}_{2}=\theta^{0}_{2}=\pi$ (then we should change signs of two Yukawa constants $\chi_{1}=\chi_{2}=-\chi_{3}$). However such system will have more larger ground state energy compared to the $\epsilon<0$ case.

Let us consider the Dirac terms in Eq.(\ref{4.4}):
\begin{eqnarray}\label{4.7}
    \mathcal{U}_{D}&=&\chi\varphi_{01}(\overline{\psi}_{L1}\psi_{R1}+\overline{\psi}_{R1}\psi_{L1})
    +\chi\phi_{1}(\overline{\psi}_{L1}\psi_{R1}+\overline{\psi}_{R1}\psi_{L1})\nonumber\\
    &+&\chi\varphi_{02}(\overline{\psi}_{L2}\psi_{R2}+\overline{\psi}_{R2}\psi_{L2})
    +\chi\phi_{2}(\overline{\psi}_{L2}\psi_{R2}+\overline{\psi}_{R2}\psi_{L2})\\
    &+&\chi\varphi_{03}(\overline{\psi}_{L3}\psi_{R3}+\overline{\psi}_{R3}\psi_{L3})
    +\chi\phi_{3}(\overline{\psi}_{L3}\psi_{R3}+\overline{\psi}_{R3}\psi_{L3}).\nonumber
\end{eqnarray}
However, as we could see in Sect.\ref{spontanU1}, the fields $\phi_{1},\phi_{2},\phi_{3}$ are not normal oscillations of the coupled scalar fields $|\varphi_{1}|,|\varphi_{2}|,|\varphi_{3}|$. Eigen frequencies (the masses of H-bosons) have been found in Eq.(\ref{2.30}), each normal oscillation mode involves all three scalar fields - Eq.(\ref{2.31}), Fig.\ref{Fig3}. Thus, we can introduce \emph{the flavour states}: each flavor state of Dirac fields interacts only with the corresponding normal mode $\phi_{e},\phi_{\mu},\phi_{\tau}$ of the scalar fields. At the same time, in Sect.\ref{higgs} we have seen, that due to the weakness of interband coupling $|\epsilon|\ll\alpha_{1,2,3}$ and the strong band asymmetry (\ref{2.29a}), the effective diagonalization (\ref{2.32}) can be realized. As a result, we obtain the flavour states of condensates (\ref{2.36}), in the sense that each normal mode $\phi_{e},\phi_{\mu},\phi_{\tau}$ is oscillations of corresponding effective band (flavour) $\varphi_{e},\varphi_{\mu},\varphi_{\tau}$. Then we can write:
\begin{eqnarray}\label{4.8}
    \mathcal{U}_{De\mu\tau}&=&\chi\varphi_{0e}(\overline{\psi}_{Le}\psi_{Re}+\overline{\psi}_{Re}\psi_{Le})
    +\chi\phi_{e}(\overline{\psi}_{Le}\psi_{Re}+\overline{\psi}_{Re}\psi_{Le})\nonumber\\
    &+&\chi\varphi_{0\mu}(\overline{\psi}_{L\mu}\psi_{R\mu}+\overline{\psi}_{R\mu}\psi_{L\mu})
    +\chi\phi_{\mu}(\overline{\psi}_{L\mu}\psi_{R\mu}+\overline{\psi}_{R\mu}\psi_{L\mu})\\
    &+&\chi\varphi_{0\tau}(\overline{\psi}_{L\tau}\psi_{R\tau}+\overline{\psi}_{R\tau}\psi_{L\tau})
    +\chi\phi_{\tau}(\overline{\psi}_{L\tau}\psi_{R\tau}+\overline{\psi}_{R\tau}\psi_{L\tau}).\nonumber
\end{eqnarray}
Thus, the masses of the Dirac fields $\psi_{e},\psi_{\mu},\psi_{\tau}$ is determined by coupling with the equilibrium modules of corresponding scalar fields $\varphi_{0e},\varphi_{0\mu},\varphi_{0\tau}$:
\begin{equation}\label{4.5a}
  m_{De}=\chi\varphi_{0e},\quad m_{D\mu}=\chi\varphi_{0\mu},\quad m_{D\tau}=\chi\varphi_{0\tau}.
\end{equation}
Since $\varphi_{0e}\approx\varphi_{01},\varphi_{0\mu}\approx\varphi_{02},\varphi_{0\tau}\approx\varphi_{03}$ and $\phi_{e}\approx\phi_{1},\phi_{\mu}\approx\phi_{2},\phi_{\tau}\approx\phi_{3}$ we can represent the Yukawa couplings in Tab.\ref{tab1} similar to Tab.\ref{tab01} for 2HDM or 3HDM models. Thus, unlike 2HDM or 3HDM models, in the three-band model the Yukawa interactions with scalar fields is distributed over generations of fermions, not over leptons and quarks apart.
\begin{center}
\begin{table}[ht]
\begin{center}
\begin{tabular}{|c|c|c|}
  \hline
    $e,u,d$ & $\mu,c,s$ & $\tau,t,b$  \\
\hline
   $\varphi_{1}$ & $\varphi_{2}$ & $\varphi_{3}$  \\
\hline
\end{tabular}
\end{center}
\caption{Yukawa interactions  for three generations of fermions (charged leptons, upper and bottom quarks).}
  \label{tab1}
\end{table}
\end{center}

However, unlike the exact diagonalization (\ref{2.37}) for potential energy of the excitations $\phi_{1,2,3}\rightarrow\phi_{e,\mu,\tau}$ of the coupled condensates (because it is a quadratic form), the diagonalization (\ref{2.32}) is approximate. Hence, the flavour states must enter into the Lagrangian with some interflavor mixings, which compensate the inaccuracy of diagonalization (\ref{2.32}). Then, the potential energy term for the flavour states $\psi_{e},\psi_{\mu},\psi_{\tau}$ take the following form:
\begin{equation}\label{4.9}
    \mathcal{U}_{De\mu\tau}+\mathcal{U}_{\mathrm{mix}}=
    \left(
        \begin{array}{ccc}
          \overline{\psi}_{Le}, & \overline{\psi}_{L\mu}, & \overline{\psi}_{L\tau} \\
        \end{array}
      \right)
      \left(
        \begin{array}{ccc}
          m_{De} & \zeta_{e\mu} & \zeta_{e\tau} \\
          \zeta_{e\mu} & m_{D\mu} & \zeta_{\mu\tau} \\
          \zeta_{e\tau} & \zeta_{\mu\tau} & m_{D\tau} \\
        \end{array}
      \right)
      \left(
        \begin{array}{c}
          \psi_{Re} \\
          \psi_{R\mu} \\
          \psi_{R\tau} \\
        \end{array}
      \right)
        +h.c.\equiv\langle\psi_{e\mu\tau}|M_{e\mu\tau}|\psi_{e\mu\tau}\rangle,
\end{equation}
where $\zeta_{ik}$ are the mixing parameters analogous to the interband coupling $\epsilon$. Thus, the coupling of $L$ and $R$ components $\overline{\psi}_{Li}\psi_{Ri}+\overline{\psi}_{Ri}\psi_{Li}$ gives the Dirac masses $m_{De},m_{D\mu},m_{D\tau}$, at the same time, each $L$ and $R$ components are mixed with the corresponding $R$ and $L$ components of other flavors $\overline{\psi}_{Li}\psi_{Rk}+\overline{\psi}_{Rk}\psi_{Li}$. As a result of diagonalization of the matrix $M_{e\mu\tau}$: $M_{123}=\mathrm{diag}(M_{e\mu\tau})$, we obtain the potential energy term in Lagrangian (\ref{4.7}) for the band states:
\begin{equation}\label{4.10}
  \mathcal{U}_{De\mu\tau}+\mathcal{U}_{\mathrm{mix}}=\mathcal{U}_{D}=
  \left(\begin{array}{ccc}
          \overline{\psi}_{L1}, & \overline{\psi}_{L2}, & \overline{\psi}_{L3} \\
        \end{array}
      \right)
      \left(
        \begin{array}{ccc}
          m_{D1} & 0 & 0 \\
          0 & m_{D2} & 0 \\
          0 & 0 & m_{D3} \\
        \end{array}
      \right)
      \left(
        \begin{array}{c}
          \psi_{R1} \\
          \psi_{R2} \\
          \psi_{R3} \\
        \end{array}
      \right)
        +h.c.\equiv\langle\psi_{123}|M_{123}|\psi_{123}\rangle .
\end{equation}
Thus, we have the system of equations for the mixing parameters $\zeta_{e\mu},\zeta_{e\tau},\zeta_{\mu\tau}$:
\begin{equation}\label{4.11}
  \left|
        \begin{array}{ccc}
          m_{De}-m_{Di} & \zeta_{e\mu} & \zeta_{e\tau} \\
          \zeta_{e\mu} & m_{D\mu}-m_{Di} & \zeta_{\mu\tau} \\
          \zeta_{e\tau} & \zeta_{\mu\tau} & m_{D\tau}-m_{Di} \\
        \end{array}
      \right|=0,\quad\mathrm{where}\quad i=1,2,3.
\end{equation}
Obviously, $m_{De}-m_{D1}\sim m_{D\mu}-m_{D2}\sim m_{D\tau}-m_{D3}\sim\zeta_{e\mu},\zeta_{e\tau},\zeta_{\mu\tau}$. 
Using Eq.(\ref{2.30}) we obtain:
\begin{equation}\label{4.12a}
  \zeta_{\alpha\beta}\sim m_{Di}\frac{\epsilon^{2}}{m_{Hi}^{2}\Delta m_{Hij}^{2}}
\end{equation}
where $\Delta m_{Hij}^{2}=m_{Hi}^{2}-m_{Hj}^{2}$. Thus, the mixing parameters $\zeta_{\alpha\beta}$ are determined with the interband coupling $\epsilon$. So, if the interband coupling is weak $|\epsilon|\lll m_{H}^{2},m_{D}^{2}$, then the mixing parameter $|\zeta|\lll m_{D}$.

It should be noted, that in SM we can write the mass matrix $M_{e\mu\tau}^{\mathrm{SM}}$ as:
\begin{eqnarray}\label{4.12b}
         M_{e\mu\tau}^{\mathrm{SM}}=\left(
        \begin{array}{ccc}
          m_{De} & \zeta_{e\mu} & \zeta_{e\tau} \\
          \zeta_{e\mu} & m_{D\mu} & \zeta_{\mu\tau} \\
          \zeta_{e\tau} & \zeta_{\mu\tau} & m_{D\tau} \\
        \end{array}
      \right)=
      \varphi_{0}\left(
        \begin{array}{ccc}
          \chi_{ee} & \chi_{e\mu} & \chi_{e\tau} \\
          \chi_{e\mu} & \chi_{\mu\mu} & \chi_{\mu\tau} \\
          \chi_{e\tau} & \chi_{\mu\tau} & \chi_{\tau\tau} \\
        \end{array}
      \right).
\end{eqnarray}
That is, both diagonal elements and off-diagonal elements are just Yukawa constants $\chi_{ij}$ due to presence single scalar field $\varphi_{0}$. However, in the three-band model we have three scalar fields $\varphi_{0e},\varphi_{0\mu},\varphi_{0\tau}$ and Eq.(\ref{4.5a}), hence we cannot write the mass matrix $M_{e\mu\tau}$ in the form of Eq.(\ref{4.12b}). This means, that the mixing coefficients $\zeta_{ij}$ are not off-diagonal Yukawa interaction. \emph{The mixing coefficients $\zeta_{ij}$ are the fermionic analog of the interband Josephson coupling}. This mixing takes place due to the interband Josephson coupling of the scalar fields $\varphi_{1},\varphi_{2},\varphi_{3}$: from Eq.(\ref{4.12a}) we can see that $\zeta_{\alpha\beta}\propto\epsilon^{2}$.


The band states and the flavour states are connected by an unitary transformation: $|\psi_{e\mu\tau}\rangle=U|\psi_{123}\rangle$, $|\psi_{123}\rangle=U^{T}|\psi_{e\mu\tau}\rangle$, where $U$ is an unitary matrix $U^{-1}=U^{T}$, which can be written via the mixing angles $\alpha_{12},\alpha_{13},\alpha_{23}$ - Eqs.(\ref{2.36a},\ref{2.36b}). The mixing angles can be found from Eq.(\ref{2.38}). So, supposing the independent mixing for each pair of bands $e\leftrightarrow \mu,e\leftrightarrow \tau,\mu\leftrightarrow \tau$, we obtain (for example, for $e\leftrightarrow \mu$ via the mixing of the band states 1 and 2):
\begin{eqnarray}\label{4.13}
  \tan2\alpha_{12}=\frac{2\zeta_{e\mu}}{m_{De}-m_{D\mu}},
\end{eqnarray}
and besides, the band masses $m_{D1},m_{D2}$ and the flavor masses $m_{De},m_{D\mu}$ are connected by the following way:
\begin{eqnarray}\label{4.14}
\begin{array}{c}
  (m_{D1}-m_{D2})^{2}=(m_{De}-m_{D\mu})^{2}+4\zeta_{e\mu}^{2} \\
  m_{De}+m_{D\mu}=m_{D1}+m_{D2}
\end{array}.
\end{eqnarray}
In the case of weak interband coupling $|\epsilon|\ll|a_{1}|,|a_{2}|,|a_{3}|$ (hence, $|\zeta_{ik}|\ll m_{D1},m_{D2},m_{D3}$) and strongly asymmetrical bands $\varphi_{01}\ll\varphi_{02}\ll\varphi_{03}$ (hence, $m_{D1}\ll m_{D2}\ll m_{D3}$) we have, that the mixing angles are very small $|\tan\alpha_{ik}|\lll 1$, hence the oscillations charged leptons (i.e electron-muon-tauon) are negligible and experimentally unobservable, unlike the neutrino oscillations.

Now, let us return to Eq.(\ref{4.2}) again. Despite the fact, that equilibrium phases are $\theta_{1}^{0}=\theta_{2}^{0}=\theta_{3}^{0}=0$, phase oscillations (\ref{2.14}) can takes place. The full interaction term has the form:
\begin{eqnarray}\label{4.15}
    \mathcal{U}_{D}&=&\chi|\varphi_{1}|(\overline{\psi}_{L1}\psi_{R1}+\overline{\psi}_{R1}\psi_{L1})
    +\chi|\varphi_{2}|(\overline{\psi}_{L2}\psi_{R2}+\overline{\psi}_{R2}\psi_{L2})
    +\chi|\varphi_{3}|(\overline{\psi}_{L3}\psi_{R3}+\overline{\psi}_{R3}\psi_{L3})\nonumber\\
    \nonumber\\
    &+&\mathrm{i}\chi\varphi_{01}(\overline{\psi}_{L1}\psi_{R1}-\overline{\psi}_{R1}\psi_{L1})\theta_{1}
    +\mathrm{i}\chi\varphi_{02}(\overline{\psi}_{L2}\psi_{R2}-\overline{\psi}_{R2}\psi_{L2})\theta_{2}
    +\mathrm{i}\chi\varphi_{03}(\overline{\psi}_{L3}\psi_{R3}-\overline{\psi}_{R3}\psi_{L3})\theta_{3},
\end{eqnarray}
where $\theta=\theta(t,\mathbf{r})$ is small phase oscillations $|\theta|\ll 1$. Unlike the interaction with the amplitudes of the scalar fields $|\varphi_{i}|$, interaction with the phase oscillations would have to violate the $P$-invariance. However, we can see, that the Dirac field $\psi_{i}=\psi_{Li}+\psi_{Ri}$ of each band interacts with corresponding phase of the scalar field $\theta_{i}$. As have been demonstrated in Sect.\ref{spontanU1gauge}, the phase oscillations $\theta_{i}(t,\mathbf{r})$ are absorbed by the gauge fields $A_{\mu}$ due to the Higgs mechanism, hence in Eq.(\ref{4.15}) the phases are equal to their equilibrium value $\theta_{i}=\theta^{0}_{i}=0$.


\section{Spontaneous breaking of $SU(2)_{I}$ gauge symmetry in the three-band system with the Josephson couplings}\label{spontanSU2}

Let the fields $\Psi_{1},\Psi_{2},\Psi_{3}$ are isospinors, where each has two complex (four real) scalar components:
\begin{equation}\label{5.1}
  \Psi=\left(\begin{array}{c}
           \varphi^{(1)} \\
           \varphi^{(2)} \\
         \end{array}\right),\quad
         \Psi^{+}=\left(\begin{array}{cc}
             \varphi^{(1)\ast}, & \varphi^{(2)\ast} \\
           \end{array}\right)
\end{equation}
being transformed at rotation in the isospace as
\begin{equation}\label{5.1a}
  \Psi=S\Psi',\quad S=e^{\mathrm{i}\frac{\vec{\tau}}{2}\vec{\vartheta}}=\left(\tau_{0}\cos\frac{\vartheta}{2}+\mathrm{i}(\vec{n}\vec{\tau})\sin\frac{\vartheta}{2}\right),
\end{equation}
where $\vec{\tau}=(\tau_{x},\tau_{y},\tau_{z})$ is a vector consisting of Pauli matrices, $\tau_{0}$ is an identity matrix, $\vec{\vartheta}=\vec{n}\vartheta$, where $\vec{n}$ is an unit vector in the direction of the axis around which the rotation is made in the isospace. Thus, the isospinor fields, corresponding for each band, can be represented in the following form:
\begin{equation}\label{5.2}
    \Psi_{1}(x)=e^{\mathrm{i}\frac{\vec{\tau}}{2}\vec{\vartheta}_{1}(x)}\left(\begin{array}{c}
           0 \\
           \varphi_{1}(x) \\
         \end{array}\right),\quad
    \Psi_{2}(x)=e^{\mathrm{i}\frac{\vec{\tau}}{2}\vec{\vartheta}_{2}(x)}\left(\begin{array}{c}
           0 \\
           \varphi_{2}(x) \\
         \end{array}\right),\quad
    \Psi_{3}(x)=e^{\mathrm{i}\frac{\vec{\tau}}{2}\vec{\vartheta}_{3}(x)}\left(\begin{array}{c}
           0 \\
           \varphi_{3}(x) \\
         \end{array}\right),
\end{equation}
where $\varphi_{1},\varphi_{2},\varphi_{3}$ are real and $\varphi_{1},\varphi_{2},\varphi_{3}>0$. Thus, \emph{we assign the third projection of isospin as} $I_{z}=-\frac{1}{2}$ \emph{to the scalar fields} $\varphi_{i}$, \emph{then hypercharge} $Y=1$ \emph{so, that the electrical charge is} $Q=I_{z}+\frac{Y}{2}=0$. At the same time, the phases $\vec{\vartheta}_{i}$ are characterized with zero charges $I_{z}=Y=Q=0$.  Corresponding Lagrangian $\mathcal{L}$ is a sum of the gauge invariant part (relative to the $SU(2)$ gauge symmetry) and the Josephson terms:
\begin{eqnarray}\label{5.4}
    \mathcal{L}&=&\partial_{\mu}\Psi_{1}\partial^{\mu}\Psi_{1}^{+}+\partial_{\mu}\Psi_{2}\partial^{\mu}\Psi_{2}^{+}
    +\partial_{\mu}\Psi_{3}\partial^{\mu}\Psi_{3}^{+}\nonumber\\
    &-&a_{1}\Psi_{1}\Psi_{1}^{+}-a_{2}\Psi_{2}\Psi_{2}^{+}-a_{3}\Psi_{3}\Psi_{3}^{+}
    -\frac{b_{1}}{2}\left(\Psi_{1}\Psi_{1}^{+}\right)^{2}-\frac{b_{2}}{2}\left(\Psi_{2}\Psi_{2}^{+}\right)^{2}
    -\frac{b_{3}}{2}\left(\Psi_{3}\Psi_{3}^{+}\right)^{2}\nonumber\\
    &-&\epsilon\left(\Psi_{1}^{+}\Psi_{2}+\Psi_{1}\Psi_{2}^{+}\right)-\epsilon\left(\Psi_{1}^{+}\Psi_{3}+\Psi_{1}\Psi_{3}^{+}\right)
    -\epsilon\left(\Psi_{2}^{+}\Psi_{3}+\Psi_{2}\Psi_{3}^{+}\right).
\end{eqnarray}
The Josephson terms $\Psi_{i}^{+}\Psi_{j}+\Psi_{j}\Psi_{i}^{+}$ are not invariant relatively to the $SU(2)$ gauge symmetry, however these terms should depend on the phase differences $\vartheta_{i}-\vartheta_{j}$ only: $\Psi_{i}\Psi_{j}^{+}+\Psi_{i}^{+}\Psi_{j}=2\varphi_{i}\varphi_{j}\cos\frac{\vartheta_{i}-\vartheta_{j}}{2}$, in order to have a physical sense as interference between condensates $\Psi_{1},\Psi_{2},\Psi_{3}$. In order to ensure such a property it is necessary:
\begin{equation}\label{5.4a}
  \vec{n}_{1}=\vec{n}_{2}=\vec{n}_{3},
\end{equation}
that is the isospinors (\ref{5.2}) must rotate around a common axis. Moreover, it is not difficult to see, that
\begin{equation}\label{5.4b}
  \Psi=e^{\mathrm{i}\frac{\vec{\tau}}{2}\vec{\vartheta}}\left(\begin{array}{c}
           0 \\
           \varphi \\
         \end{array}\right)=
         \left(\begin{array}{c}
           [\mathrm{i}\varphi n_{x}+\varphi n_{y}]\sin\frac{\vartheta}{2} \\
           \varphi\cos\frac{\vartheta}{2}-\mathrm{i}n_{z}\varphi\sin\frac{\vartheta}{2} \\
         \end{array}\right),
\end{equation}
then,
\begin{eqnarray}\label{5.4d}
  \Psi_{k}\Psi_{j}^{+}+\Psi_{j}^{+}\Psi_{k}&=&
  \left(
    \begin{array}{cc}
      [-\mathrm{i}\varphi_{j}n_{x}+\varphi_{j} n_{y}]\sin\frac{\vartheta_{j}}{2}, & \varphi_{j}\cos\frac{\vartheta_{j}}{2}+\mathrm{i}n_{z}\varphi_{j}\sin\frac{\vartheta_{j}}{2} \\
    \end{array}
  \right)
         \left(\begin{array}{c}
           [\mathrm{i}\varphi_{k} n_{x}+\varphi_{k} n_{y}]\sin\frac{\vartheta_{k}}{2} \\
           \varphi_{k}\cos\frac{\vartheta_{k}}{2}-\mathrm{i}n_{z}\varphi_{k}\sin\frac{\vartheta_{k}}{2} \\
         \end{array}\right)\nonumber\\
         &+&
         \left(
    \begin{array}{cc}
      [-\mathrm{i}\varphi_{k} n_{x}+\varphi_{k} n_{y}]\sin\frac{\vartheta_{k}}{2}, & \varphi_{k}\cos\frac{\vartheta_{k}}{2}+\mathrm{i}n_{z}\varphi_{k}\sin\frac{\vartheta_{k}}{2} \\
    \end{array}
  \right)
         \left(\begin{array}{c}
           [\mathrm{i}\varphi_{j} n_{x}+\varphi_{j} n_{y}]\sin\frac{\vartheta_{j}}{2} \\
           \varphi_{j}\cos\frac{\vartheta_{j}}{2}-\mathrm{i}n_{z}\varphi_{j}\sin\frac{\vartheta_{j}}{2} \\
         \end{array}\right)\nonumber\\
         &=&
         2\varphi_{j}\varphi_{k}\left[\cos\frac{\vartheta_{j}}{2}\cos\frac{\vartheta_{k}}{2}
         +n_{z}^{2}\sin\frac{\vartheta_{j}}{2}\sin\frac{\vartheta_{k}}{2}\right]\nonumber\\
         &+&2\varphi_{j}\varphi_{k}\left[n_{x}^{2}+n_{y}^{2}\right]\sin\frac{\vartheta_{j}}{2}\sin\frac{\vartheta_{k}}{2}.
\end{eqnarray}
Therefore, it must be
\begin{equation}\label{5.4c}
  n_{x}=n_{y}=0\Rightarrow n_{z}=\pm 1
\end{equation}
Then, substituting representation (\ref{5.2}) in the Lagrangian (\ref{5.4}) we obtain:
\begin{eqnarray}\label{5.5}
    \mathcal{L}&=&\partial_{\mu}\varphi_{1}\partial^{\mu}\varphi_{1}
    +\partial_{\mu}\varphi_{2}\partial^{\mu}\varphi_{2}+\partial_{\mu}\varphi_{3}\partial^{\mu}\varphi_{3}\nonumber\\
    &+&\varphi_{1}^{2}\partial_{\mu}\frac{\vartheta_{1}}{2}\partial^{\mu}\frac{\vartheta_{1}}{2}
    +\varphi_{2}^{2}\partial_{\mu}\frac{\vartheta_{2}}{2}\partial^{\mu}\frac{\vartheta_{2}}{2}
    +\varphi_{3}^{2}\partial_{\mu}\frac{\vartheta_{3}}{2}\partial^{\mu}\frac{\vartheta_{3}}{2}\nonumber\\
    &-&a_{1}\varphi_{1}^{2}-\frac{b_{1}}{2}\varphi_{1}^{4}-a_{2}\varphi_{2}^{2}-\frac{b_{2}}{2}\varphi_{2}^{4}
    -a_{3}\varphi_{3}^{2}-\frac{b_{3}}{2}\varphi_{3}^{4}\nonumber\\
    &-&2\epsilon\varphi_{1}\varphi_{2}\cos\frac{\vartheta_{1}-\vartheta_{2}}{2}
    -2\epsilon\varphi_{1}\varphi_{3}\cos\frac{\vartheta_{1}-\vartheta_{3}}{2}
    -2\epsilon\varphi_{2}\varphi_{3}\cos\frac{\vartheta_{2}-\vartheta_{3}}{2}.
\end{eqnarray}
Let us consider stationary and spatially homogeneous case, i.e. $\partial_{t}\varphi=0,\nabla\varphi=0$, $\partial_{t}\vartheta=0,\nabla\vartheta=0$. Then we obtain equations for the equilibrium values of the fields $\varphi_{0i}$ and $\vartheta_{i}^{0}-\vartheta_{j}^{0}$:
\begin{equation}\label{5.6}
\left\{\begin{array}{c}
  a_{1}\varphi_{01}+\epsilon\varphi_{02}\cos\frac{\vartheta_{2}^{0}-\vartheta_{1}^{0}}{2}
  +\epsilon\varphi_{03}\cos\frac{\vartheta_{3}^{0}-\vartheta_{1}^{0}}{2}+b_{1}\varphi_{01}^{3}=0 \\
  a_{2}\varphi_{02}+\epsilon\varphi_{01}\cos\frac{\vartheta_{1}^{0}-\vartheta_{2}^{0}}{2}
  +\epsilon\varphi_{03}\cos\frac{\vartheta_{3}^{0}-\vartheta_{2}^{0}}{2}+b_{1}\varphi_{02}^{3}=0 \\
  a_{3}\varphi_{03}+\epsilon\varphi_{01}\cos\frac{\vartheta_{1}^{0}-\vartheta_{3}^{0}}{2}
  +\epsilon\varphi_{02}\cos\frac{\vartheta_{2}^{0}-\vartheta_{3}^{0}}{2}+b_{1}\varphi_{03}^{3}=0 \\
  \varphi_{02}\sin\frac{\vartheta_{2}^{0}-\vartheta_{1}^{0}}{2}+\varphi_{03}\sin\frac{\vartheta_{3}^{0}-\vartheta_{1}^{0}}{2}=0 \\
  \varphi_{01}\sin\frac{\vartheta_{1}^{0}-\vartheta_{2}^{0}}{2}+\varphi_{03}\sin\frac{\vartheta_{3}^{0}-\vartheta_{2}^{0}}{2}=0 \\
  \varphi_{01}\sin\frac{\vartheta_{1}^{0}-\vartheta_{3}^{0}}{2}+\varphi_{02}\sin\frac{\vartheta_{2}^{0}-\vartheta_{3}^{0}}{2}=0\\
\end{array}\right\}.
\end{equation}
If the interband coupling is weak, i.e. $|\epsilon|\ll|a_{1}|,|a_{2}|,|a_{3}|$, then we can assume $\varphi_{0i}=\sqrt{\frac{a_{i}}{b_{i}}}$.

From the other hand, let us consider three Dirac spinor fields $\psi_{1},\psi_{2},\psi_{3}$ as we have considered them in Sect.\ref{dirac}. However, now the Lagrangian (\ref{4.1}) has the form:
\begin{eqnarray}\label{5.7}
    \mathcal{L}&=&\sum_{i=1}^{3}\left[\mathrm{i}\overline{\psi}_{Li}\gamma^{\mu}\overset{\leftrightarrow}{\partial}_{\mu}\psi_{Li}
    +\mathrm{i}\overline{\psi}_{Ri}\gamma^{\mu}\overset{\leftrightarrow}{\partial}_{\mu}\psi_{Ri}
    -\chi(\overline{\psi}_{Li}\Psi_{i}\psi_{Ri}+\overline{\psi}_{Ri}\Psi_{i}^{+}\psi_{Li})\right].
\end{eqnarray}
In order for the terms of interaction of Dirac fields with isospinor fields $\mathcal{U}_{D}=\chi(\overline{\psi}_{L}\Psi\psi_{R}+\overline{\psi}_{R}\Psi^{+}\psi_{L})$ to take the form of the mass term of Dirac type $\mathcal{U}_{D}=m_{D}(\overline{\psi}_{L}\psi_{R}+\overline{\psi}_{R}\psi_{L})$ the following conditions must be satisfied:
\begin{enumerate}
  \item Since $\mathcal{L}$ must be a scalar and $\Psi$ is a doublet (\ref{5.1}), then the spinor $\psi_{L}$ must be a dublet (singlet) and $\psi_{R}$ must be a singlet (dublet). That is, for example, $\psi_{L}=\left(
                                                                              \begin{array}{c}
                                                                                \nu_{L} \\
                                                                                l_{L} \\
                                                                              \end{array}
                                                                            \right)$ and $\psi_{R}=l_{R}$.
  This means violation of the spatial parity symmetry. Here $l_{L,R}$ are electrically charged leptons: $Q_{L}=Q_{R}=-1$, $I_{zL}=-\frac{1}{2},Y_{L}=-1,I_{zR}=0,Y_{R}=-2$, at the same time, $\nu_{L}$ is electrically neutral leptons (neutrino): $Q=0$, $I_{zL}=\frac{1}{2}$, $Y_{L}=-1$.
  \item As in Sect.\ref{dirac}, the coupling terms $\mathcal{U}_{D}$ in Eq.(\ref{5.7}) can be reduced to the Dirac mass terms, only when three condensates (\ref{5.1}) have the same equilibrium phases, which must be supposed as $\vartheta^{0}_{1}=\vartheta^{0}_{2}=\vartheta^{0}_{3}=0$. This is possible only in the case of attractive interband coupling $\epsilon<0$.
 \item   If we use an isospinor $\Psi=\left(\begin{array}{c}
           \varphi^{(1)} \\
           \varphi^{(2)} \\
         \end{array}\right)$ the coupling terms $\mathcal{U}_{D}=\chi(\overline{\psi}_{L}\Psi\psi_{R}+\overline{\psi}_{R}\Psi^{+}\psi_{L})$ takes the form: $\mathcal{U}_{D}=\chi\left(\overline{\nu}_{L}\varphi^{(1)}l_{R}+\overline{l}_{R}\varphi^{(1)+}\nu_{L}\right)
         +\chi\left(\overline{l}_{L}\varphi^{(2)}l_{R}+\overline{l}_{R}\varphi^{(2)+}l_{L}\right)$. We can see, that we must suppose $\varphi^{(1)}=0$ to avoid mixing of neutrinos with the charged leptons, and $\varphi^{(2)+}=\varphi^{(2)}>0$ . This corresponds to our selection of the vacuum as (\ref{5.4c}).
 \item neutrino masses are supposed zero: $m_{\nu1}=m_{\nu2}=m_{\nu3}=0$. We postpone discussion of this issue until Sect.\ref{mixing}.
\end{enumerate}
At the same time, except the band states $\psi_{1},\psi_{2},\psi_{3}$ (i.e. the states, which interact with the corresponding isospinor fields), the flavor states $\psi_{e},\psi_{\mu},\psi_{\tau}$ (i.e. the states, which interact with normal oscillation modes of the coupled isospinor fields) must exist:
\begin{equation}\label{5.24}
  \psi_{Le}=\left(\begin{array}{c}
    \nu_{Le} \\
    e_{L} \\
    \end{array}\right),\psi_{Re}=e_{R},\quad
    \psi_{L\mu}=\left(\begin{array}{c}
    \nu_{L\mu} \\
    \mu_{L} \\
    \end{array}\right),\psi_{R\mu}=\mu_{R},\quad
    \psi_{L\tau}=\left(\begin{array}{c}
    \nu_{L\tau} \\
    \tau_{L} \\
    \end{array}\right),\psi_{R\tau}=\tau_{R}
\end{equation}
The relationship between the band states and the flavor states (i.e. the lepton oscillations) has been considered in Sect.\ref{dirac} and will be considered else in Sect.\ref{mixing}.

Let us consider movement of the phases $\vartheta_{1,2,3}$. Corresponding Lagrange equations for Lagrangian (\ref{5.5}) are:
\begin{eqnarray}\label{5.8}
  &&\varphi_{01}^{2}\partial_{\mu}\partial^{\mu}\vartheta_{1}
  -2\varphi_{01}\varphi_{02}\epsilon\sin\frac{\vartheta_{1}-\vartheta_{2}}{2}-2\varphi_{01}\varphi_{03}\epsilon\sin\frac{\vartheta_{1}-\vartheta_{3}}{2}=0\nonumber\\
  &&\varphi_{02}^{2}\partial_{\mu}\partial^{\mu}\vartheta_{2}
  +2\varphi_{01}\varphi_{02}\epsilon\sin\frac{\vartheta_{1}-\vartheta_{2}}{2}-2\varphi_{02}\varphi_{03}\epsilon\sin\frac{\vartheta_{2}-\vartheta_{3}}{2}=0\\
  &&\varphi_{03}^{2}\partial_{\mu}\partial^{\mu}\vartheta_{3}
  +2\varphi_{01}\varphi_{03}\epsilon\sin\frac{\vartheta_{1}-\vartheta_{3}}{2}+2\varphi_{02}\varphi_{03}\epsilon\sin\frac{\vartheta_{2}-\vartheta_{3}}{2}=0\nonumber
\end{eqnarray}
As we have seen above, the coupling terms $\mathcal{U}_{D}$ in Eq.(\ref{5.7}) can be reduced to the Dirac mass terms, only when three condensates $\langle\Psi_{1}\rangle,\langle\Psi_{2}\rangle,\langle\Psi_{3}\rangle$ have the same equilibrium phases $\vartheta^{0}_{1}=\vartheta^{0}_{2}=\vartheta^{0}_{3}=0$. This is possible only in the case of attractive interband coupling $\epsilon<0$. Considering small variations, i.e. $|\vartheta|\ll\pi$, we can linearise Eq.(\ref{5.8}):
\begin{eqnarray}\label{5.9}
  &&\varphi_{01}^{2}\partial_{\mu}\partial^{\mu}\vartheta_{1}
  -\varphi_{01}\varphi_{02}\epsilon(\vartheta_{1}-\vartheta_{2})-\varphi_{01}\varphi_{03}\epsilon(\vartheta_{1}-\vartheta_{3})=0\nonumber\\
  &&\varphi_{02}^{2}\partial_{\mu}\partial^{\mu}\vartheta_{2}
  +\varphi_{01}\varphi_{02}\epsilon(\vartheta_{1}-\vartheta_{2})-\varphi_{02}\varphi_{03}\epsilon(\vartheta_{2}-\vartheta_{3})=0\\
  &&\varphi_{03}^{2}\partial_{\mu}\partial^{\mu}\vartheta_{3}
  +\varphi_{01}\varphi_{03}\epsilon(\vartheta_{1}-\vartheta_{3})+\varphi_{02}\varphi_{03}\epsilon(\vartheta_{2}-\vartheta_{3})=0,\nonumber
\end{eqnarray}
that coincides with Eq.(\ref{2.15}) for the phases $\theta_{1},\theta_{2},\theta_{3}$ when $\epsilon<0$, i.e. all equilibrium phase differences are $\theta_{ij}^{0}=0\Rightarrow\cos\theta_{ij}^{0}=1$. Hence, the spectrum of Goldstone modes due to the spontaneous breaking of $SU(2)$ gauge symmetry in the three-band system with the interband coupling coincides with the spectrum (\ref{2.17}) of Goldstone modes due to the spontaneous breaking of $U(1)$ gauge symmetry in the three-band system with the interband coupling.

Let us consider oscillations of $\varphi_{i}$ only. Then, at $\epsilon<0$ (i.e. equilibrium phase differences are $\vartheta_{ij}^{0}=0$) the Lagrangian (\ref{5.5}) takes the form:
\begin{eqnarray}\label{5.10}
    \mathcal{L}&=&\partial_{\mu}\varphi_{1}\partial^{\mu}\varphi_{1}
    +\partial_{\mu}\varphi_{2}\partial^{\mu}\varphi_{2}+\partial_{\mu}\varphi_{3}\partial^{\mu}\varphi_{3}\nonumber\\
    &-&a_{1}\varphi_{1}^{2}-\frac{b_{1}}{2}\varphi_{1}^{4}-a_{2}\varphi_{2}^{2}-\frac{b_{2}}{2}\varphi_{2}^{4}
    -a_{3}\varphi_{3}^{2}-\frac{b_{3}}{2}\varphi_{3}^{4}\nonumber\\
    &-&2\epsilon\varphi_{1}\varphi_{2}-2\epsilon\varphi_{1}\varphi_{3}-2\epsilon\varphi_{2}\varphi_{3},
\end{eqnarray}
that coincides with the Lagrangian (\ref{2.10}) for the fields $|\varphi_{1}|,|\varphi_{2}|,|\varphi_{3}|$ when $\epsilon<0$. Hence, the spectrum of Higgs modes due to the spontaneous breaking of $SU(2)$ gauge symmetry in the three-band system with the interband coupling coincides with the spectrum (\ref{2.27}) of Higgs modes due to the spontaneous breaking of $U(1)$ gauge symmetry in the three-band system with the interband couplings.

Let us consider interaction of the isospinor fields $\Psi_{1,2,3}$ (\ref{5.2}), breaking the gauge $SU(2)$ symmetry each, with the gauge Yang–Mills field $\vec{A}_{\mu}$. Corresponding gauge invariant Lagrangian has the form:
\begin{eqnarray}\label{5.11}
    \mathcal{L}&=&D_{\mu}\Psi_{1}\left(D^{\mu}\Psi_{1}\right)^{+}+D_{\mu}\Psi_{2}\left(D^{\mu}\Psi_{2}\right)^{+}
    +D_{\mu}\Psi_{3}\left(D^{\mu}\Psi_{3}\right)^{+}\nonumber\\
    &-&a_{1}\Psi_{1}\Psi_{1}^{+}-a_{2}\Psi_{2}\Psi_{2}^{+}-a_{3}\Psi_{3}\Psi_{3}^{+}
    -\frac{b_{1}}{2}\left(\Psi_{1}\Psi_{1}^{+}\right)^{2}-\frac{b_{2}}{2}\left(\Psi_{2}\Psi_{2}^{+}\right)^{2}
    -\frac{b_{3}}{2}\left(\Psi_{3}\Psi_{3}^{+}\right)^{2}\nonumber\\
    &-&\epsilon\left(\Psi_{1}^{+}\Psi_{2}+\Psi_{1}\Psi_{2}^{+}\right)-\epsilon\left(\Psi_{1}^{+}\Psi_{3}+\Psi_{1}\Psi_{3}^{+}\right)
    -\epsilon\left(\Psi_{2}^{+}\Psi_{3}+\Psi_{2}\Psi_{3}^{+}\right)-\frac{1}{16\pi}\vec{F}_{\mu\nu}\vec{F}^{\mu\nu},
\end{eqnarray}
where
\begin{equation}\label{5.12}
  D_{\mu}\equiv\tau_{0}\partial_{\mu}-\mathrm{i}g\frac{\vec{\tau}}{2}\vec{A}_{\mu}
\end{equation}
is the covariant derivation,
\begin{equation}\label{5.13}
  \vec{F}_{\mu\nu}=\partial_{\mu}\vec{A}_{\nu}-\partial_{\nu}\vec{A}_{\mu}+g\left[\vec{A}_{\mu}\times\vec{A}_{\nu}\right]
\end{equation}
is the tensor of the Yang–Mills field. Using Eq.(\ref{5.2}), where it can be supposed $|\vartheta|\ll 1\Rightarrow S=\tau_{0}+\mathrm{i}\frac{\vec{\tau}}{2}\vec{\vartheta}$, and using a property of the Pauli matrixes $-\mathrm{i}\left[\frac{\vec{\tau}}{2}\cdot\vec{\vartheta}^{\mu},\frac{\vec{\tau}}{2}\cdot\vec{A}^{\mu}\right]=
\left[\vec{\vartheta}\times\vec{A}^{\mu}\right]\cdot\frac{\vec{\tau}}{2}$ \cite{sad}, the Lagrangian (\ref{5.11}) can be rewritten in the following form:
\begin{eqnarray}\label{5.14}
    \mathcal{L}&=&\sum_{i=1}^{3}\left(\begin{array}{cc}
                      0, & \varphi_{i} \\
                    \end{array}\right)
                    \left(\tau_{0}\partial^{\mu}-i\frac{\vec{\tau}}{2}\partial^{\mu}\vec{\vartheta}_{i}
                    +\mathrm{i}g\frac{\vec{\tau}}{2}\vec{A}^{\mu}+\mathrm{i}g\left[\frac{\vec{\tau}}{2}\times\vec{\vartheta}_{i}\right]\vec{A}^{\mu}\right)
                    \left(\tau_{0}\partial_{\mu}+\mathrm{i}\frac{\vec{\tau}}{2}\partial_{\mu}\vec{\vartheta}_{i}
                    -\mathrm{i}g\frac{\vec{\tau}}{2}\vec{A}_{\mu}-\mathrm{i}g\left[\frac{\vec{\tau}}{2}\times\vec{\vartheta}_{i}\right]\vec{A}_{\mu}\right)
                    \left(\begin{array}{c}
                        0 \\
                        \varphi_{i}\\
                      \end{array}\right)\nonumber\\
    &-&a_{1}\varphi_{1}^{2}-a_{2}\varphi_{2}^{2}-a_{3}\varphi_{3}^{2}
    -\frac{b_{1}}{2}\varphi_{1}^{4}-\frac{b_{2}}{2}\varphi_{2}^{4}-\frac{b_{3}}{2}\varphi_{3}^{4}\nonumber\\
    &-&2\epsilon\varphi_{1}\varphi_{2}\cos\frac{\vartheta_{1}-\vartheta_{2}}{2}
    -2\epsilon\varphi_{1}\varphi_{3}\cos\frac{\vartheta_{1}-\vartheta_{3}}{2}
    -2\epsilon\varphi_{2}\varphi_{3}\cos\frac{\vartheta_{2}-\vartheta_{3}}{2}-\frac{1}{16\pi}\vec{F}_{\mu\nu}\vec{F}^{\mu\nu}.
\end{eqnarray}
Corresponding Lagrange equation
\begin{equation}\label{5.15}
  \partial_{\nu}\frac{\partial\mathcal{L}}{\partial(\partial_{\nu}\vec{A}_{\mu})}-\frac{\partial\mathcal{L}}{\partial \vec{A}_{\mu}}=0
\end{equation}
with the Yang–Mills equation $\partial_{\nu}\vec{F}^{\mu\nu}+g\left[\vec{A}_{\nu}\times\vec{F}^{\mu\nu}\right]=4\pi\vec{J}^{\mu}$ give the current:
\begin{eqnarray}\label{5.16}
  \vec{J}^{\mu}&=&\frac{g}{2}\sum_{i=1}^{3}\varphi_{0i}^{2}\left(\partial^{\mu}\vec{\vartheta}_{i}
                    -g\vec{A}^{\mu}-g\left[\vec{\vartheta}_{i}\times\vec{A}^{\mu}\right]\right).
\end{eqnarray}
The gauge field can be transformed as
\begin{equation}\label{5.17}
    \vec{A}_{\mu}'=\vec{A}_{\mu}-\alpha\left(\frac{1}{g}\partial_{\mu}\vec{\vartheta}_{1}-\left[\vec{\vartheta}_{1}\times\vec{A}_{\mu}\right]\right)
    -\beta\left(\frac{1}{g}\partial_{\mu}\vec{\vartheta}_{2}-\left[\vec{\vartheta}_{2}\times\vec{A}_{\mu}\right]\right)
    -\gamma\left(\frac{1}{g}\partial_{\mu}\vec{\vartheta}_{3}-\left[\vec{\vartheta}_{3}\times\vec{A}_{\mu}\right]\right),
\end{equation}
where
\begin{eqnarray}\label{5.18}
\alpha=\frac{\varphi_{01}^{2}}{\varphi_{01}^{2}+\varphi_{02}^{2}+\varphi_{03}^{2}}\quad
\beta=\frac{\varphi_{02}^{2}}{\varphi_{01}^{2}+\varphi_{02}^{2}+\varphi_{03}^{2}},\quad
\gamma=\frac{\varphi_{03}^{2}}{\varphi_{01}^{2}+\varphi_{02}^{2}+\varphi_{03}^{2}},
\end{eqnarray}
which are analogous to Eqs.(\ref{3.4},\ref{3.5}). Then, neglecting the second order of smallness in the phase $\vartheta\partial_{\mu}\vartheta$, Eq.(\ref{5.16}) can be reduced to the "London law":
\begin{equation}\label{5.19}
  \vec{J}^{\mu}=-\frac{g^{2}}{2}\left(\varphi_{01}^{2}+\varphi_{02}^{2}+\varphi_{03}^{2}\right)\vec{A}^{\mu}\equiv-\frac{1}{4\pi\lambda^{2}}\vec{A}^{\mu},
\end{equation}
where
\begin{equation}\label{5.20}
  \lambda=\frac{1}{\sqrt{2\pi g^{2}\left(\varphi_{01}^{2}+\varphi_{02}^{2}+\varphi_{03}^{2}\right)}}
\end{equation}
is the "penetration depth" - the length of interaction mediated by the gauge bosons $\vec{A}_{\mu}$.

Applying the transformation (\ref{5.17}) we can transform the Lagrangian (\ref{5.11}) to the following form:
\begin{eqnarray}\label{5.21}
    \mathcal{L}&=&\partial_{\mu}\varphi_{1}\partial^{\mu}\varphi_{1}+\partial_{\mu}\varphi_{2}\partial^{\mu}\varphi_{2}+
    \partial_{\mu}\varphi_{3}\partial^{\mu}\varphi_{3}+
    \frac{g^{2}}{4}\left(\varphi_{1}^{2}+\varphi_{2}^{2}+\varphi_{3}^{2}\right)\vec{A}_{\mu}\vec{A}^{\mu}\nonumber\\
    &-&2\epsilon\varphi_{1}\varphi_{2}\cos\frac{\vartheta_{1}-\vartheta_{2}}{2}
    -2\epsilon\varphi_{1}\varphi_{3}\cos\frac{\vartheta_{1}-\vartheta_{3}}{2}
    -2\epsilon\varphi_{2}\varphi_{3}\cos\frac{\vartheta_{2}-\vartheta_{3}}{2}\nonumber\\
    &+&\left(\varphi_{1}^{2}\beta^{2}+\varphi_{2}^{2}\alpha^{2}\right)
    \partial_{\mu}\frac{\vartheta_{1}-\vartheta_{2}}{2}\partial^{\mu}\frac{\vartheta_{1}-\vartheta_{2}}{2}\nonumber\\
    &+&\left(\varphi_{1}^{2}\gamma^{2}+\varphi_{3}^{2}\alpha^{2}\right)
    \partial_{\mu}\frac{\vartheta_{1}-\vartheta_{3}}{2}\partial^{\mu}\frac{\vartheta_{1}-\vartheta_{3}}{2}\nonumber\\
    &+&\left(\varphi_{2}^{2}\gamma^{2}+\varphi_{3}^{2}\beta^{2}\right)
    \partial_{\mu}\frac{\vartheta_{2}-\vartheta_{3}}{2}\partial^{\mu}\frac{\vartheta_{2}-\vartheta_{3}}{2}\nonumber\\
    &-&\varphi_{1}^{2}2\gamma\beta\partial_{\mu}\frac{\vartheta_{1}-\vartheta_{2}}{2}\partial^{\mu}\frac{\vartheta_{1}-\vartheta_{3}}{2}
    -\varphi_{2}^{2}2\alpha\gamma\partial_{\mu}\frac{\vartheta_{1}-\vartheta_{2}}{2}\partial^{\mu}\frac{\vartheta_{2}-\vartheta_{3}}{2}
    -\varphi_{3}^{2}2\alpha\beta\partial_{\mu}\frac{\vartheta_{1}-\vartheta_{3}}{2}\partial^{\mu}\frac{\vartheta_{2}-\vartheta_{3}}{2}\nonumber\\
    &-&a_{1}\varphi_{1}^{2}-a_{2}\varphi_{2}^{2}-a_{3}\varphi_{3}^{2}
    -\frac{b_{1}}{2}\varphi_{1}^{4}-\frac{b_{2}}{2}\varphi_{2}^{4}-\frac{b_{3}}{2}\varphi_{3}^{4}-\frac{1}{16\pi}\vec{F}_{\mu\nu}\vec{F}^{\mu\nu},
\end{eqnarray}
which is analogous to the calibrated Lagrangian (\ref{3.11}) with the spontaneous breaking of $U(1)$ symmetry. We can see, that the phases $\vartheta_{1},\vartheta_ {2},\vartheta_ {3}$ have been excluded from the Lagrangian individually leaving only their differences: $\vartheta_{1}-\vartheta_{2},\vartheta_{1}-\vartheta_{3},\vartheta_{2}-\vartheta_{3}$. Thus, the gauge field $\vec{A}_{\mu}$ absorbs the Goldstone boson (i.e. the common mode oscillations, where $\nabla\vartheta_{1}=\nabla\vartheta_{2}=\nabla\vartheta_{3}$) with the acoustic spectrum (\ref{2.19}). At the same time, the Leggett bosons (i.e. the oscillations of the relative phases $\vartheta_{i}-\vartheta_{j}$) with massive spectrum (\ref{2.20a},\ref{2.20b}) "survive".

Substituting the calibrated Lagrangian (\ref{5.21}) in the Eq.(\ref{5.15}) we obtain the equation for the field $\vec{A}_{\mu}$:
\begin{equation}\label{5.22}
   \partial_{\nu}\vec{F}^{\nu\mu}+g\left[\vec{A}_{\nu}\times\vec{F}^{\nu\mu}\right]+\frac{1}{\lambda^{2}}\vec{A}^{\mu}=0,
\end{equation}
where
\begin{equation}\label{5.23}
   \frac{1}{\lambda^{2}}=2\pi g^{2}\left(\varphi_{01}^{2}+\varphi_{02}^{2}+\varphi_{03}^{2}\right)\equiv m_{A}^{2}
\end{equation}
is the squared mass of the gauge boson $\vec{A}^{\mu}$, which is the squared reciprocal "penetration depth" (\ref{5.20}) in the London law (\ref{5.19}).

\section{Spontaneous breaking of $SU(2)_{I}\otimes U(1)_{Y}$ gauge symmetry in the three-band system with the Josephson couplings}\label{spontanSU2U1}

It is not difficult to notice that the scalar product of the isospinors (\ref{5.1}) $\Psi\Psi^{+}$ is invariant under both the $SU(2)$ transformation and the $U(1)$ transformation. Thus we can write by analogy with Eq.(\ref{5.2}):
\begin{equation}\label{6.1}
    \Psi_{1}(x)=e^{\mathrm{i}\theta_{1}(x)}e^{\mathrm{i}\frac{\vec{\tau}}{2}\vec{\vartheta}_{1}(x)}\left(\begin{array}{c}
           0 \\
           \varphi_{1}(x) \\
         \end{array}\right),\quad
    \Psi_{2}(x)=e^{\mathrm{i}\theta_{2}(x)}e^{\mathrm{i}\frac{\vec{\tau}}{2}\vec{\vartheta}_{2}(x)}\left(\begin{array}{c}
           0 \\
           \varphi_{2}(x) \\
         \end{array}\right),\quad
    \Psi_{3}(x)=e^{\mathrm{i}\theta_{3}(x)}e^{\mathrm{i}\frac{\vec{\tau}}{2}\vec{\vartheta}_{3}(x)}\left(\begin{array}{c}
           0 \\
           \varphi_{3}(x) \\
         \end{array}\right).
\end{equation}
Lagrangian (\ref{5.4}) is a sum of the gauge invariant part (relative to the $SU(2)\otimes U(1)$ gauge symmetry) and the Josephson terms. As in the previous case, the Josephson terms are not invariant relatively to this gauge symmetry due to the terms $\Psi_{i}^{+}\Psi_{j}+\Psi_{j}\Psi_{i}^{+}$, however these terms depend on the phase differences $\vartheta_{i}-\vartheta_{j}$ and $\theta_{i}-\theta_{j}$ only, but not on the single phases $\vartheta_{i},\theta_{i}$, if the conditions (\ref{5.4a},\ref{5.4c}) are satisfied. Then, substituting representation (\ref{6.1}) in the Lagrangian (\ref{5.4}) in consideration of (\ref{5.4c}), we obtain:
\begin{eqnarray}\label{6.2}
    \mathcal{L}&=&\sum_{i=1}^{3}\left[\partial_{\mu}\varphi_{i}\partial^{\mu}\varphi_{i}
    +\varphi_{i}^{2}\left(\partial_{\mu}\frac{\vartheta_{i}}{2}\partial^{\mu}\frac{\vartheta_{i}}{2}
    +\partial_{\mu}\theta_{i}\partial^{\mu}\theta_{i}\right)-a_{i}\varphi_{i}^{2}-\frac{b_{i}}{2}\varphi_{i}^{4}\right]\nonumber\\
    &-&2\epsilon\varphi_{1}\varphi_{2}\left[\cos\frac{\vartheta_{1}-\vartheta_{2}}{2}\cos(\theta_{1}-\theta_{2})
    +n_{z}\sin\frac{\vartheta_{1}-\vartheta_{2}}{2}\sin(\theta_{1}-\theta_{2})\right]\nonumber\\
    &-&2\epsilon\varphi_{1}\varphi_{3}\left[\cos\frac{\vartheta_{1}-\vartheta_{3}}{2}\cos(\theta_{1}-\theta_{3})
    +n_{z}\sin\frac{\vartheta_{1}-\vartheta_{3}}{2}\sin(\theta_{1}-\theta_{3})\right]\nonumber\\
    &-&2\epsilon\varphi_{2}\varphi_{3}\left[\cos\frac{\vartheta_{2}-\vartheta_{3}}{2}\cos(\theta_{2}-\theta_{3})
    +n_{z}\sin\frac{\vartheta_{2}-\vartheta_{3}}{2}\sin(\theta_{2}-\theta_{3})\right].
\end{eqnarray}
Considering small variations of the phases from their equilibrium values $\vartheta_{0i}=\theta_{0i}=0$ we can rewrite this Lagrangian in the form:
\begin{eqnarray}\label{6.3}
    \mathcal{L}&=&\sum_{i=1}^{3}\left[\partial_{\mu}\varphi_{i}\partial^{\mu}\varphi_{i}
    +\varphi_{i}^{2}\left(\partial_{\mu}\frac{\vartheta_{i}}{2}\partial^{\mu}\frac{\vartheta_{i}}{2}
    +\partial_{\mu}\theta_{i}\partial^{\mu}\theta_{i}\right)-a_{i}\varphi_{i}^{2}-\frac{b_{i}}{2}\varphi_{i}^{4}\right]\nonumber\\
    &-&2\epsilon\varphi_{1}\varphi_{2}\left[1-\frac{(\vartheta_{1}-\vartheta_{2})^{2}}{8}-\frac{(\theta_{1}-\theta_{2})^{2}}{2}
    +n_{z}\frac{\vartheta_{1}-\vartheta_{2}}{2}(\theta_{1}-\theta_{2})\right]\nonumber\\
    &-&2\epsilon\varphi_{1}\varphi_{3}\left[1-\frac{(\vartheta_{1}-\vartheta_{3})^{2}}{8}-\frac{(\theta_{1}-\theta_{3})^{2}}{2}
    +n_{z}\frac{\vartheta_{1}-\vartheta_{3}}{2}(\theta_{1}-\theta_{3})\right]\nonumber\\
    &-&2\epsilon\varphi_{2}\varphi_{3}\left[1-\frac{(\vartheta_{2}-\vartheta_{3})^{2}}{8}-\frac{(\theta_{2}-\theta_{3})^{2}}{2}
    +n_{z}\frac{\vartheta_{2}-\vartheta_{3}}{2}(\theta_{2}-\theta_{3})\right].
\end{eqnarray}
We can see that the Goldstone modes corresponding to $U(1)$ gauge symmetry (oscillations of the phases $\theta_{1},\theta_{2},\theta_{3}$) are coupled with the Goldstone modes corresponding to $SU(2)$ symmetry (oscillations of the phases $\vartheta_{1},\vartheta_{2},\vartheta_{3}$) by the  component $n_{z}=\pm 1$ of the unit vector $\vec{n}=\mathbf{k}n_{z}$ in the direction of the axis around which the rotation is made in isospace $\vec{\vartheta}=\vec{n}\vartheta$.


In presence of the abelian field $B_{\mu}$, corresponding to the local gauge $U(1)$ symmetry, and the nonabelian field $\vec{A}_{\mu}$, corresponding to the local gauge $SU(2)$ symmetry, we must apply the covariant derivative:
\begin{equation}\label{6.4}
  D_{\mu}\equiv\tau_{0}\partial_{\mu}-\mathrm{i}\tau_{0}\frac{f}{2}B_{\mu}-\mathrm{i}g\frac{\vec{\tau}}{2}\vec{A}_{\mu},
\end{equation}
where $f$ and $g$ are corresponding coupling constants. Using the gauge transformations (\ref{3.4}) and (\ref{5.17}), the Lagrangian (\ref{5.11}) can be presented in the form:
\begin{eqnarray}\label{6.5}
    \mathcal{L}&=&\partial_{\mu}\varphi_{1}\partial^{\mu}\varphi_{1}+\partial_{\mu}\varphi_{2}\partial^{\mu}\varphi_{2}+
    \partial_{\mu}\varphi_{3}\partial^{\mu}\varphi_{3}\nonumber\\
    &+&\left(\varphi_{1}^{2}\beta^{2}+\varphi_{2}^{2}\alpha^{2}\right)
    \left[\partial_{\mu}\frac{\vartheta_{1}-\vartheta_{2}}{2}\partial^{\mu}\frac{\vartheta_{1}-\vartheta_{2}}{2}
    +\partial_{\mu}\left(\theta_{1}-\theta_{2}\right)\partial^{\mu}\left(\theta_{1}-\theta_{2}\right)\right]\nonumber\\
    &+&\left(\varphi_{1}^{2}\gamma^{2}+\varphi_{3}^{2}\alpha^{2}\right)
    \left[\partial_{\mu}\frac{\vartheta_{1}-\vartheta_{3}}{2}\partial^{\mu}\frac{\vartheta_{1}-\vartheta_{3}}{2}
    +\partial_{\mu}\left(\theta_{1}-\theta_{3}\right)\partial^{\mu}\left(\theta_{1}-\theta_{3}\right)\right]\nonumber\\
    &+&\left(\varphi_{2}^{2}\gamma^{2}+\varphi_{3}^{2}\beta^{2}\right)
    \left[\partial_{\mu}\frac{\vartheta_{2}-\vartheta_{3}}{2}\partial^{\mu}\frac{\vartheta_{3}-\vartheta_{3}}{2}
    +\partial_{\mu}\left(\theta_{2}-\theta_{3}\right)\partial^{\mu}\left(\theta_{2}-\theta_{3}\right)\right]\nonumber\\
    &-&\varphi_{1}^{2}2\gamma\beta\left[\partial_{\mu}\frac{\vartheta_{1}-\vartheta_{2}}{2}\partial^{\mu}\frac{\vartheta_{1}-\vartheta_{3}}{2}
    +\partial_{\mu}(\theta_{1}-\theta_{2})\partial^{\mu}(\theta_{1}-\theta_{3})\right]\nonumber\\
    &-&\varphi_{2}^{2}2\alpha\gamma\left[\partial_{\mu}\frac{\vartheta_{1}-\vartheta_{2}}{2}\partial^{\mu}\frac{\vartheta_{2}-\vartheta_{3}}{2}
    +\partial_{\mu}(\theta_{1}-\theta_{2})\partial^{\mu}(\theta_{2}-\theta_{3})\right]\nonumber\\
    &-&\varphi_{3}^{2}2\alpha\beta\left[\partial_{\mu}\frac{\vartheta_{1}-\vartheta_{3}}{2}\partial^{\mu}\frac{\vartheta_{2}-\vartheta_{3}}{2}
    +\partial_{\mu}(\theta_{1}-\theta_{3})\partial^{\mu}(\theta_{2}-\theta_{3})\right]\nonumber\\
    &-&2\epsilon\varphi_{1}\varphi_{2}\left[1-\frac{(\vartheta_{1}-\vartheta_{2})^{2}}{8}-\frac{(\theta_{1}-\theta_{2})^{2}}{2}
    +n_{z}\frac{\vartheta_{1}-\vartheta_{2}}{2}(\theta_{1}-\theta_{2})\right]\nonumber\\
    &-&2\epsilon\varphi_{1}\varphi_{3}\left[1-\frac{(\vartheta_{1}-\vartheta_{3})^{2}}{8}-\frac{(\theta_{1}-\theta_{3})^{2}}{2}
    +n_{z}\frac{\vartheta_{1}-\vartheta_{3}}{2}(\theta_{1}-\theta_{3})\right]\nonumber\\
    &-&2\epsilon\varphi_{2}\varphi_{3}\left[1-\frac{(\vartheta_{2}-\vartheta_{3})^{2}}{8}-\frac{(\theta_{2}-\theta_{3})^{2}}{2}
    +n_{z}\frac{\vartheta_{2}-\vartheta_{3}}{2}(\theta_{2}-\theta_{3})\right]\nonumber\\
    &-&a_{1}\varphi_{1}^{2}-a_{2}\varphi_{2}^{2}-a_{3}\varphi_{3}^{2}
    -\frac{b_{1}}{2}\varphi_{1}^{4}-\frac{b_{2}}{2}\varphi_{2}^{4}-\frac{b_{3}}{2}\varphi_{3}^{4}\nonumber\\
    &+&\frac{g^{2}}{4}\left(\varphi_{1}^{2}+\varphi_{2}^{2}+\varphi_{3}^{2}\right)
    (A_{\mathrm{x}\mu}A^{\mu}_{\mathrm{x}}+A_{\mathrm{y}\mu}A^{\mu}_{\mathrm{y}})
    +\frac{1}{4}\left(\varphi_{1}^{2}+\varphi_{2}^{2}+\varphi_{3}^{2}\right)
    \left(g^{2}A_{\mathrm{z}\mu}A^{\mu}_{\mathrm{z}}+f^{2}B_{\mu}B^{\mu}\right)\nonumber\\
    &-&\frac{1}{16\pi}\vec{F}_{\mu\nu}\vec{F}^{\mu\nu}-\frac{1}{16\pi}G_{\mu\nu}G^{\mu\nu},
\end{eqnarray}
where
\begin{equation}\label{6.6}
  G_{\mu\nu}=\partial_{\mu}B_{\nu}-\partial_{\nu}B_{\mu}
\end{equation}
is the field tensor for the abelian gauge field $B_{\mu}$. In GWS theory the linear combinations
\begin{eqnarray}
  W_{\mu} &=& \frac{1}{\sqrt{2}}\left(A_{\mathrm{x}\mu}+\mathrm{i}A_{\mathrm{y}\mu}\right) \label{6.7}\\
  \nonumber \\
  \begin{array}{c}
    Z_{\mu} = A_{\mathrm{z}\mu}\cos\alpha-B_{\mu}\sin\alpha \\
    A_{\mu} = A_{\mathrm{z}\mu}\sin\alpha+B_{\mu}\cos\alpha
  \end{array} &\Rightarrow&
  \begin{array}{c}
    A_{\mathrm{z}\mu} = Z_{\mu}\cos\alpha+A_{\mu}\sin\alpha \\
    B_{\mu} = -Z_{\mu}\sin\alpha+A_{\mu}\cos\alpha
  \end{array},\label{6.8}
\end{eqnarray}
where
\begin{equation}\label{6.9}
  \cos\alpha=\frac{g}{\widetilde{g}},\quad \sin\alpha=\frac{f}{\widetilde{g}},\quad \widetilde{g}=\sqrt{g^{2}+f^{2}},
\end{equation}
allow to make the transformation:
\begin{equation}\label{6.10}
  \frac{g^{2}}{4}(A_{\mathrm{x}\mu}A^{\mu}_{\mathrm{x}}+A_{\mathrm{y}\mu}A^{\mu}_{\mathrm{y}})
    +\frac{1}{4}\left(g^{2}A_{\mathrm{z}\mu}A^{\mu}_{\mathrm{z}}+f^{2}B_{\mu}B^{\mu}\right)=
  \frac{g^{2}}{2}W_{\mu}W^{\ast\mu}+\frac{\widetilde{g}^{2}}{4}Z_{\mu}Z^{\mu}.
\end{equation}
Thus, the masses of charged $W$-boson and neutral $Z$-boson are
\begin{equation}\label{6.11}
  m_{W}=g\sqrt{2\pi\left(\varphi_{01}^{2}+\varphi_{02}^{2}+\varphi_{03}^{2}\right)},\quad
  m_{Z}=\widetilde{g}\sqrt{2\pi\left(\varphi_{01}^{2}+\varphi_{02}^{2}+\varphi_{03}^{2}\right)}=\frac{m_{W}}{\cos\alpha},
\end{equation}
but the field $A_{\mu}$ (photon) remains massless (with the interaction constant - electrical charge $e=g\sin\alpha$). However, separation of the components $A_{\mathrm{x}\mu}, A_{\mathrm{y}\mu}$ from the component $A_{\mathrm{z}\mu}$ (which is mixed with the Abelian field $B_{\mu}$) takes place only in the London gauge, where we excludes the single phases $\theta,\vartheta$ from Lagrangian, see Eq.(\ref{6.5}). Then, let us consider the gauge transformation (\ref{5.17}):
\begin{equation}\label{6.12}
    \vec{A}_{\mu}'=\vec{A}_{\mu}-\frac{1}{g}\partial_{\mu}\vec{\vartheta}+\left[\vec{\vartheta}\times\vec{A}_{\mu}\right]=
    \vec{A}_{\mu}-\frac{1}{g}\partial_{\mu}\vec{\vartheta}+
    \vartheta\cdot\left|\begin{array}{ccc}
            \vec{i} & \vec{j} & \vec{k} \\
            n_{x} & n_{y} & n_{z} \\
            A_{\mathrm{x}\mu} & A_{\mathrm{y}\mu} & A_{\mathrm{z}\mu}
          \end{array}
    \right|.
\end{equation}
we can see that the gauge transformation mixes the components $A_{\mathrm{x}\mu}$ and $A_{\mathrm{y}\mu}$ with the component $A_{\mathrm{z}\mu}$. From the other hand, the separation of the field $W_{\mu}$ (\ref{6.7}) from the component $A_{\mathrm{z}\mu}$ has physical sense then and only then, when the fields $W_{\mu}=(A_{\mu1}+\mathrm{i}A_{\mu2})/\sqrt{2}$ and $A_{\mathrm{z}\mu}$ are transformed by itself each. From Eq.(\ref{6.12}) we can see, that it is possible only when $n_{x}=n_{y}=0$:
\begin{eqnarray}
&&\begin{array}{c}
  A_{\mathrm{x}\mu}' = A_{\mathrm{x}\mu}+\vartheta n_{z}A_{\mathrm{y}\mu} \\
  A_{\mathrm{y}\mu}' = A_{\mathrm{y}\mu}-\vartheta n_{z}A_{\mathrm{x}\mu}
\end{array},\label{6.16a}\\
&&\begin{array}{c}
  A_{\mathrm{z}\mu}' = A_{\mathrm{z}\mu}-n_{z}\frac{1}{g}\partial_{\mu}\vartheta
\end{array}.\label{6.16c}
\end{eqnarray}
Hence, $n_{z}=\pm 1$, that coincides with Eq.(\ref{5.4c}) as condition for interference of condensates  $\Psi_{1},\Psi_{2},\Psi_{3}$. Thus, the separation of the field $W_{\mu}$ selects a direction in the isospace also. The spectrum of excitations depends only on $n_{z}^{2},n_{z}^{4}$, so that the signum of $n_{z}$ is not important.

As and before, we must suppose $\epsilon<0$, hence the equilibrium phases are such that $\cos\theta_{ij}=\cos\vartheta_{ij}=1$. Then, from Lagrangian (\ref{6.2}) we can see that the spectrum of Higgs oscillations coincides with the spectrum (\ref{2.27}). At the same time, spectrum of the Goldstone modes takes the form:
\begin{equation}\label{6.13}
  \left(q_{\mu}q^{\mu}\right)^{4}\left(\left(q_{\mu}q^{\mu}\right)^{2}+\left(q_{\mu}q^{\mu}\right)b+c\right)=0,
\end{equation}
where
\begin{eqnarray}\label{6.14}
  b &=& 2\epsilon\frac{\varphi_{01}^{2}(\varphi_{02}+\varphi_{03})+\varphi_{02}^{2}(\varphi_{01}+\varphi_{03})+\varphi_{03}^{2}(\varphi_{01}+\varphi_{02})}
  {\varphi_{01}\varphi_{02}\varphi_{03}} \nonumber\\
  c &=& 4\epsilon^{2}\frac{\varphi_{01}^{3}+\varphi_{02}^{3}+\varphi_{03}^{3}+\varphi_{01}^{2}(\varphi_{02}+\varphi_{03})+\varphi_{02}^{2}(\varphi_{01}+\varphi_{03})+\varphi_{03}^{2}(\varphi_{01}+\varphi_{02})}
  {\varphi_{01}\varphi_{02}\varphi_{03}}.
\end{eqnarray}
From Eq.(\ref{6.13}) we can see, that one of dispersion relations is $q_{\mu}q^{\mu}=0$. This relation corresponds to the twofold degenerated common mode oscillations, which are absorbed by the gauge fields $W_{\mu},W_{\mu}^{\ast},Z_{\mu}$ and to the twofold degenerated massless Leggett mode. Remaining quadratic equation determines two Leggett modes with massive spectrums: $m_{L1,2}^{2}=q_{\mu}q^{\mu}=\frac{1}{2}(-b\mp\sqrt{b^{2}-4c})$. As we could see above, the L-bosons are not absorbed by the gauge fields. Thus, if all bands are independent, i.e $\epsilon=0$, then we have two massless Goldstone modes per band (independet oscillations of the phase $\vartheta$ and $\theta$), a total of six independent Goldstone modes. Due to the internal proximity effect, i.e $\epsilon\neq 0$, the Goldstone modes from each band transform into the following normal oscillations for all bands: twofold degenerated common mode oscillations with the acoustic spectrum, the twofold degenerated massless Leggett mode and two Leggett modes with the energy gaps. Squared masses of the L-bosons are proportional to the interband coupling $m_{L1,2}^{2}\sim|\epsilon|$. For symmetrical three-band system, i.e. $\varphi_{01}=\varphi_{02}=\varphi_{03}$, masses of both massive L-bosons are equal:
\begin{equation}\label{6.15}
    m_{L1}=m_{L2}=\sqrt{6|\epsilon|}.
\end{equation}
Thus, we can see, that, unlike the cases of $U(1)$ and $SU(2)$ symmetries, for the case $SU(2)\otimes U(1)$ symmetry we have two massless L-bosons and two massive L-bosons. However, the massless bosons lose their energy in the process of space expansion, like relic photons. Hence, the role of these bosons can be neglected. In contrast to them, the massive L-bosons are able to form stable gravitationally bound structures (clusters, halo). Moreover, the L-bosons are sterile. Therefore, the massive L-bosons are suitable candidate for the "dark matter".

It should be noted, that if we suppose nonsymmetrical Josephson coupling $\epsilon_{12}\neq\epsilon_{13}\neq\epsilon_{23}$ instead the uniform coefficient $\epsilon$, then the twofold degenerated massless Leggett mode splits into one massless mode and one massive mode. However, in what follows, we will consider only the minimal model with the uniform coefficient $\epsilon$.

\section{Lepton mixing and the mass states of neutrinos}\label{mixing}

From Eq.(\ref{5.7}) we can see, that the band states of the Dirac fields, i.e $\psi_{1},\psi_{2},\psi_{3}$, are determined by the coupling between the corresponding Dirac field $\psi_{j}$ and the scalar field $\varphi_{j}$ (isospinor field $\Psi_{j}$). Then, the gauge invariant Dirac Lagrangian for the lepton fields has the form:
\begin{eqnarray}\label{7.1}
L_{D}=\sum_{j=1}^{3}\mathrm{i}\overline{\psi}_{Lj}\gamma^{\mu}\overset{\leftrightarrow}D_{\mu}^{\vec{A},B}\psi_{Lj}
  +\mathrm{i}\overline{\psi}_{Rj}\gamma^{\mu}\overset{\leftrightarrow}D_{\mu}^{B}\psi_{Rj}
-\chi\left[\overline{\psi}_{Lj}\Psi_{j}\psi_{Rj}+\overline{\psi}_{Rj}\Psi^{+}_{j}\psi_{Lj}\right],
\end{eqnarray}
where
\begin{equation}\label{7.1a}
  D^{\vec{A},B}_{\mu}\equiv \tau_{0}\partial_{\mu}
-\mathrm{i}g\frac{\vec{\tau}}{2}\vec{A}_{\mu}+\mathrm{i}\tau_{0}\frac{f}{2}B_{\mu},\quad
D^{B}_{\mu}\equiv \tau_{0}\partial_{\mu}+\mathrm{i}\tau_{0}fB_{\mu}
\end{equation}
are covariant derivations. Thus, each band state $\psi_{1},\psi_{2},\psi_{3}$, emitting or absorbing the gauge bosons $\vec{A}_{\mu},B_{\mu}$, transforms into itself only, i.e. $1\leftrightarrow 1,2\leftrightarrow 2,3\leftrightarrow 3$. Analogously, for the flavour states $\psi_{e},\psi_{\mu},\psi_{\tau}$ (\ref{5.24}):
\begin{eqnarray}\label{7.2}
L_{D}=\mathrm{i}\overline{\psi}_{Le}\gamma^{\mu}\overset{\leftrightarrow}D_{\mu}^{\vec{A},B}\psi_{Le}
  +\mathrm{i}\overline{e}_{R}\gamma^{\mu}\overset{\leftrightarrow}D_{\mu}^{B}e_{R}
-\chi\left[\overline{\psi}_{L}\Psi_{e} e_{R}+\overline{e}_{R}\Psi^{+}_{e}\psi_{L}\right]+L_{\mu}+L_{\tau}.
\end{eqnarray}
We can conditionally call "e" - electron $e$ and electron neutrino $\nu_{e}$, $"\mu"$ - muon $\mu$ and muon neutrino $\nu_{\mu}$, $"\tau"$ - tauon $\tau$ and tauon neutrino $\nu_{\tau}$. If $m_{e}\ll m_{\mu}\ll m_{\tau}$, then the bands should be strongly asymmetrical: $\varphi_{01}\ll\varphi_{02}\ll\varphi_{03}$. As for the band states, each flavor state $\psi_{e},\psi_{\mu},\psi_{\tau}$, emitting or absorbing the gauge bosons $\vec{A}_{\mu},B_{\mu}$, transforms into itself only, i.e. $e\leftrightarrow e,\mu\leftrightarrow \mu,\tau\leftrightarrow \tau$.

As we could see in Sect.\ref{dirac} each $L$ and $R$ components should mix with the corresponding $R$ and $L$ components of other flavors $\overline{\psi}_{Li}\psi_{Rk}+\overline{\psi}_{Rk}\psi_{Li}$, which is the fermionic analog of the interband Josephson coupling, unlike SM, where the mixing coefficients are off-diagonal Yukawa interactions. Thus, we can take the $SU(2)$-symmetric mixing term for the Dirac fields in the following form:
\begin{eqnarray}\label{7.3}
  \mathcal{U}_{mix} &=&  \overline{\psi}_{Le}\left(\begin{array}{c}
                  0 \\
                  \zeta_{e\mu} \\
                \end{array}\right)\psi_{R\mu}+
                \overline{\psi}_{Le}\left(\begin{array}{c}
                  0 \\
                  \zeta_{e\tau} \\
                \end{array}\right)\psi_{R\tau}+
                \overline{\psi}_{L\mu}\left(\begin{array}{c}
                  0 \\
                  \zeta_{\mu\tau} \\
                \end{array}\right)\psi_{R\tau}+h.c.,
\end{eqnarray}
where $\zeta_{ik}$ are mixing parameters determined by the interband coupling $\epsilon$ of the scalar fields -  Eq.(\ref{4.12a}),
$\psi_{Le}=\left(\begin{array}{c}
\nu_{e_{L}} \\
e_{L} \\
\end{array}\right)$ is the left-handed bispinor, $\psi_{Re}=e_{R}$ is the right-handed spinor. The band masses  $m_{D1},m_{D2},m_{D3}$ are determined by Yukawa interaction of the Dirac fields with the corresponding band states of scalar fields $\varphi_{01},\varphi_{02},\varphi_{03}$, in turn the flavor masses $m_{De},m_{D\mu},m_{D\tau}$ are determined by Yukawa interaction of the Dirac fields with the corresponding flavour states of scalar fields $\varphi_{0e},\varphi_{0\mu},\varphi_{0\tau}$ which are result of the diagonalization (\ref{2.32}). The mixing $R\leftrightarrow L$ results transition from the flavour masses to the band masses via diagonalization of the matrix $M_{e\mu\tau}$ as demonstrated in Sect.\ref{dirac}:
\begin{eqnarray}\label{7.4}
    \mathcal{U}_{De\mu\tau}+\mathcal{U}_{\mathrm{mix}}&=&
    \left(
        \begin{array}{ccc}
          \overline{e}_{L}, & \overline{\mu}_{L}, & \overline{\tau}_{L} \\
        \end{array}
      \right)
      \left(
        \begin{array}{ccc}
          m_{De} & \zeta_{e\mu} & \zeta_{e\tau} \\
          \zeta_{e\mu} & m_{D\mu} & \zeta_{\mu\tau} \\
          \zeta_{e\tau} & \zeta_{\mu\tau} & m_{D\tau} \\
        \end{array}
      \right)
      \left(
        \begin{array}{c}
          e_{R} \\
          \mu_{R} \\
          \tau_{R} \\
        \end{array}
      \right)+h.c\nonumber\\
      &=&\left(
        \begin{array}{ccc}
          \overline{l}_{1L}, & \overline{l}_{2L}, & \overline{l}_{3L} \\
        \end{array}
      \right)
      \left(
        \begin{array}{ccc}
          m_{D1} & 0 & 0 \\
          0 & m_{D2} & 0 \\
          0 & 0 & m_{D3} \\
        \end{array}
      \right)
      \left(
        \begin{array}{c}
          l_{1R} \\
          l_{2R} \\
          l_{2R} \\
        \end{array}
      \right)+h.c.
\end{eqnarray}
The mixing takes place due to the interband Josephson coupling of the scalar fields $\varphi_{1},\varphi_{2},\varphi_{3}$: from Eq.(\ref{4.12a}) we can see that $\zeta_{\alpha\beta}\propto \epsilon^{2}$. As will be demonstrated in Sect.\ref{darkmatter}, the interband coupling is extremely small $\epsilon\sim 10^{-40}\mathrm{eV}^{2}$. Taking masses of H-bosons as $\Delta m_{H}^{2}\sim 10^{2}\mathrm{GeV}^{2}$ (see Sect.\ref{particle}), we can see that the mixing angles, determined by Eqs.(\ref{4.12a},\ref{4.13}), are extremely small: $\alpha_{ij}\sim 10^{-100}$. Probability of interflavor transition is $P_{ik}\sim\sin^{2}(2\alpha_{ik})$ \cite{bilen1,bilen2,gersh,bilen3,bilen4,giunti}, hence, for the massive leptons (electrons, muons, tauons) effect of mixing is negligible $P\sim 10^{-200}$.  Thus, the mixing of charged leptons is negligible and it lies beyond the sensitivity of any experiment.

In SM masses of neutrino are zero. However, observation of the neutrino oscillations in vacuum means presence mass of neutrinos \cite{bilen1,bilen2,gersh,bilen3,bilen4,giunti}, and the differences in the squares of the masses have been measured: $|\Delta m^{2}_{23}|\equiv|m^{2}_{3}-m^{2}_{2}|\approx 2.51\cdot 10^{-3}\mathrm{eV}^{2}$, $|\Delta m^{2}_{12}|\approx 7.41\cdot 10^{-5}\mathrm{eV}^{2}$ \cite{NuFIT,ester}. Formally, we can write the Dirac mass term (Yukawa interaction) for both charged lepton and neutrino in the form (\ref{1.1}), assigning for neutrinos a small but non-zero Yukawa constant $\chi_{\nu}$ and introducing the sterile right-handed neutrino. Thus, the neutrino mass becomes similar to mass of charged leptons. However, in the proposed three-band model the problem of mass is fundamental. As we could see, interaction of Dirac fields with corresponding scalar fields leads to lepton oscillations as consequence of Josephson coupling between scalar fields. Hence, the mixing angles are extremely small: $\alpha_{ij}\sim 10^{-100}$. At the same time, the experimental mixing angles for neutrinos are large: $\alpha_{12}=33.4^{\circ}$, $\alpha_{23}=42.2\ldots49.5^{\circ}$, $\alpha_{13}=8.6^{\circ}$ \cite{NuFIT,ester}.

However, within the framework of the three-band model, the presence of mixing alone without interaction with scalar fields can lead to mass generation. Let us suppose existence massless \emph{sterile} right-handed neutrinos $\nu_{Re},\nu_{R\mu},\nu_{R\tau}$, i.e which are characterized by zero isospin and hypercharge: $I_{z}=0, Y=0$, unlike the \emph{active} left-handed neutrinos $\nu_{Le},\nu_{L\mu},\nu_{L\tau}$ which are characterized by $I_{z}=\frac{1}{2}, Y=-1$. Then, the $SU(2)$-symmetric mixing term for neutrinos should have the following form:
\begin{eqnarray}\label{7.5d}
  \mathcal{U}_{mix}^{\nu} &=&  \overline{\psi}_{Le}\left(\begin{array}{c}
                  \varsigma_{e\mu} \\
                  0 \\
                \end{array}\right)\nu_{R\mu}+
                \overline{\psi}_{Le}\left(\begin{array}{c}
                  \varsigma_{e\tau} \\
                  0 \\
                \end{array}\right)\nu_{R\tau}+
                \overline{\psi}_{L\mu}\left(\begin{array}{c}
                  \varsigma_{\mu\tau} \\
                  0 \\
                \end{array}\right)\nu_{R\tau}+h.c.,
\end{eqnarray}
where $\varsigma_{ik}\neq\zeta_{ik}$ are mixing parameter specially for neutrinos,
$\psi_{Le}=\left(\begin{array}{c}
\nu_{e_{L}} \\
e_{L} \\
\end{array}\right)$ is the left-handed bispinor, $\nu_{Re}$ is the right-handed neutrino. Then, the corresponding neutrino Lagrangian has the form:
\begin{eqnarray}\label{7.5}
    \mathcal{L}_{\nu}&=&
    \mathrm{i}\left(\overline{\nu}_{Le}\gamma^{\sigma}\overset{\leftrightarrow}{\partial}_{\sigma}\nu_{Le}
    +\overline{\nu}_{L\mu}\gamma^{\sigma}\overset{\leftrightarrow}{\partial}_{\sigma}\nu_{L\mu}
    +\overline{\nu}_{L\tau}\gamma^{\sigma}\overset{\leftrightarrow}{\partial}_{\sigma}\nu_{L\tau}\right)\nonumber\\
    &-&\varsigma_{e\mu}(\overline{\nu}_{Le}\nu_{R\mu}+\overline{\nu}_{L\mu}\nu_{Re})
    -\varsigma_{e\tau}(\overline{\nu}_{Le}\nu_{R\tau}+\overline{\nu}_{L\tau}\nu_{Re})
    -\varsigma_{\mu\tau}(\overline{\nu}_{L\mu}\nu_{R\tau}+\overline{\nu}_{L\tau}\nu_{R\mu})\nonumber\\
    &+&\mathrm{i}\left(\overline{\nu}_{Re}\gamma^{\sigma}\overset{\leftrightarrow}{\partial}_{\sigma}\nu_{Re}
    +\overline{\nu}_{R\mu}\gamma^{\sigma}\overset{\leftrightarrow}{\partial}_{\sigma}\nu_{R\mu}
    +\overline{\nu}_{R\tau}\gamma^{\sigma}\overset{\leftrightarrow}{\partial}_{\sigma}\nu_{R\tau}\right)\nonumber\\
    &-&\varsigma_{e\mu}(\overline{\nu}_{Re}\nu_{L\mu}+\overline{\nu}_{R\mu}\nu_{Le})
    -\varsigma_{e\tau}(\overline{\nu}_{Re}\nu_{L\tau}+\overline{\nu}_{R\tau}\nu_{Le})
    -\varsigma_{\mu\tau}(\overline{\nu}_{R\mu}\nu_{L\tau}+\overline{\nu}_{R\tau}\nu_{L\mu}).
\end{eqnarray}
We can diagonalize the matrix $M_{e\mu\tau}$ as
\begin{equation}\label{7.5a}
    \left(
        \begin{array}{ccc}
          \overline{\nu}_{Le}, & \overline{\nu}_{L\mu}, & \overline{\nu}_{L\tau} \\
        \end{array}
      \right)
      \left(
        \begin{array}{ccc}
          0 & \varsigma_{e\mu} & \varsigma_{e\tau} \\
          \varsigma_{e\mu} & 0 & \varsigma_{\mu\tau} \\
          \varsigma_{e\tau} & \varsigma_{\mu\tau} & 0 \\
        \end{array}
      \right)
      \left(
        \begin{array}{c}
          \nu_{Re} \\
          \nu_{R\mu} \\
          \nu_{R\tau} \\
        \end{array}
      \right)+h.c=
      \left(
        \begin{array}{ccc}
          \overline{\nu}_{1L}, & \overline{\nu}_{2L}, & \overline{\nu}_{3L} \\
        \end{array}
      \right)
      \left(
        \begin{array}{ccc}
          m_{\nu1} & 0 & 0 \\
          0 & m_{\nu2} & 0 \\
          0 & 0 & m_{\nu3} \\
        \end{array}
      \right)
      \left(
        \begin{array}{c}
          \nu_{1R} \\
          \nu_{2R} \\
          \nu_{2R} \\
        \end{array}
      \right)+h.c.
\end{equation}
The corresponding characteristic equation is
\begin{eqnarray}\label{7.5b}
  m_{\nu}^{3}-(\varsigma_{e\mu}^{2}+\varsigma_{e\tau}^{2}+\varsigma_{\mu\tau}^{2})m_{\nu}
  -2\varsigma_{e\mu}\varsigma_{e\tau}\varsigma_{\mu\tau}=0.
\end{eqnarray}
For the symmetrical interband mixing $\varsigma_{e\mu}=\varsigma_{e\tau}=\varsigma_{\mu\tau}\equiv\varsigma$ we obtain the following solutions of Eq.(\ref{7.5b}):
\begin{equation}\label{7.5c}
  m_{\nu1}=m_{\nu2}=-\varsigma,\quad  m_{\nu3}=2\varsigma.
\end{equation}
Obviously, the right-handed (sterile) neutrinos and left-handed (active) neutrinos have exactly the same masses: $m_{\nu Ri}=m_{\nu Li}$. \emph{We can see, that neutrino masses can take both positive and negative magnitudes. This means, that the mass states of neutrino $\nu_{1},\nu_{2},\nu_{3}$ are quasiparticles (unlike the band state of charged leptons $l_{1},l_{2},l_{3}$, which are determined by the Yukawa coupling with scalar fields of corresponding bands). Respectively, that the masses} (\ref{7.5c}) \emph{are effective masses of the quasiparticles. Only square root of squares of masses $\sqrt{m_{\nu1}^{2}},\sqrt{m_{\nu2}^{2}},\sqrt{m_{\nu3}^{2}}$ have physical sense, because only the differences $|\Delta m^{2}_{23}|\equiv|m^{2}_{3}-m^{2}_{2}|$, $|\Delta m^{2}_{12}|\equiv|m^{2}_{2}-m^{2}_{1}|$ are measured in experiments about neutrino oscillations, and the upper limits of the masses $\sqrt{m_{\nu e}^{2}},\sqrt{m_{\nu \mu}^{2}},\sqrt{m_{\nu \tau}^{2}}$ have been experimentally determined from $\beta$-decay of tritium, pion decay, $\tau$-decays into multi-pion final states accordingly, where  the spectral distribution of leptons is determined by $m_{\nu}^{2}$, but not by $m_{\nu}$, moreover $m_{\nu\alpha}=\sqrt{\sum_{i=1}^{3}|U_{\alpha i}|^{2}m_{\nu i}^{2}}$ where $U_{\alpha i}$ is PMNS matrix \cite{gersh,formag,nucc}.}


In view of the above, we should deal with equations that include only squares of the effective masses. Lagrange equations for Lagrangian (\ref{7.5}) are:
\begin{eqnarray}\label{7.6}
  \mathrm{i}\gamma^{\sigma}\partial_{\sigma}\nu_{Le}-\varsigma_{e\mu}\nu_{R\mu}-\varsigma_{e\tau}\nu_{R\tau}&=&0\nonumber\\
  \mathrm{i}\gamma^{\sigma}\partial_{\sigma}\nu_{L\mu}-\varsigma_{e\mu}\nu_{Re}-\varsigma_{\mu\tau}\nu_{R\tau}&=&0\nonumber\\
  \mathrm{i}\gamma^{\sigma}\partial_{\sigma}\nu_{L\tau}-\varsigma_{e\tau}\nu_{Re}-\varsigma_{\mu\tau}\nu_{R\mu}&=&0\nonumber\\
  \mathrm{i}\gamma^{\sigma}\partial_{\sigma}\nu_{Re}-\varsigma_{e\mu}\nu_{L\mu}-\varsigma_{e\tau}\nu_{L\tau}&=&0\nonumber\\
  \mathrm{i}\gamma^{\sigma}\partial_{\sigma}\nu_{R\mu}-\varsigma_{e\mu}\nu_{Le}-\varsigma_{\mu\tau}\nu_{L\tau}&=&0\nonumber\\
  \mathrm{i}\gamma^{\sigma}\partial_{\sigma}\nu_{R\tau}-\varsigma_{e\tau}\nu_{Le}-\varsigma_{\mu\tau}\nu_{L\mu}&=&0,
\end{eqnarray}
Then Eq.(\ref{7.6}) can be transformed to the system of Klein–Gordon-like equations for the left-handed fields separately:
\begin{eqnarray}\label{7.7}
  \partial^{\sigma}\partial_{\sigma}\nu_{Le}+\left(\varsigma_{e\mu}^{2}+\varsigma_{e\tau}^{2}\right)\nu_{Le}
  +\varsigma_{e\tau}\varsigma_{\mu\tau}\nu_{L\mu}+\varsigma_{e\mu}\varsigma_{\mu\tau} \nu_{L\tau}&=&0\nonumber\\
  \partial^{\sigma}\partial_{\sigma}\nu_{L\mu}+\varsigma_{e\tau}\varsigma_{\mu\tau}\nu_{Le}+\left(\varsigma_{e\mu}^{2}+\varsigma_{\mu\tau}^{2}\right)\nu_{L\mu}
  +\varsigma_{e\mu}\varsigma_{e\tau}\nu_{L\tau}&=&0\\
  \partial^{\sigma}\partial_{\sigma}\nu_{L\tau}+\varsigma_{e\mu}\varsigma_{\mu\tau}\nu_{Le}+\varsigma_{e\tau}\varsigma_{e\mu}\nu_{L\mu}
  +\left(\varsigma_{e\tau}^{2}+\varsigma_{\mu\tau}^{2}\right)\nu_{L\tau}&=&0,\nonumber
\end{eqnarray}
where we have used $(\gamma^{\nu}\partial_{\nu})(\gamma^{\mu}\partial_{\mu})
=\frac{1}{2}\partial_{\mu}\partial_{\nu}(\gamma^{\mu}\gamma^{\nu}+\gamma^{\nu}\gamma^{\mu})
=\frac{1}{2}\partial_{\mu}\partial_{\nu}2g^{\mu\nu}=\partial_{\mu}\partial^{\mu}$. Thus, we obtain Lorentz covariant equations of motion for the left-handed neutrinos $\nu_{Le},\nu_{L\mu},\nu_{L\tau}$ only. Analogously we can obtain such equations for the right-handed fields.

Let us consider the spinors $\nu_{Le,\mu,\tau}$ in the form of plane waves: $\nu_{e,\mu,\tau}=u_{e,\mu,\tau}e^{-ip_{\sigma}x^{\sigma}}$, where $u_{e,\mu,\tau}$ are corresponding spinor amplitudes. Then Eq.(\ref{7.7}) takes the form:
\begin{eqnarray}\label{7.8}
  \left(\varsigma_{e\mu}^{2}+\varsigma_{e\tau}^{2}-p^{\sigma}p_{\sigma}\right)u_{e}
  +\varsigma_{e\tau}\varsigma_{\mu\tau}u_{\mu}+\varsigma_{e\mu}\varsigma_{\mu\tau}u_{\tau}&=&0\nonumber\\
  \varsigma_{e\tau}\varsigma_{\mu\tau}u_{e}+\left(\varsigma_{e\mu}^{2}+\varsigma_{\mu\tau}^{2}-p^{\sigma}p_{\sigma}\right)u_{\mu}
  +\varsigma_{e\mu}\varsigma_{e\tau}u_{\tau}&=&0\\
  \varsigma_{e\mu}\varsigma_{\mu\tau}u_{e}+\varsigma_{e\tau}\varsigma_{e\mu}u_{\mu}
  +\left(\varsigma_{e\tau}^{2}+\varsigma_{\mu\tau}^{2}-p^{\sigma}p_{\sigma}\right)u_{\tau}&=&0.\nonumber
\end{eqnarray}
The corresponding characteristic equation is
\begin{eqnarray}\label{7.9}
  &&(p^{\sigma}p_{\sigma})^{3}-(2\varsigma_{e\mu}^{2}+2\varsigma_{e\tau}^{2}+2\varsigma_{\mu\tau}^{2})(p^{\sigma}p_{\sigma})^{2}\nonumber\\
  &&+(\varsigma_{e\mu}^{4}+\varsigma_{e\tau}^{4}+\varsigma_{\mu\tau}^{4}
  +2\varsigma_{e\mu}^{2}\varsigma_{e\tau}^{2}+2\varsigma_{e\mu}^{2}\varsigma_{\mu\tau}^{2}+2\varsigma_{e\tau}^{2}\varsigma_{\mu\tau}^{2})(p^{\sigma}p_{\sigma})
  -4\varsigma_{e\mu}^{2}\varsigma_{e\tau}^{2}\varsigma_{\mu\tau}^{2}=0.
\end{eqnarray}
This equation has three positive real solutions: $m_{\nu1}^{2}=(p^{\sigma}p_{\sigma})_{1}$, $m_{\nu2}^{2}=(p^{\sigma}p_{\sigma})_{2}$, $m_{\nu3}^{2}=(p^{\sigma}p_{\sigma})_{3}$, which can be associated with the mass states of neutrinos $\nu_{L1},\nu_{L2},\nu_{L3}$:
\begin{eqnarray}\label{7.10}
  \partial^{\sigma}\partial_{\sigma}\nu_{L1}+m_{\nu1}^{2}\nu_{L1}&=&0\nonumber\\
  \partial^{\sigma}\partial_{\sigma}\nu_{L2}+m_{\nu2}^{2}\nu_{L2}&=&0\\
  \partial^{\sigma}\partial_{\sigma}\nu_{L3}+m_{\nu3}^{2}\nu_{L3}&=&0,\nonumber
\end{eqnarray}
and we can suppose the hierarchy of the masses as $m_{\nu1}^{2}\leq m_{\nu2}^{2}\leq m_{\nu3}^{2}$.
So, for the symmetrical interband mixing $\varsigma_{e\mu}=\varsigma_{e\tau}=\varsigma_{\mu\tau}\equiv\varsigma$ we obtain the following solutions of Eq.(\ref{7.9}):
\begin{equation}\label{7.11}
  m_{\nu1}^{2}=m_{\nu2}^{2}=\varsigma^{2},\quad  m_{\nu3}^{2}=4\varsigma^{2}.
\end{equation}
Thus, the effective masses of neutrinos is of order of the interband mixing parameters. It should be noted, that the masses $m_{\nu1}, m_{\nu2}, m_{\nu3}$ are result of the interband mixing, unlike the electron-muon-tauon masses, which are result of the coupling with corresponding scalar fields $\varphi_{e,\mu,\tau}$. Obviously, the flavour states $\nu_{Le},\nu_{L\mu},\nu_{L\tau}$ must be linear combinations of the mass states $\nu_{L1},\nu_{L2},\nu_{L3}$ and vice versa, that can be written by the following way:
\begin{eqnarray}
  \left(
    \begin{array}{c}
      \nu_{Le} \\
      \nu_{L\mu} \\
      \nu_{L\tau} \\
    \end{array}
  \right)=
   U\cdot
  \left(
    \begin{array}{c}
      \nu_{L1} \\
      \nu_{L2} \\
      \nu_{L3} \\
    \end{array}
  \right),\quad
  \left(
    \begin{array}{c}
      \nu_{L1} \\
      \nu_{L2} \\
      \nu_{L3} \\
    \end{array}
  \right)=
  U^{T}\cdot
  \left(
    \begin{array}{c}
      \nu_{Le} \\
      \nu_{L\mu} \\
      \nu_{L\tau} \\
    \end{array}
  \right),\label{7.12b}
\end{eqnarray}
where $U$ and $U^{T}$ are mixing matrices (\ref{2.36a},\ref{2.36b}). Let us find a relation which the angles $\alpha_{12},\alpha_{13},\alpha_{23}$ must satisfy. At first, let us introduce the designations in Eq.(\ref{7.7}):
\begin{eqnarray}\label{7.13}
  &&\partial^{\sigma}\partial_{\sigma}\nu_{Le}+A\nu_{Le}+B\nu_{L\mu}+C\nu_{L\tau}=0\nonumber\\
  &&\partial^{\sigma}\partial_{\sigma}\nu_{L\mu}+B\nu_{Le}+E\nu_{L\mu}+D\nu_{L\tau}=0\\
  &&\partial^{\sigma}\partial_{\sigma}\nu_{L\tau}+C\nu_{Le}+D\nu_{L\mu}+F\nu_{L\tau}=0.\nonumber
\end{eqnarray}
Using Eqs.(\ref{7.10},\ref{7.12b}) we can write Eq.(\ref{7.13}) as
\begin{eqnarray}\label{7.14}
  c_{12}c_{13}m_{1}^{2}\nu_{L1}+c_{13}s_{12}m_{2}^{2}\nu_{L2}+s_{13}m_{3}^{2}\nu_{L3}&=&A\nu_{Le}+B\nu_{L\mu}+C\nu_{L\tau}\nonumber\\
  (-s_{13}s_{23}c_{12}-c_{23}s_{12})m_{1}^{2}\nu_{L1}+(-s_{12}s_{13}s_{23}+c_{23}c_{12})m_{2}^{2}\nu_{L2}+c_{13}s_{23}m_{3}^{2}\nu_{L3}
  &=&B\nu_{Le}+E\nu_{L\mu}+D\nu_{L\tau}\\
  (-c_{23}c_{12}s_{13}+s_{23}s_{12})m_{1}^{2}\nu_{L1}+(-c_{23}s_{12}s_{13}-s_{23}c_{12})m_{2}^{2}\nu_{L2}+c_{13}c_{23}m_{3}^{2}\nu_{L3}
  &=&C\nu_{Le}+D\nu_{L\mu}+F\nu_{L\tau}.\nonumber
\end{eqnarray}
The right side can be transformed like this:
\begin{eqnarray}\label{7.15}
  &&\left(\begin{array}{c}
      A\nu_{Le}+B\nu_{L\mu}+C\nu_{L\tau} \\
      B\nu_{Le}+E\nu_{L\mu}+D\nu_{L\tau} \\
      C\nu_{Le}+D\nu_{L\mu}+F\nu_{L\tau}
    \end{array}\right)
  \equiv\left(
    \begin{array}{ccc}
      A & B & C \\
      B & E & D \\
      C & D & F \\
    \end{array}
  \right)
  \left(
    \begin{array}{c}
      \nu_{Le} \\
      \nu_{L\mu} \\
      \nu_{L\tau} \\
    \end{array}
  \right)=
  \left(
    \begin{array}{ccc}
      A & B & C \\
      B & E & D \\
      C & D & F \\
    \end{array}
  \right)\cdot\widehat{U}\cdot
  \left(
    \begin{array}{c}
      \nu_{L1} \\
      \nu_{L2} \\
      \nu_{L3} \\
    \end{array}
  \right)\nonumber\\
  &&=\left(
       \begin{array}{c}
         \begin{array}{c}
             [Ac_{13}c_{12}-B(s_{23}s_{13}c_{12}+c_{23}s_{12})-C(c_{23}s_{13}c_{12}-s_{23}s_{12})]\nu_{L1} \\
            +[Ac_{13}s_{12}-B(s_{23}s_{13}s_{12}-c_{23}c_{12})-C(c_{23}s_{13}s_{12}+s_{23}c_{12})]\nu_{L2}\\
            +[As_{13}+Bs_{23}c_{13}+Cc_{23}c_{13}]\nu_{L3}
          \end{array}\\
          \\
         \begin{array}{c}
             [Bc_{13}c_{12}-E(s_{23}s_{13}c_{12}+c_{23}s_{12})-D(c_{23}s_{13}c_{12}-s_{23}s_{12})]\nu_{L1} \\
            +[Bc_{13}s_{12}-E(s_{23}s_{13}s_{12}-c_{23}c_{12})-D(c_{23}s_{13}s_{12}+s_{23}c_{12})]\nu_{L2}\\
            +[Bs_{13}+Es_{23}c_{13}+Dc_{23}c_{13}]\nu_{L3}
          \end{array} \\
         \\
         \begin{array}{c}
             [Cc_{13}c_{12}-D(s_{23}s_{13}c_{12}+c_{23}s_{12})-F(c_{23}s_{13}c_{12}-s_{23}s_{12})]\nu_{L1} \\
            +[Cc_{13}s_{12}-D(s_{23}s_{13}s_{12}-c_{23}c_{12})-F(c_{23}s_{13}s_{12}+s_{23}c_{12})]\nu_{L2}\\
            +[Cs_{13}+Ds_{23}c_{13}+Fc_{23}c_{13}]\nu_{L3}
          \end{array} \\
       \end{array}
     \right).
\end{eqnarray}
Then, the angles $\alpha_{12},\alpha_{13},\alpha_{23}$ satisfy the following equation:
\begin{equation}\label{7.16}
  \Delta\equiv\left|
    \begin{array}{ccc}
      \Delta_{11} & \Delta_{12} & \Delta_{13} \\
      \Delta_{21} & \Delta_{22} & \Delta_{23} \\
      \Delta_{31} & \Delta_{32} & \Delta_{33} \\
    \end{array}
  \right|=0,
\end{equation}
where
\begin{equation}\label{7.17}
\begin{array}{ccc}
  \Delta_{11}&=&(A-m_{1}^{2})c_{13}c_{12}-B(s_{23}s_{13}c_{12}+c_{23}s_{12})-C(c_{23}s_{13}c_{12}-s_{23}s_{12}) \\
  \Delta_{12}&=&(A-m_{2}^{2})c_{13}s_{12}-B(s_{23}s_{13}s_{12}-c_{23}c_{12})-C(c_{23}s_{13}s_{12}+s_{23}c_{12})\\
  \Delta_{13}&=&(A-m_{3}^{2})s_{13}+Bs_{23}c_{13}+Cc_{23}c_{13}\\
  \Delta_{21}&=&Bc_{13}c_{12}-(E-m_{1}^{2})(s_{23}s_{13}c_{12}+c_{23}s_{12})-D(c_{23}s_{13}c_{12}-s_{23}s_{12})\\
  \Delta_{22}&=&Bc_{13}s_{12}-(E-m_{2}^{2})(s_{23}s_{13}s_{12}-c_{23}c_{12})-D(c_{23}s_{13}s_{12}+s_{23}c_{12})\\
  \Delta_{23}&=&Bs_{13}+(E-m_{3}^{2})s_{23}c_{13}+Dc_{23}c_{13}\\
  \Delta_{31}&=&Cc_{13}c_{12}-D(s_{23}s_{13}c_{12}+c_{23}s_{12})-(F-m_{1}^{2})(c_{23}s_{13}c_{12}-s_{23}s_{12})\\
  \Delta_{32}&=&Cc_{13}s_{12}-D(s_{23}s_{13}s_{12}-c_{23}c_{12})-(F-m_{2}^{2})(c_{23}s_{13}s_{12}+s_{23}c_{12})\\
  \Delta_{33}&=&Cs_{13}+Ds_{23}c_{13}+(F-m_{3}^{2})c_{23}c_{13}
\end{array}.
\end{equation}
It is noteworthy, that for the case of two-band system we obtain:
\begin{eqnarray}\label{7.20}
\begin{array}{c}
  \mathrm{i}\gamma^{\sigma}\partial_{\sigma}\nu_{Le}-\varsigma_{e\mu}\nu_{R\mu}=0 \\
  \mathrm{i}\gamma^{\sigma}\partial_{\sigma}\nu_{L\mu}-\varsigma_{e\mu}\nu_{Re}=0 \\
  \mathrm{i}\gamma^{\sigma}\partial_{\sigma}\nu_{Re}-\varsigma_{e\mu}\nu_{L\mu}=0 \\
  \mathrm{i}\gamma^{\sigma}\partial_{\sigma}\nu_{R\mu}-\varsigma_{e\mu}\nu_{Le}=0
\end{array}\Rightarrow
\begin{array}{c}
  \partial^{\sigma}\partial_{\sigma}\nu_{Le}+\varsigma_{e\mu}^{2}\nu_{Le}=0 \\
  \partial^{\sigma}\partial_{\sigma}\nu_{L\mu}+\varsigma_{e\mu}^{2}\nu_{L\mu}=0
\end{array}.
\end{eqnarray}
Thus we can see, that, unlike the three-band system, in the two-band system the flavour states coincides with the mass states.  This means, that \emph{neutrino oscillations in the two-band system are impossible}.

We can see that the mixing of massive (charged) leptons and the mixing of neutrinos have completely different nature. From Eq.(\ref{4.12a}) we can see, that the lepton mixing parameters $\zeta_{\alpha\beta}$ are determined with the interband coupling $\epsilon$, since masses of electron, muon and tauon $m_{Di}$ are determined by coupling with scalar fields $\varphi_{1},\varphi_{2},\varphi_{3}$ accordingly - Eq.(\ref{4.5}), and, in its turn, these scalar fields are mixed by the interband coupling $\epsilon$ - Eq.(\ref{2.32}). Thus, if we turn off the interband interaction, i.e. $\epsilon=0$ is supposed, then the lepton mixing will be absent. On the contrary, neutrinos do not interact with the scalar fields, hence the neutrino mixing parameters $\varsigma_{\alpha\beta}$ are not determined with the interband coupling $\epsilon$. Thus, \emph{the neutrino mixing parameters} $\varsigma_{\alpha\beta}$ \emph{remain as free parameters of the theory}. Cosmological data (anisotropy of cosmic microwave background radiation, formation of structures, etc.) impose restrictions on the masses: $\sum_{\nu}m_{\nu}<0.19\mathrm{eV}$ \cite{lorenz}, $\sum_{\nu}m_{\nu}<0.28\mathrm{eV}$ \cite{thomas}. Since $m_{\nu}\sim|\varsigma|$, then $|\varsigma|\sim 0.1\mathrm{eV}$.

The matrix $U$ (\ref{2.36a}), which is determined by the three mixing angles $\alpha_{12},\alpha_{13},\alpha_{23}$, are unitary. The unitarity property remains in the presence of one more parameter - the phase $\delta$, so that:
\begin{eqnarray}
  \left(
    \begin{array}{c}
      \nu_{Le} \\
      \nu_{L\mu} \\
      \nu_{L\tau} \\
    \end{array}
  \right)&=&
  \left(
    \begin{array}{ccc}
      1 & 0 & 0 \\
      0 & c_{23} & s_{23} \\
      0 & -s_{23} & c_{23} \\
    \end{array}
  \right)
  \left(
    \begin{array}{ccc}
      c_{13} & 0 & s_{13}e^{-\mathrm{i}\delta} \\
      0 & 1 & 0 \\
      -s_{13}e^{\mathrm{i}\delta} & 0 & c_{13} \\
    \end{array}
  \right)
  \left(
    \begin{array}{ccc}
      c_{12} & s_{12} & 0 \\
      -s_{12} & c_{12} & 0 \\
      0 & 0 & 1 \\
    \end{array}
  \right)
  \left(
    \begin{array}{c}
      \nu_{L1} \\
      \nu_{L2} \\
      \nu_{L3} \\
    \end{array}
  \right)\label{7.21a}\\
  \left(
    \begin{array}{c}
      \nu_{L1} \\
      \nu_{L2} \\
      \nu_{L3} \\
    \end{array}
  \right)&=&
  \left(
    \begin{array}{ccc}
      c_{12} & -s_{12} & 0 \\
      s_{12} & c_{12} & 0 \\
      0 & 0 & 1 \\
    \end{array}
  \right)
  \left(
    \begin{array}{ccc}
      c_{13} & 0 & -s_{13}e^{-\mathrm{i}\delta} \\
      0 & 1 & 0 \\
      s_{13}e^{\mathrm{i}\delta} & 0 & c_{13} \\
    \end{array}
  \right)
  \left(
    \begin{array}{ccc}
      1 & 0 & 0 \\
      0 & c_{23} & -s_{23} \\
      0 & s_{23} & c_{23} \\
    \end{array}
  \right)
  \left(
    \begin{array}{c}
      \nu_{Le} \\
      \nu_{L\mu} \\
      \nu_{L\tau} \\
    \end{array}
  \right),\label{7.21b}
\end{eqnarray}
As well known, the complex multipliers $e^{i\delta},e^{-i\delta}$ produce the violation of CP-invariance \cite{branco1,esteban,bilen3,bilen4,giunti}. Then, instead Eq.(\ref{7.16},\ref{7.17}) we obtain:
\begin{equation}\label{7.22}
  \Delta\equiv\left|
    \begin{array}{ccc}
      \Delta_{11} & \Delta_{12} & \Delta_{13} \\
      \Delta_{21} & \Delta_{22} & \Delta_{23} \\
      \Delta_{31} & \Delta_{32} & \Delta_{33} \\
    \end{array}
  \right|=\Re(\Delta)+\mathrm{i}\Im(\Delta)=0
  \Longrightarrow
  \left\{
    \begin{array}{c}
      \Re(\Delta)=0 \\
      \Im(\Delta)=0 \\
    \end{array}
  \right\},
\end{equation}
\begin{equation}\label{7.23}
\begin{array}{ccc}
  \Delta_{11}&=&(A-m_{1}^{2})c_{13}c_{12}-B(s_{23}s_{13}c_{12}e^{\mathrm{i}\delta}+c_{23}s_{12})-C(c_{23}s_{13}c_{12}e^{\mathrm{i}\delta}-s_{23}s_{12})\\
  \Delta_{12}&=&(A-m_{2}^{2})c_{13}s_{12}-B(s_{23}s_{13}s_{12}e^{\mathrm{i}\delta}-c_{23}c_{12})-C(c_{23}s_{13}s_{12}e^{\mathrm{i}\delta}+s_{23}c_{12})\\
  \Delta_{13}&=&(A-m_{3}^{2})s_{13}e^{-\mathrm{i}\delta}+Bs_{23}c_{13}+Cc_{23}c_{13}\\
  \Delta_{21}&=&Bc_{13}c_{12}-(E-m_{1}^{2})(s_{23}s_{13}c_{12}e^{\mathrm{i}\delta}+c_{23}s_{12})-D(c_{23}s_{13}c_{12}e^{\mathrm{i}\delta}-s_{23}s_{12})\\
  \Delta_{22}&=&Bc_{13}s_{12}-(E-m_{2}^{2})(s_{23}s_{13}s_{12}e^{\mathrm{i}\delta}-c_{23}c_{12})-D(c_{23}s_{13}s_{12}e^{\mathrm{i}\delta}+s_{23}c_{12})\\
  \Delta_{23}&=&Bs_{13}e^{-\mathrm{i}\delta}+(E-m_{3}^{2})s_{23}c_{13}+Dc_{23}c_{13}\\
  \Delta_{31}&=&Cc_{13}c_{12}-D(s_{23}s_{13}c_{12}e^{\mathrm{i}\delta}+c_{23}s_{12})-(F-m_{1}^{2})(c_{23}s_{13}c_{12}e^{\mathrm{i}\delta}-s_{23}s_{12})\\
  \Delta_{32}&=&Cc_{13}s_{12}-D(s_{23}s_{13}s_{12}e^{\mathrm{i}\delta}-c_{23}c_{12})-(F-m_{2}^{2})(c_{23}s_{13}s_{12}e^{\mathrm{i}\delta}+s_{23}c_{12})\\
  \Delta_{33}&=&Cs_{13}e^{-\mathrm{i}\delta}+Ds_{23}c_{13}+(F-m_{3}^{2})c_{23}c_{13}\\
\end{array}.
\end{equation}
In the case of symmetrical interband mixing $\varsigma_{e\mu}=\varsigma_{e\tau}=\varsigma_{\mu\tau}\equiv\varsigma$ it is not difficult to see, that any magnitudes of the mixing angles $\alpha_{12},\alpha_{13},\alpha_{23}$ and the CP-violation phase $\delta$ satisfy Eq.(\ref{7.22}). Thus, \emph{the asymmetry of interband neutrino mixing selects values of the mixing angles $\alpha_{ik}$ and the CP-violation phase $\delta$}.

At present time it is known from experiments: $|\Delta m^{2}_{23}|\equiv|m^{2}_{3}-m^{2}_{2}|\approx 2.51\cdot 10^{-3}\mathrm{eV}^{2}$, $|\Delta m^{2}_{12}|\approx 7.41\cdot 10^{-5}\mathrm{eV}^{2}$, $\alpha_{12}=33.4^{\circ}$, $\alpha_{23}=42.2\ldots49.5^{\circ}$, $\alpha_{13}=8.6^{\circ}$, $\delta/^{\circ}=195^{+51}_{-25}$ \cite{NuFIT,ester}. Thus, to find masses of neutrinos, we should solve an inverse problem: knowing the mixing angles $\alpha_{ik}$, the CP-violation phase $\delta$ and the mass differences $|\Delta m^{2}_{ik}|$ find the mixing parameters $\varsigma_{e\mu},\varsigma_{e\tau},\varsigma_{\mu\tau}$. However, such a problem is very difficult for calculation. At the same time, we can see, that two angles $\alpha_{12},\alpha_{23}$ are close to $\pi/4$, i.e this mixing is close to full mixing. On the other hand, the mass differences are strongly asymmetric $\Delta m^{2}_{12}\ll\Delta m^{2}_{23}$. In the case of symmetrical mixing, i.e $\varsigma_{e\mu}=\varsigma_{e\tau}=\varsigma_{\mu\tau}\equiv\varsigma$ we have effective masses (\ref{7.11}). We can see, that there is a tendency $\Delta m^{2}_{23}\gg\Delta m^{2}_{12}\rightarrow 0$. Thus, we can estimate the mixing parameters as
\begin{equation}\label{7.24}
  \varsigma^{2}\sim\frac{1}{3}\Delta m^{2}_{23}\sim 8.369\cdot 10^{-4}\mathrm{eV}^{2}.
\end{equation}
Hence, the band masses of neutrinos can be estimated as:
\begin{equation}\label{7.25}
  \sqrt{m_{\nu1}^{2}}\approx\sqrt{m_{\nu2}^{2}}\approx|\varsigma|=0.0289\mathrm{eV},\quad  \sqrt{m_{\nu3}^{2}}\approx 2|\varsigma|=0.0579\mathrm{eV}.
\end{equation}
Magnitudes of the band masses (\ref{7.25}) are result of very rough approximation of the symmetric mixing $\varsigma_{e\mu}=\varsigma_{e\tau}=\varsigma_{\mu\tau}$, in reality $m_{\nu1}^{2}\neq m_{\nu2}^{2}$, although $|\Delta m_{12}^{2}|\ll|\Delta m_{23}^{2}|$. Then, we can choose the mixing parameters $\varsigma_{e\mu},\varsigma_{e\tau},\varsigma_{\mu\tau}$ so, to obtain the experimentally observed difference of squared masses $\Delta m^{2}_{12},\Delta m^{2}_{23}$ by slightly changing parameter $\varsigma$ from Eq.(\ref{7.24}) (by module):
\begin{equation}\label{7.26}
  \varsigma_{e\mu}=2.988\cdot 10^{-2}\mathrm{eV},\quad\varsigma_{e\tau}=\varsigma_{\mu\tau}=2.893\cdot 10^{-2}\mathrm{eV}.
\end{equation}
Then, using Eq.(\ref{7.9}) the band masses of neutrinos can be estimated as:
\begin{equation}\label{7.27}
  \sqrt{m_{\nu1}^{2}}=0.0286\mathrm{eV},\quad \sqrt{m_{\nu2}^{2}}=0.0299\mathrm{eV}\quad \sqrt{m_{\nu3}^{2}}=0.0585\mathrm{eV}.
\end{equation}
Unfortunately, Eq.(\ref{7.22}) is extremely sensitive to parameters $m_{\nu i},c_{ik}$, hence we can make only some estimations. Let us suppose $\alpha_{13}\rightarrow 0$ in Eqs.(\ref{7.22},\ref{7.23}), then we should take
\begin{equation}\label{7.28}
  \alpha_{13}\rightarrow 0\Rightarrow \alpha_{12}\approx\alpha_{23}\approx 38^{\circ},
\end{equation}
that is close to the \emph{tribimaximal mixing} $\alpha_{12}=35.3^{\circ},\alpha_{23}=45^{\circ},\alpha_{13}=0$. Then $\sqrt{m_{\nu1}^{2}}+\sqrt{m_{\nu2}^{2}}+\sqrt{m_{\nu3}^{2}}\approx 0.12\mathrm{eV}$, that is in consistent with current cosmological data $\sum_{\nu}m_{\nu}<0.19\mathrm{eV}$ \cite{lorenz,thomas} (where all $m_{\nu}>0$).

\section{Systematics of elementary particles, masses of Higgs bosons and "dark matter"}\label{particle}

\begin{figure}[ht]
\begin{center}
\includegraphics[width=12cm]{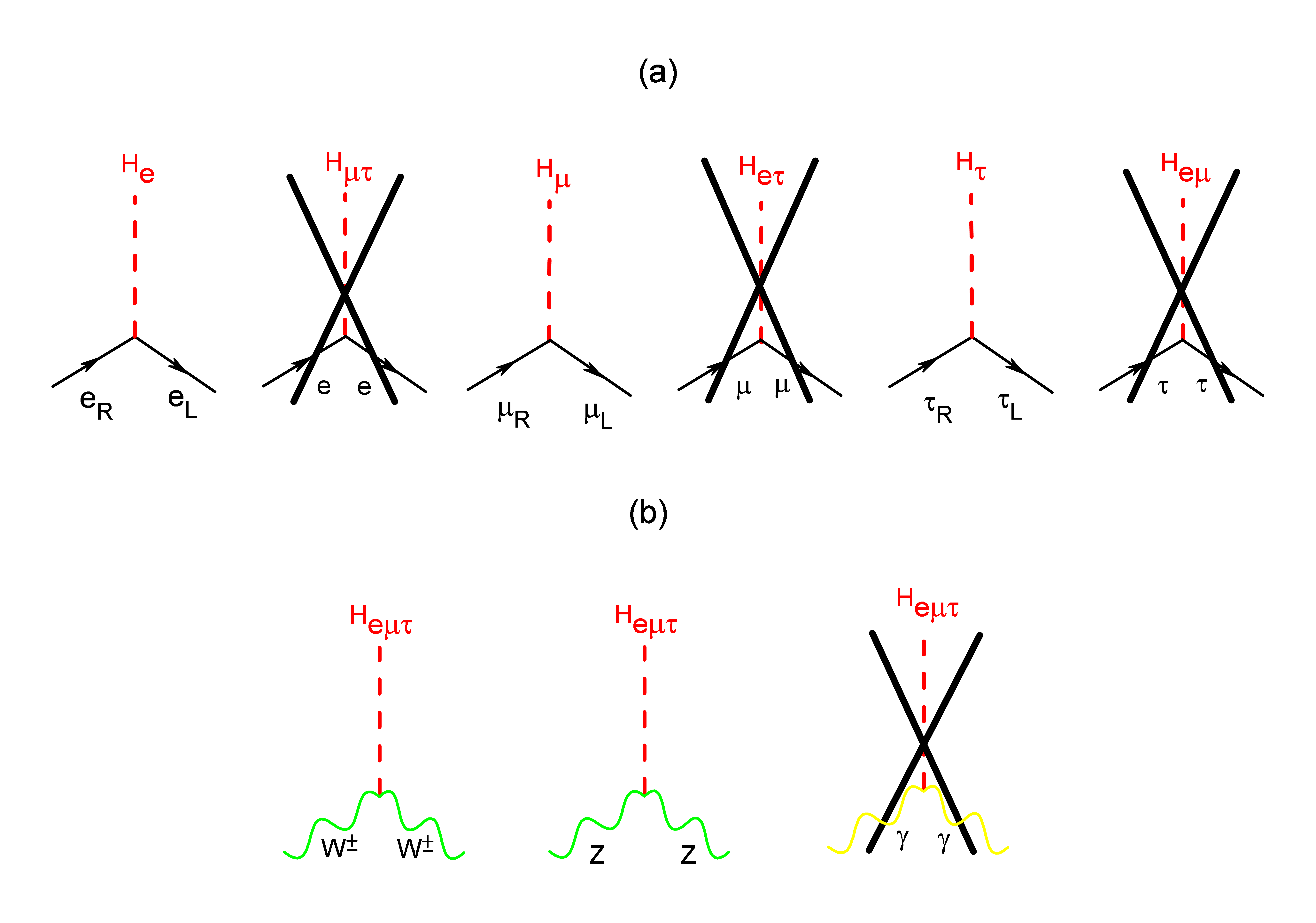}
\end{center}
\caption{The Higgs-lepton vertices (a), and the Higgs-gauge boson vertices (b). Leptons of each flavor can interact only with the H-bosons of corresponding flavor. $W^{\pm}$ and $Z$ gauge bosons can interact with H-bosons of all flavors, but photon $\gamma$ does not interact with the Higgs fields.}
\label{Fig4}
\end{figure}

Summarizing the results of previous sections we can make Tab.\ref{tab2} of elementary particles in the three-band GWS theory (except quarks). We can see, that, unlike the single-band theory, in the three-band case we have three H-bosons with somewhat different masses. In the limit of weak interband coupling $|\epsilon|\ll |a_{1,2,3}|$ we can write their flavour masses via the band parameters:
\begin{equation}\label{8.1}
  m_{He}=\sqrt{2|a_{1}|}<m_{H\mu}=\sqrt{2|a_{2}|}<m_{H\tau}=\sqrt{2|a_{3}|}\sim100\mathrm{GeV}.
\end{equation}
All H-bosons have zero electrical charge $Q=0$, zero lepton charges $l_{e},l_{\mu},l_{\tau}=0$, hypercharge $Y=1$ and the third projection of isospin $I_{3}=-1/2$. At the same time, the bosons $H_{e},H_{\mu},H_{\tau}$ interact only with the corresponding leptons $e,\mu,\tau$ changing their chirality according to Eq.(\ref{4.8}) as shown in Fig.\ref{Fig4}a. The masses of leptons are:
\begin{equation}\label{8.2}
  m_{e}=\chi\varphi_{01},\quad m_{\mu}=\chi\varphi_{02}, \quad m_{\tau}=\chi\varphi_{03},
\end{equation}
where
\begin{equation}\label{8.3}
  \varphi_{01}=\sqrt{\frac{|a_{1}|}{b_{1}}}=\frac{m_{He}}{\sqrt{2b_{1}}},\quad
  \varphi_{02}=\sqrt{\frac{|a_{2}|}{b_{2}}}=\frac{m_{H\mu}}{\sqrt{2b_{2}}}, \quad
  \varphi_{03}=\sqrt{\frac{|a_{3}|}{b_{3}}}=\frac{m_{H\tau}}{\sqrt{2b_{3}}},
\end{equation}
are the equilibrium values of the scalar fields, $\chi$ is the dimensionless coupling constant between the corresponding Dirac fields and the scalar fields (Yukawa coupling).

\begin{center}
\begin{table}[ht]
 \begin{center}
 \begin{tabular}{|c|c|c|c|}
  \hline
  & electron flavor & muon flavor & tauon flavor \\\hline
  Higgs bosons & $H_{e}$ & $H_{\mu}$ & $H_{\tau}$ \\\hline
  charged leptons & $e_{L,R}$ & $\mu_{L,R}$ & $\tau_{L,R}$ \\
  active neutrinos & $\nu_{Le}$ & $\nu_{L\mu}$ & $\nu_{L\tau}$ \\
  sterile neutrinos & $\nu_{Re}$ & $\nu_{R\mu}$ & $\nu_{R\tau}$ \\\hline
\end{tabular}\\
 \end{center}
\bigskip
\begin{center}
\begin{tabular}{|c|c|c|}
\hline
\multicolumn{3}{|c|}{Leggett bosons}\\
  \hline
  massive  & $L_{1}$ & $L_{2}$  \\
  \hline
  massless & \multicolumn{2}{c|}{$L_{3}\leftrightarrow L_{4}$}  \\\hline
\end{tabular}\\
\end{center}
\bigskip
\begin{center}
\begin{tabular}{|c|c|c|}
\hline
\multicolumn{3}{|c|}{gauge bosons}\\
  \hline
  massive  & $W^{\pm}$ & $Z$  \\
  \hline
  massless & \multicolumn{2}{c|}{$\gamma$}  \\\hline
\end{tabular}\\
\end{center}
\caption{Elementary particles in the three-band GWS theory: leptons, Higgs bosons (scalar), Leggett bosons (scalar), gauge bosons (vector). Each flavor of leptons can interact with the Higgs field of only corresponding flavor. The Leggett bosons are sterile particles, hence the massive modes form the so-called "ultra-light dark matter". The sterile right-handed neutrinos have exactly the same effective masses to the corresponding active left-handed neutrinos. Each charged lepton can be both left-handed and right-handed.} 
  \label{tab2}
\end{table}
\end{center}

According to Eqs.(\ref{6.5},\ref{6.10},\ref{6.11}) the gauge fields $W^{\pm}$ and $Z$ interact with all scalar fields as shown in Fig.\ref{Fig4}b. At the same time, photon $\gamma$ does not interact with the scalar fields and remains massless. The masses of charged $W$-boson and neutral $Z$-boson are
\begin{equation}\label{8.4}
  m_{W}=\frac{e}{\sin\alpha}\sqrt{2\pi\left(\varphi_{01}^{2}+\varphi_{02}^{2}+\varphi_{03}^{2}\right)}
  =\frac{e}{\sin\alpha}\sqrt{2\pi\left(m_{e}^{2}+m_{\mu}^{2}+m_{\tau}^{2}\right)}\frac{1}{\chi},\quad
  m_{Z}=\frac{m_{W}}{\cos\alpha},
\end{equation}
where $\sin\alpha=0.4721$ is the Weinberg angle, $e=1/\sqrt{128}$ is the electromagnetic coupling constant at energy $\sim 100\mathrm{GeV}$. Using masses of the gauge boson $m_{W}=80.377\mathrm{GeV}$, lepton masses $m_{e}=0.51\cdot10^{-3}\mathrm{GeV}$, $m_{\mu}=0.1057\mathrm{GeV}$, $m_{\tau}=1.7768\mathrm{GeV}$ we obtain the coupling constant $\chi$:
\begin{equation}\label{8.5}
  \chi=0.0104,
\end{equation}
and the amplitudes of the scalar fields $\varphi_{0i}=m_{i}/\chi$:
\begin{equation}\label{8.6}
  \varphi_{0e}\approx\varphi_{01}=0.05\mathrm{GeV},\quad \varphi_{0\mu}\approx\varphi_{02}=10.17\mathrm{GeV}, \quad \varphi_{0\tau}\approx\varphi_{03}=170.98\mathrm{GeV},
\end{equation}
In standard representation of the isospinor field $\Psi=\frac{1}{\sqrt{2}}\left(\begin{array}{c}
                                                                              0 \\
                                                                              \varphi
                                                                            \end{array}
\right)$ we have $\varphi_{0e}=0.07\mathrm{GeV}$, $\varphi_{0\mu}=14.38\mathrm{GeV}$, $\varphi_{0\tau}=241.80\mathrm{GeV}$, so that the effective amplitude of scalar field is $\varphi_{\mathrm{eff}}=\sqrt{\left(\varphi_{01}^{2}+\varphi_{02}^{2}+\varphi_{03}^{2}\right)}=242\mathrm{GeV}$. If we take the electromagnetic coupling constant at energy $\sim 1\mathrm{GeV}$: $e\approx1/\sqrt{132}$, then we obtain $\varphi_{\mathrm{eff}}=246\mathrm{GeV}$.

Unfortunately both the single-band GWS theory and three-band GWS theory do not allow to calculate the masses of H-bosons (\ref{8.1}). We only know one H-boson with mass $m_{H}=125.10\mathrm{GeV}$. Since the H-boson mediated interactions between leptons (as illustrated in Fig.\ref{Fig4}a) are interactions of common nature and are characterized by the same coupling constant (\ref{8.5}) in our model, then these interactions should have approximately the same effective interaction constants $\sim\frac{\chi^{2}}{m_{He}^{2}}\approx\frac{\chi^{2}}{m_{H\mu}^{2}}\approx\frac{\chi^{2}}{m_{H\tau}^{2}}$ and radii $\sim\frac{1}{m_{He}}\approx\frac{1}{m_{H\mu}}\approx\frac{1}{m_{H\tau}}$, similar to the weak interactions which have approximately equal interaction constants $\sim\frac{g^{2}}{m_{W}^{2}}\approx\frac{\widetilde{g}^{2}}{m_{Z}^{2}}$ and radii $\sim\frac{1}{m_{W}}\approx\frac{1}{m_{Z}}$, since the masses of mediators are of the same order: $m_{W}=80.4\mathrm{GeV}\sim m_{Z}=91.2\mathrm{GeV}$. Therefore, the masses of H-bosons should be of the same order too: $m_{He}\sim m_{H\mu}\sim m_{H\tau}$. At the same time, different Dirac masses of leptons $m_{e}\ll m_{\mu}\ll m_{\tau}$ are caused by different amplitudes of scalar fields $\varphi_{01}\ll \varphi_{02}\ll \varphi_{03}$. The amplitudes of scalar fields $\varphi_{01},\varphi_{02},\varphi_{03}$ (\ref{8.3},\ref{8.6}) differ from each other by orders, namely $\varphi_{01}:\varphi_{02}:\varphi_{03}=m_{e}:m_{\mu}:m_{\tau}$. Thus, the small changes of mass of H-bosons $m_{He}<m_{H\mu}<m_{H\tau}$ should be accompanied by significant changes of the scalar fields $\varphi_{01}\ll \varphi_{02}\ll \varphi_{03}$.

\begin{figure}[ht]
\begin{center}
\includegraphics[width=9cm]{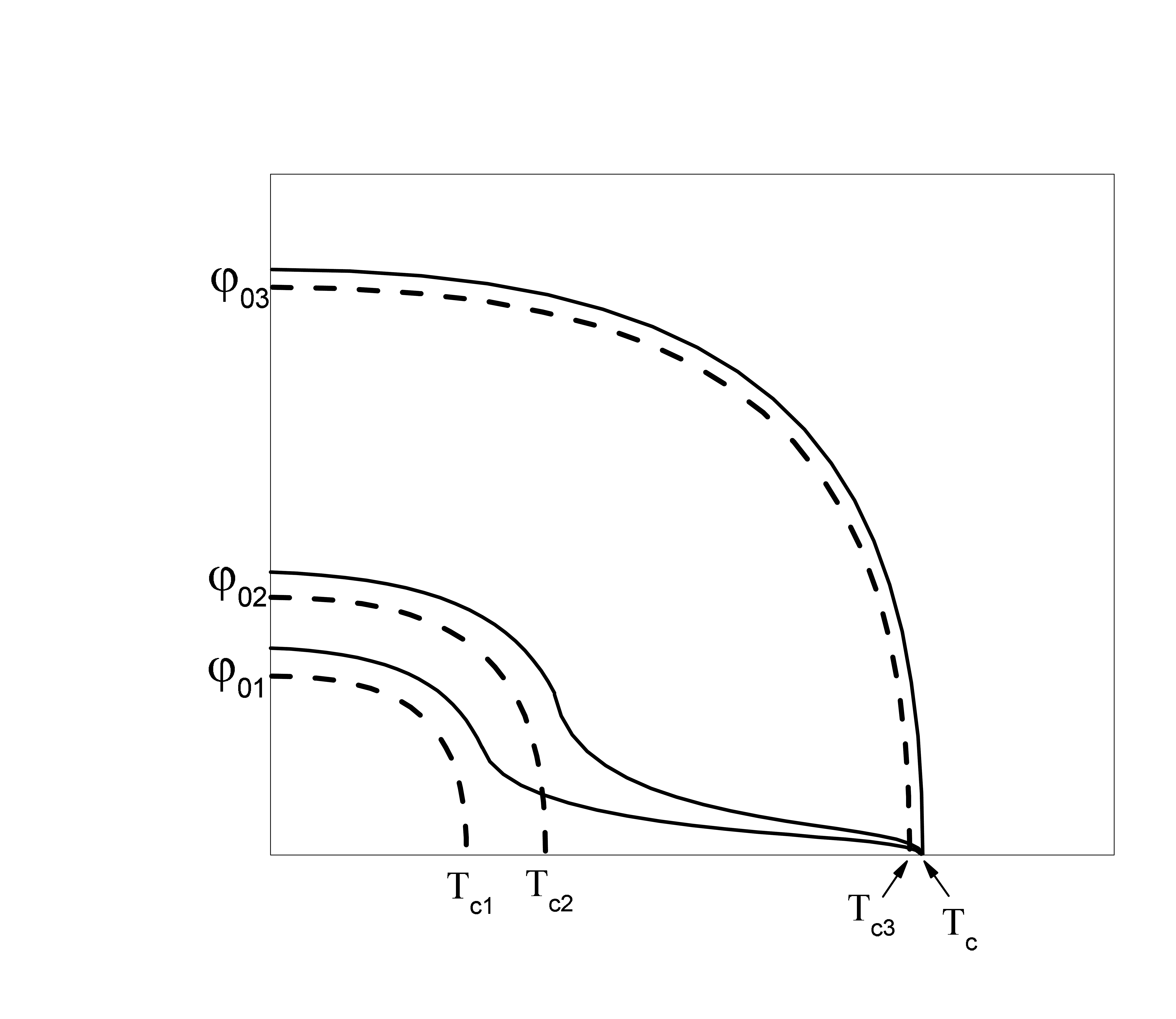}
\end{center}
\caption{The scalar fields $\varphi_{01}(T),\varphi_{02}(T),\varphi_{03}(T)$ as solutions of Eq.(\ref{2.12a4}), if the interband coupling is absent, i.e $\epsilon=0$ (dash lines), and if the weak interband interaction takes place, i.e. $|\epsilon|\ll |a_{i}(0)|$ (solid lines). Applying the weak interband coupling drags the smaller parameters $\varphi_{01,02}$ up to new critical temperature $T_{c}\gg T_{c1,c2}$. The effect on the larger parameter $\varphi_{03}$ is not so significant. The magnitudes of the scalar fields $\varphi_{01,02,03}$ at $T=0$ change very little.}
\label{Fig5}
\end{figure}

In a single-band case the critical temperature is determined by the equilibrium magnitude of the scalar field at $T=0$: $T_{c}=2\varphi_{0}$, at the same time at nonzero temperatures we have $\varphi(T)=\varphi(0)\sqrt{1-\frac{T^{2}}{T_{c}^{2}}}$ \cite{linde}. Let us write coefficients $a(T)$ and $b$ in the following manner:
\begin{equation}\label{8.7}
  a=\mathcal{N}\left(\frac{T^{2}}{T_{c}^{2}}-1\right),\quad b=\frac{4\mathcal{N}}{T_{c}^{2}}.
\end{equation}
Then, the coefficient $\mathcal{N}$ does not take part in the condensate density: $\varphi_{0}(T)=\sqrt{\frac{|a(T)|}{b}}=\frac{T_{c}}{2}\sqrt{1-\frac{T^{2}}{T_{c}^{2}}}$. Since $m_{H}=\sqrt{2|a|}$, at $T=0$ we have:
\begin{equation}\label{8.9}
  \mathcal{N}=\frac{m_{H}^{2}}{2}.
\end{equation}
For three-band system we can write the coefficients $a_{1,2,3}$ and $b_{1,2,3}$ as:
\begin{eqnarray}
  &&a_{1}=\mathcal{N}_{1}\left(\frac{T^{2}}{T_{c1}^{2}}-1\right),\quad
    a_{2}=\mathcal{N}_{2}\left(\frac{T^{2}}{T_{c2}^{2}}-1\right),\quad
    a_{3}=\mathcal{N}_{3}\left(\frac{T^{2}}{T_{c3}^{2}}-1\right),\label{8.9a}\\
  && b_{1}=\frac{4\mathcal{N}_{1}}{T_{c1}^{2}},\quad
     b_{2}=\frac{4\mathcal{N}_{2}}{T_{c2}^{2}},\quad
     b_{3}=\frac{4\mathcal{N}_{3}}{T_{c3}^{2}}.\label{8.9b}
\end{eqnarray}
Here $T_{c1},T_{c2},T_{c3}$ are critical temperatures of corresponding bands, if the bands were independent, i.e $\epsilon=0$. In presence of interband coupling $\epsilon\neq 0$ the system is characterised with the single critical temperature $T_{c}$, which can be calculated using linearized Eq.(\ref{2.12a4}) as the condition of existence of nonzero solutions at $T_{c}$:
\begin{equation}\label{8.10}
  \left|\begin{array}{ccc}
    a_{1}(T_{c}) & \epsilon & \epsilon \\
    \epsilon & a_{2}(T_{c}) & \epsilon \\
    \epsilon & \epsilon & a_{3}(T_{c})
  \end{array}\right|=0\Rightarrow
  a_{1}(T_{c})a_{2}(T_{c})a_{3}(T_{c})+2\epsilon^{3}-\epsilon^{2}\left(a_{1}(T_{c})+a_{2}(T_{c})+a_{3}(T_{c})\right)=0.
\end{equation}
It should be noted that the coefficient $d$ in Eq.(\ref{2.28}) is such that $d(T_{c})=0$ (here $\alpha_{i}(T_{c})=a_{i}(T_{c})>0$ as follows from Eq.(\ref{2.24})). The solutions of Eq.(\ref{2.12a4}) are illustrated in Fig.\ref{Fig5} for the case of strongly asymmetrical bands $T_{c1,c2}\ll T_{c3}$. Effect of interband coupling $\epsilon\neq 0$, even if the coupling is weak $|\epsilon|\ll |a_{i}(0)|$, is non-perturbative for the smaller scalar fields $\varphi_{1,2}$ - applying of the interband coupling drags the smaller amplitudes up to new critical temperature $T_{c}\gg T_{c1,c2}$. At the same time, the effect on the largest scalar fields $\varphi_{3}$ is not so significant - applying of the interband coupling slightly increases the critical temperature $T_{c}\gtrsim T_{c3}$ only. If the interband coupling is weak, then the magnitude of the scalar fields $\varphi_{01,02,03}$ at $T=0$ change very little \cite{grig2,grig3}, for example:
\begin{equation}\label{8.11a}
  \varphi_{01}(0)=\sqrt{\frac{|a_{1}(0)|}{b_{1}}}+\frac{|\epsilon|}{|a_{1}(0)|}\left(\sqrt{\frac{|a_{2}(0)|}{b_{2}}}
+\sqrt{\frac{|a_{3}(0)|}{b_{3}}}\right)\approx\sqrt{\frac{|a_{1}(0)|}{b_{1}}},
\end{equation}
i.e $\varphi_{0i}(0)$ is determined with the intraband coefficients $a_{i}(0),b_{i}$ predominantly.

The coefficients $\mathcal{N}_{1},\mathcal{N}_{2},\mathcal{N}_{3}$, i.e Higgs masses as generalization Eq.(\ref{8.9}) in a sense
\begin{equation}\label{8.9c}
  m_{He}^{2}=2\mathcal{N}_{1},\quad m_{H\mu}^{2}=2\mathcal{N}_{2},\quad m_{H\tau}^{2}=2\mathcal{N}_{3},
\end{equation}
cannot be calculated at present time. In SM the mass of H-boson $m_{H}=125.10\mathrm{GeV}$ is taken from experiment as a parameter of the theory. In superconductors the coefficient $\mathcal{N}$ plays role of the density of electron states on Fermi surface. The critical temperature $T_{c}$ depends exponentially on $\mathcal{N}$, $T_{c}\sim\varphi_{0}(0)\sim\Omega\exp\left(-\frac{1}{g\mathcal{N}}\right)$, at weak coupling, where $\Omega$ is the phonon frequency and $g$ is the constant of electron-phonon interaction. The larger the parameter $\mathcal{N}$, the higher the critical temperature $T_{c}$. In our model $\mathcal{N}_{1}<\mathcal{N}_{2}<\mathcal{N}_{3}$ and $\varphi_{01}\ll\varphi_{02}\ll\varphi_{03}$, that is small changes of parameter $\mathcal{N}$ cause large (exponential) changes of the scalar field $\varphi_{0}$. Then, we can suppose, that the amplitudes of the scalar fields $\varphi_{0i}$ at $T=0$ are determined with the corresponding parameters $\mathcal{N}_{i}$ by analogy with the BCS theory:
\begin{equation}\label{8.11}
  \varphi_{01}=\Omega\exp\left(-\frac{1}{g\mathcal{N}_{1}}\right),\quad
  \varphi_{02}=\Omega\exp\left(-\frac{1}{g\mathcal{N}_{2}}\right),\quad
  \varphi_{03}=\Omega\exp\left(-\frac{1}{g\mathcal{N}_{3}}\right),
\end{equation}
where the parameters $g,\Omega$ are some common parameters for all three bands. Thus, the change of $\varphi$ ($T_{c}$) is accompanied by the logarithmic change of $\mathcal{N}$. Moreover, if the "interaction constant" is zero, i.e $g\mathcal{N}=0$, then the "condensate" is absent $\varphi_{0}=0$. Thus, the scalar fields $\varphi_{i}$ can be result of the Cooper pairing of some more fundamental fermions, as, for example, in models with the top-quark condensation \cite{cvet,volov} or with the technicolor \cite{lane}.

\begin{figure}[ht]
\begin{center}
\includegraphics[width=8.5cm]{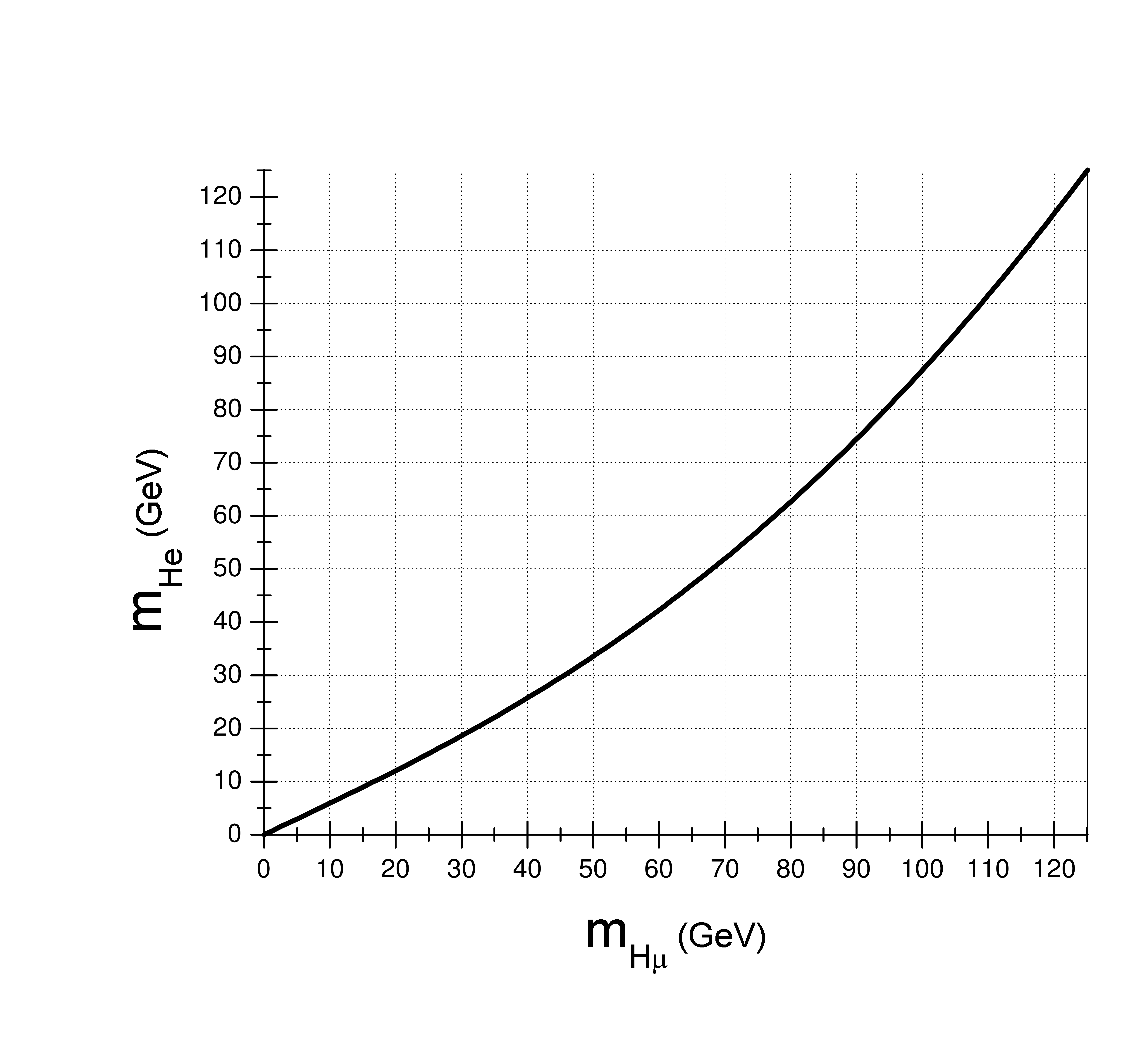}
\end{center}
\caption{The mass of $e$-Higgs boson $m_{He}$ as function of the mass of $\mu$-Higgs boson $m_{H\mu}$, which are limited above by the mass of $\tau$-Higgs boson $m_{H\tau}=125.10\mathrm{GeV}$.}
\label{Fig6}
\end{figure}

We can get rid of the parameter $\Omega$:
\begin{equation}\label{8.12}
  \ln\frac{\varphi_{02}}{\varphi_{01}}=\frac{1}{g}\left(\frac{1}{\mathcal{N}_{1}}-\frac{1}{\mathcal{N}_{2}}\right),\quad
  \ln\frac{\varphi_{03}}{\varphi_{01}}=\frac{1}{g}\left(\frac{1}{\mathcal{N}_{1}}-\frac{1}{\mathcal{N}_{3}}\right),\quad
  \ln\frac{\varphi_{03}}{\varphi_{02}}=\frac{1}{g}\left(\frac{1}{\mathcal{N}_{2}}-\frac{1}{\mathcal{N}_{3}}\right).
\end{equation}
Getting rid of the parameter $g$ we obtain an expression connecting the parameters $\mathcal{N}_{1},\mathcal{N}_{2},\mathcal{N}_{3}$ between themselves:
\begin{equation}\label{8.13}
\frac{\mathcal{N}_{2}-\mathcal{N}_{1}}{\mathcal{N}_{3}-\mathcal{N}_{1}}\frac{\mathcal{N}_{3}}{\mathcal{N}_{2}}=\frac{A}{B}
  \Rightarrow \frac{m_{H\mu}^{2}-m_{He}^{2}}{m_{H\tau}^{2}-m_{He}^{2}}\frac{m_{H\tau}^{2}}{m_{H\mu}^{2}}=\frac{A}{B},
\end{equation}
where we have used Eq.(\ref{8.9c}), $m_{He}\neq m_{H\mu}\neq m_{H\tau}$, and we have denoted:
\begin{equation}\label{8.14}
  A\equiv\frac{\ln\frac{\varphi_{02}}{\varphi_{01}}}{\ln\frac{\varphi_{03}}{\varphi_{02}}}=1.89,
  \quad\quad B\equiv\frac{\ln\frac{\varphi_{03}}{\varphi_{01}}}{\ln\frac{\varphi_{03}}{\varphi_{02}}}=A+1=2.89.
\end{equation}
Thus, due to three-band system the magnitudes of Higgs masses $m_{He}<m_{H\mu}<m_{H\tau}$ are related by Eq.(\ref{8.13}). \emph{We suppose (as will be demonstrated below), that $\tau$-Higgs boson coincides with the observed H-boson of mass $m_{H}=125.10\mathrm{GeV}$, i.e the masses of $m_{H\mu}$ and $m_{He}$ are limited from above by the mass $125.10\mathrm{GeV}$}. Using Eq.(\ref{8.13}) we can find mass of the lightest H-boson $m_{He}$ as function of the boson of medium mass $m_{H\mu}$ at known mass of the heaviest H-boson $m_{H\tau}=125.10\mathrm{GeV}$:
\begin{equation}\label{8.14a}
m_{He}=m_{H\mu}\sqrt{\frac{Bm_{H\tau}^{2}-Am_{H\tau}^{2}}{Bm_{H\tau}^{2}-Am_{H\mu}^{2}}}
\end{equation}
that is illustrated in Fig.\ref{Fig6}.

\begin{figure}[ht]
\begin{center}
\includegraphics[width=12.0cm]{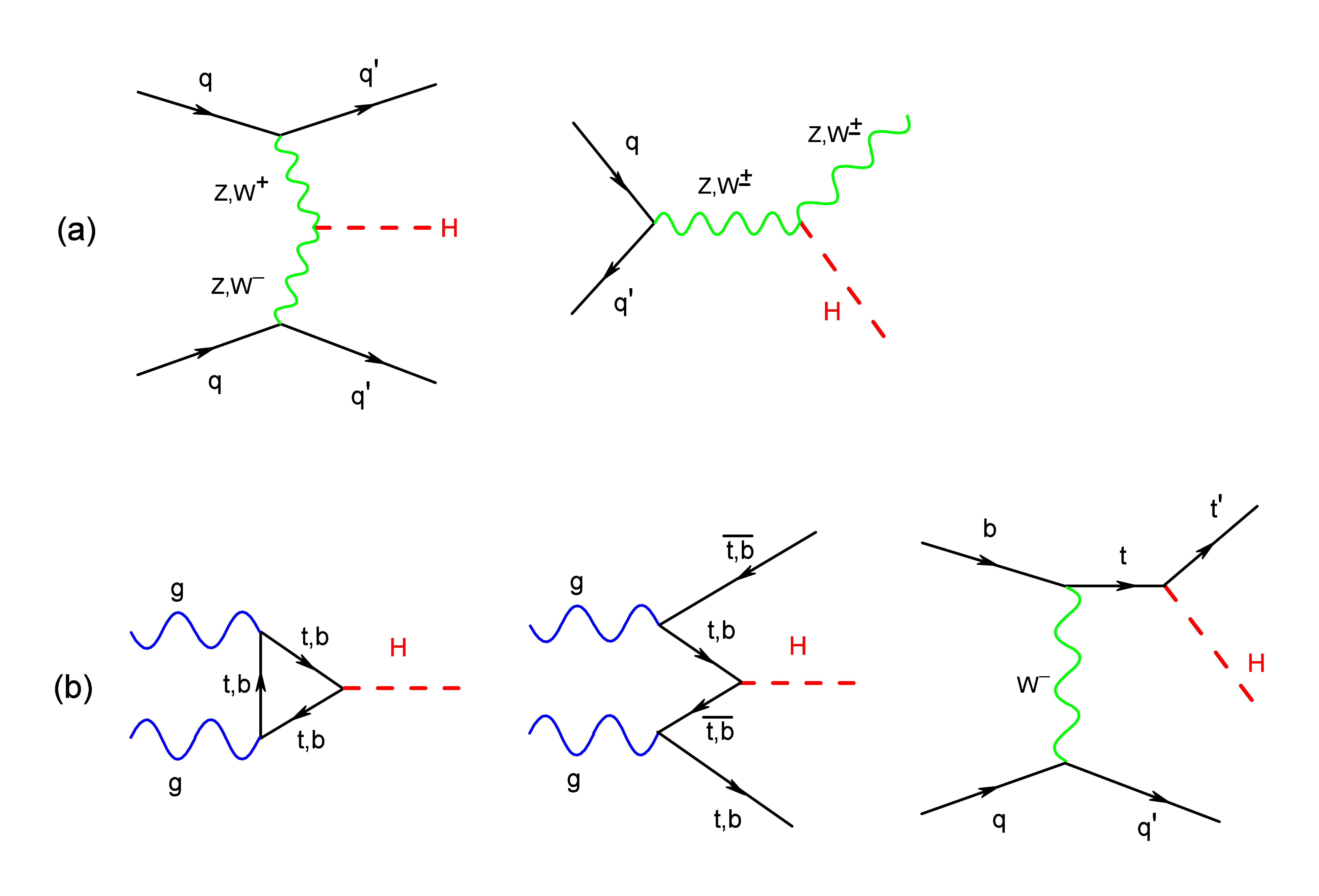}
\end{center}
\caption{Some processes of Higgs boson production: (a) production by the vector bosons $W^{\pm},Z$, (b) production by $t$ and $b$ quarks (here the blue lines $g$ are gluons).}
\label{Fig7}
\end{figure}

Thus, in the proposed model we have H-bosons of three flavors (generations) $H_{e},H_{\mu},H_{\tau}$ which should be characterised by some quantum numbers similar to, for example, lepton numbers or quark flavors. However, only one H-boson of mass $125\mathrm{GeV}$ is experimentally observed at present time. Let us consider processes of the H-boson production \cite{atlas,cms,atlascms}. These processes can be categorized into two types: (a) production by the vector bosons - Fig.\ref{Fig7}a due to interaction (\ref{3.17}), (b) production by the heaviest quarks ($t$ and $b$) - Fig.\ref{Fig7}b due to the Yukawa interaction similar to Eq.(\ref{4.8}).

At first, let us comparable the constants for coupling between gauge bosons and H-bosons for each flavor (generation) from Eq.(\ref{3.17}) using Eq.(\ref{8.6}):
\begin{equation}\label{8.15}
  2e^{2}\varphi_{0e}:2e^{2}\varphi_{0\mu}:2e^{2}\varphi_{0\tau}=0.00028:0.059:1.
\end{equation}
Thus, \emph{gauge bosons $W^{\pm},Z$ most efficiently radiate $H_{\tau}$ bosons}. $H_{\mu}$ and $H_{e}$ bosons must be radiated too, but extremely inefficient compared with $H_{\tau}$ bosons.

\begin{figure}[ht]
\begin{center}
\includegraphics[width=8.0cm]{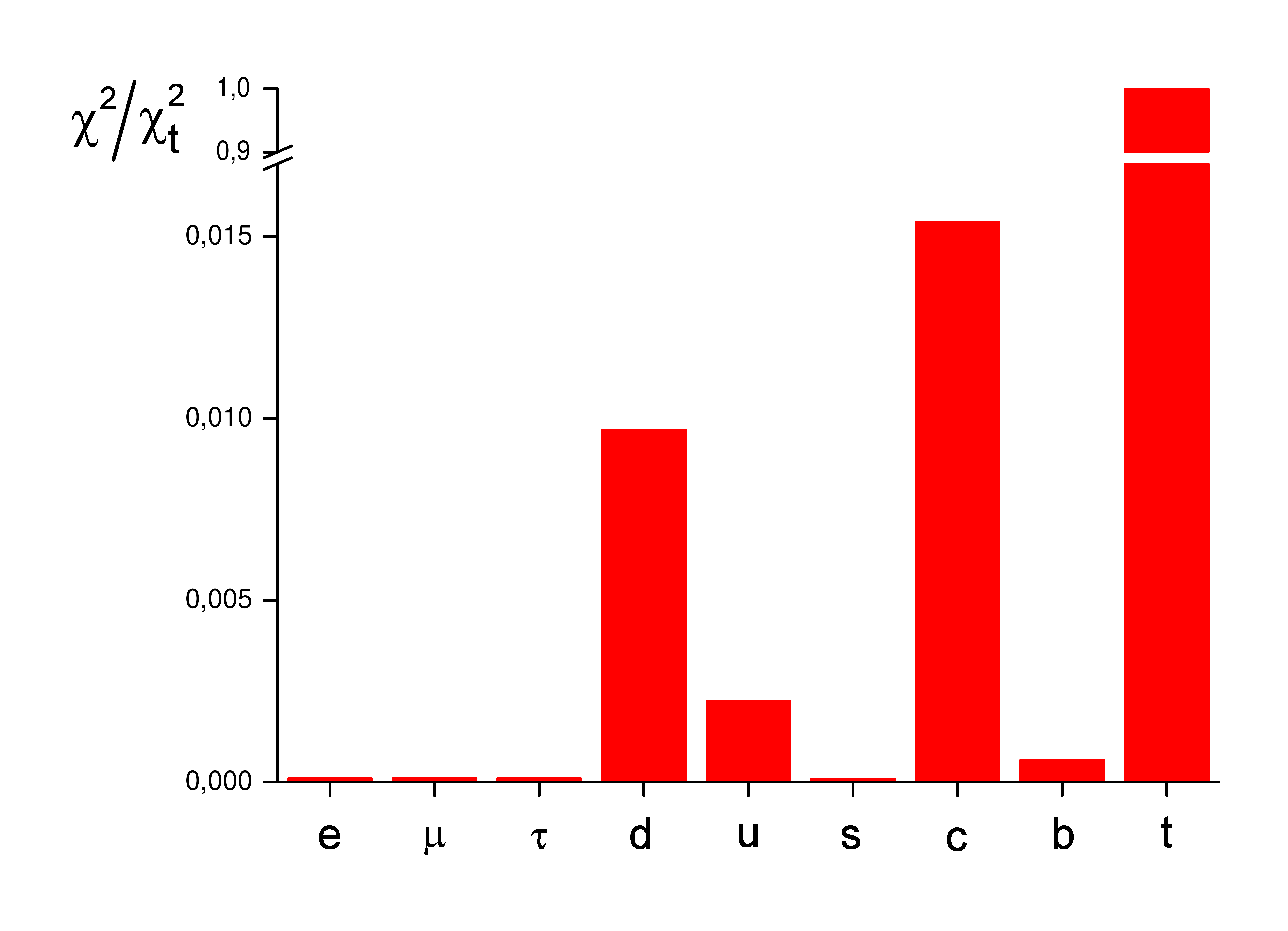}
\end{center}
\caption{Squared Yukawa constants related to the $t$-quark coupling constant: $\frac{\chi^{2}}{\chi_{t}^{2}}$ in the three-band GWS model.}
\label{Fig8}
\end{figure}

\begin{figure}[ht]
\begin{center}
\includegraphics[width=8.0cm]{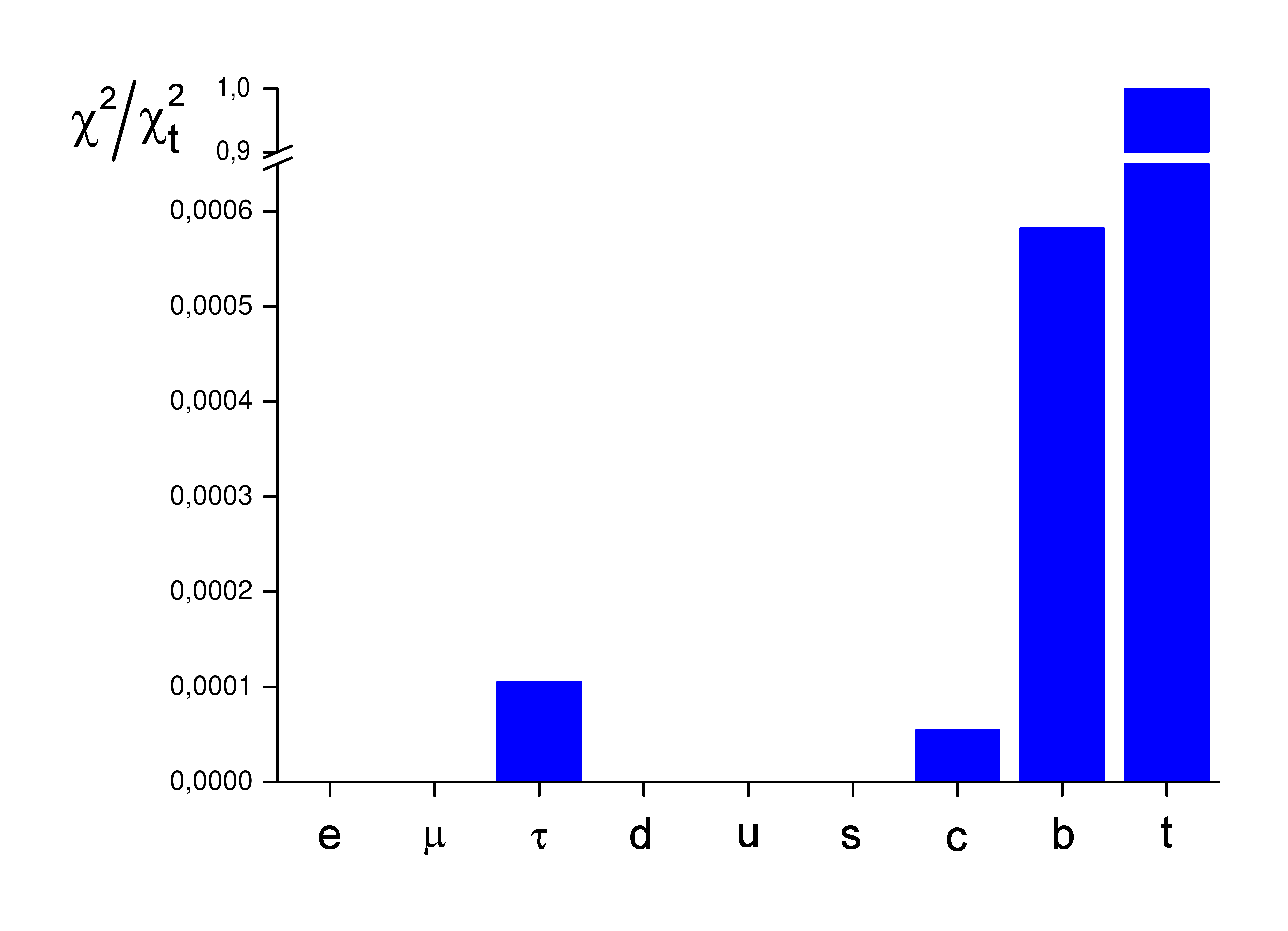}
\end{center}
\caption{Squared Yukawa constants related to the $t$-quark coupling constant: $\frac{\chi^{2}}{\chi_{t}^{2}}=\frac{m^{2}}{m_{t}^{2}}$ in the single-band GWS theory (Standard model).}
\label{Fig9}
\end{figure}

Now, let us consider producing of H-bosons by quarks (or leptons). We should calculate constants of Yukawa coupling $\chi$ as $\chi=\frac{m_{i}}{\varphi_{0i}}$, where an index $i$ means flavor. Amplitudes of the scalar fields $\varphi_{0i}$ are taken from Eq.(\ref{8.6}). Results of this calculation are presented in Tab.\ref{tab3}. Probability of producing or decay of the H-bosons is $\Gamma\propto\chi^{2}$. Squared Yukawa constants related to the $t$-quark coupling constant: $\chi^{2}/\chi_{t}^{2}$ are shown in Fig.\ref{Fig8}. For comparison, the Yukawa constants for SM $\frac{\chi^{2}}{\chi_{t}^{2}}=\frac{m^{2}}{m_{t}^{2}}$ are shown in Fig.\ref{Fig9}. In the single-band GWS-theory (i.e in SM) the masses of fermions are controlled by $\chi$ only, because the scalar field $\varphi$ is single. In multi-band GWS model the masses of fermions are controlled by both $\chi$ and corresponding (for each generation) amplitudes of the scalar fields $\varphi_{0i}$. Thus, differences between Yukawa constants for different flavors are some smoothed out. So, as we could see before, for leptons of all flavors $\chi=0.01$. However, from Tab.\ref{tab3} and Fig.\ref{Fig8} (and also in Fig.\ref{Fig9}) we can see, that $\chi_{t}$ is giant, moreover $m_{t}>m_{H}$ (but $m_{c,b}\ll m_{H}$). This means, that $H_{\tau}$\emph{-bosons are produced in the vast majority of cases}, as in above-described producing by the vector bosons $W^{\pm},Z$.

\begin{center}
\begin{table}[ht]
\begin{center}
 \begin{tabular}{|c|c|c|c|}
  \hline
  & electron flavour & muon flavour & tauon flavour \\\hline
  scalar fields & $\varphi_{0e}=0.05\mathrm{GeV}$ & $\varphi_{0\mu}=10.51\mathrm{GeV}$ & $\varphi_{0\tau}=176.70\mathrm{GeV}$ \\\hline
  charged leptons & $m_{e}=0.0005\mathrm{GeV}$, $\chi=0.010$ & $m_{\mu}=0.1057\mathrm{GeV}$, $\chi=0.010$ & $m_{\tau}=1.7768\mathrm{GeV}$, $\chi=0.010$ \\\hline
  the up quarks & $m_{u}=0.0023\mathrm{GeV}$, $\chi=0.046$ & $m_{c}=1.275\mathrm{GeV}$, $\chi=0.121$ & $m_{t}=173.210\mathrm{GeV}$, $\mathbf{\chi=0.975}$ \\\hline
  the down quarks & $m_{d}=0.0048\mathrm{GeV}$, $\chi=0.096$ & $m_{s}=0.095\mathrm{GeV}$, $\chi=0.009$ & $m_{b}=4.180\mathrm{GeV}$, $\chi=0.024$ \\\hline
\end{tabular}\\
\end{center}
\bigskip
\caption{Masses (experimental) and Yukawa constants of elementary fermions $\chi$ calculated in the three-band GWS theory using amplitudes of the scalar fields $\varphi_{0i}$ from Eq.(\ref{8.6}) for corresponding "flavours" (generations).}
\label{tab3}
\end{table}
\end{center}


\begin{figure}[ht]
\begin{center}
\includegraphics[width=8.0cm]{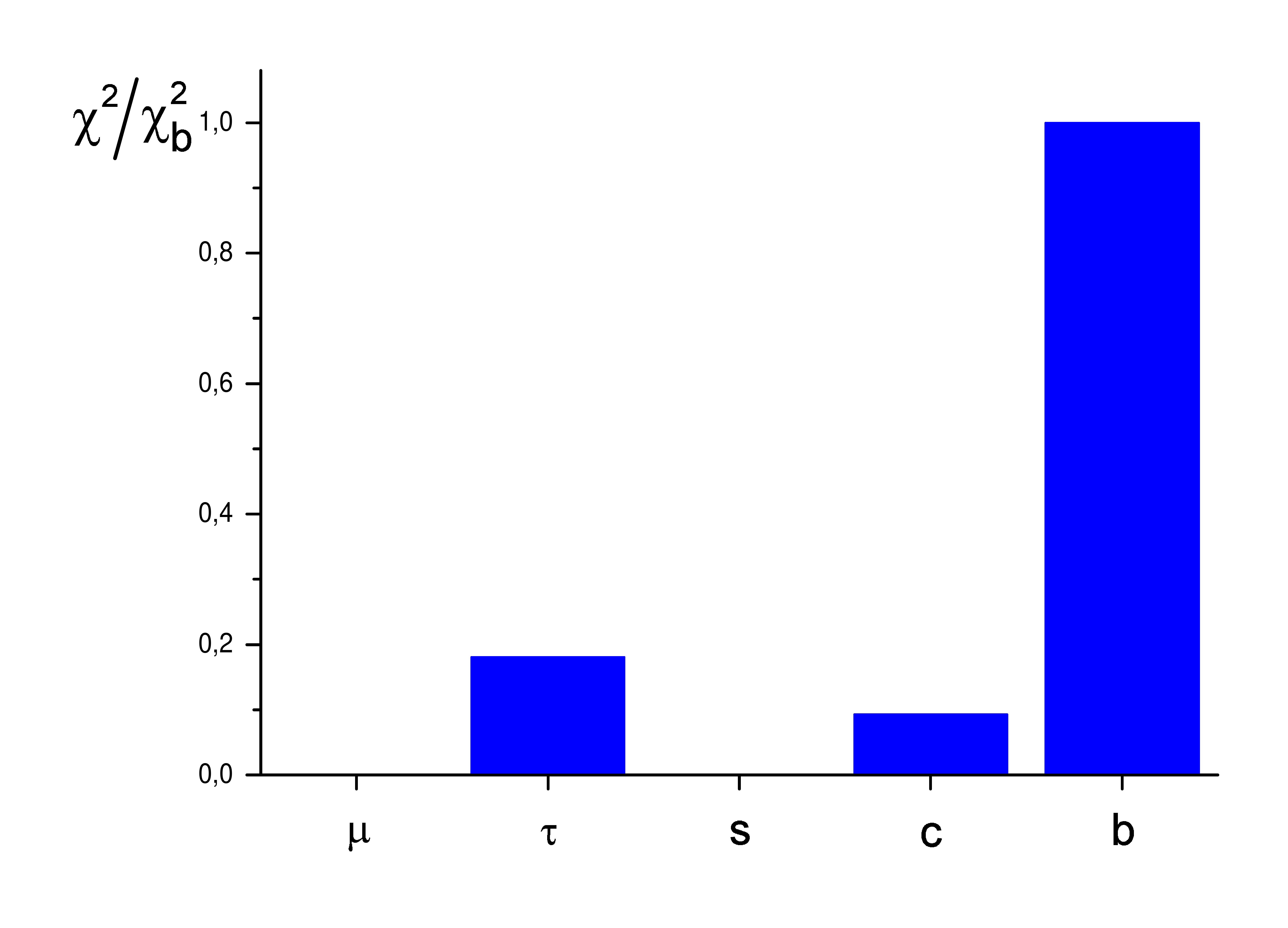}
\end{center}
\caption{Squared Yukawa constants for fermions of the second and third generations (muon, tauon, $s$-quark, $c$-quark, $b$-quark) related to the $b$-quark coupling constant: $\frac{\chi^{2}}{\chi_{b}^{2}}=\frac{m^{2}}{m_{b}^{2}}$ in the single-band GWS theory (Standard model).}
\label{Fig10}
\end{figure}
\begin{figure}[ht]
\begin{center}
\includegraphics[width=8.0cm]{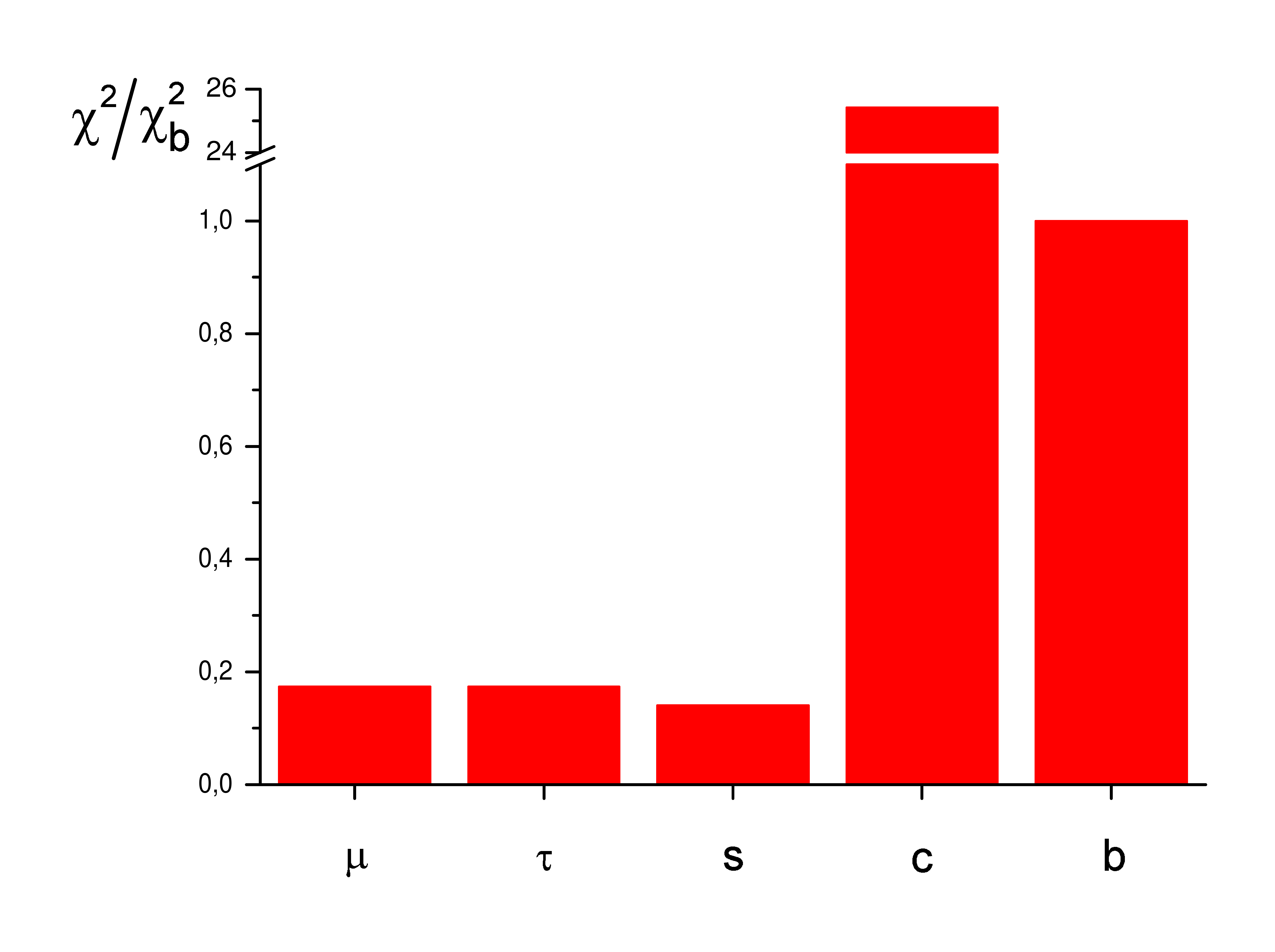}
\end{center}
\caption{Squared Yukawa constants for fermions of the second and third generations (muon, tauon, $s$-quark, $c$-quark, $b$-quark) related to the $b$-quark coupling constant: $\frac{\chi^{2}}{\chi_{b}^{2}}$ in the three-band GWS model.}
\label{Fig11}
\end{figure}

Let us consider decays of the H-boson into quarks or leptons shown in Fig.\ref{Fig02}. According to SM the H-boson should decay as $H\rightarrow b\bar{b}$ with probability $57.5\%$, $H\rightarrow\tau\bar{\tau}$ with probability $6.30\%$, $H\rightarrow c\bar{c}$ with probability $2.90\%$, $H\rightarrow\mu\bar{\mu}$ with probability $\lesssim0.022\%$ \cite{atlascms}. As illustration, the squared Yukawa constants for fermions of the second and third generations (muon, tauon, $s$-quark, $c$-quark, $b$-quark) related to the $b$-quark coupling constant $\frac{\chi^{2}}{\chi_{b}^{2}}=\frac{m^{2}}{m_{b}^{2}}$ calculated in SM are shown in Fig.\ref{Fig10}. Thus, in SM the H-boson interacts most strongly with the third generation, which is the most massive, hence the decays $H\rightarrow b\bar{b}$ and $H\rightarrow\tau\bar{\tau}$ are dominant. However, the decays into the second generation fermions $H\rightarrow c\bar{c}$ should also be noticeable.

At the same time, there has been no experimental evidence found in direct searches by the ATLAS and CMS collaborations \cite{atlas1,cms1} of the H-boson decaying into charm quark–antiquark, into strange quark–antiquark, and into electron–positron. The decay into muon–antimuon has been detected with a significance of 3$\sigma$ \cite{cms4}, that is clearly not enough for an experimental fact (i.e more than 5$\sigma$), moreover there are similar decays $H\rightarrow \gamma\mu\overline{\mu},\gamma e\overline{e}$, but which occur through many intermediate channels due to various interactions (via virtual photon, Z,W bosons, quarks) with a significance of 3.2$\sigma$ \cite{atlas4}. This fact (absence of observations of decays of H-bosons into fermions of the first and second generations) is usually associated with the small Yukawa constant of the first and second generations. However, from Fig.\ref{Fig10} we can see that the decay rate into a pair of $c$-quarks is not much less the decay rate into a pair of $\tau$-leptons, i.e $\chi_{c}^{2}\lesssim\chi_{\tau}^{2}$ (the decay probabilities are $2.9\%$ and $6.4\%$ accordingly). On the other hand, such rare decays as two-photon decay $H\rightarrow\gamma\gamma$ with probabilities $\approx0.2\%$ have been detected.

If we turn to the three-band GWS model, then we have squared Yukawa constants for fermions of the second and third generations (muon, tauon, $s$-quark, $c$-quark, $b$-quark) related to the $b$-quark coupling constant: $\frac{\chi^{2}}{\chi_{b}^{2}}$ shown in Fig.\ref{Fig11}. Let us compare these relations with Fig.\ref{Fig10}. It is not difficult to see, that
\begin{equation}\label{8.16}
  \frac{\Gamma(H_{\tau}\rightarrow\tau\overline{\tau})}{\Gamma(H_{\tau}\rightarrow b\overline{b})}\approx
  \frac{\Gamma(H\rightarrow\tau\overline{\tau})}{\Gamma(H\rightarrow b\overline{b})}\approx 0.2.
\end{equation}
That is, probability of decay of $H_{\tau}$-boson into $\tau$-leptons regarding the decay into $b$-quarks in the three-band GWS model and probability of decay of $H$-boson into $\tau$-leptons regarding the decay into $b$-quarks in SM model are almost equal. However, in the three-band model the decays $H_{\tau}\rightarrow c\overline{c},H_{\tau}\rightarrow s\overline{s},H_{\tau}\rightarrow \mu\overline{\mu}$ are prohibited. But the decays $H_{\mu}\rightarrow c\overline{c}, s\overline{s}, \mu\overline{\mu}$ are allowed, where the decay $H_{\mu}\rightarrow c\overline{c}$ sharply dominates. However, as have been demonstrated before, $H_{\mu}$-boson is emitted extremely inefficiently. Thus, due to the inefficiency of production of $H_{e}$ and $H_{\mu}$ by gauge bosons, the anomalously large Yukawa constant of $t$-quark and the huge background from QCD, searching for $H_{e}$ and $H_{\mu}$ by hadron-hadron collisions in LHC requires to probe the Higgs decays to a deeper level with sufficient accuracy. At the same time, it is planning to build Future Circular electron-positron Collider (FCC-ee) that would provide unprecedented precision measurements and potentially point the way to physics beyond SM \cite{blondel}. It would allow to look for $H_{e}$ in direct collisions $e^{-}e^{+}$ at hight energies, because then the background from QCD should be absent, electron-positron pairs can directly annihilate to $H_{e}$-bosons (similar to muon-antimuon pairs could annihilate to $H_{\mu}$-bosons and tauon-antitauon pairs can annihilate to $H_{\tau}$-bosons).


Proceeding from aforesaid, we should identify $H_{\tau}$-boson with experimentally observed $H$-boson:
\begin{equation}\label{8.17}
  H_{\tau}\equiv H_{\mathrm{observed}}.
\end{equation}
Other generations (flavours) of H-bosons $H_{\mu}$ and $H_{e}$ require detection. So, the H-boson of the electron flavor (the first generation) should decay as $H_{e}\rightarrow d\overline{d}, u\overline{u}, e\overline{e}$.

It should be noted, that in recent years observation of so called "multi-lepton anomalies" \cite{crivel,budden1} in Large Hadron Collider is interpreted (with a local significance of $\lesssim 3\sigma$) as the existence of beyond the SM Higgs bosons: a new scalar particle $S$ with a mass $m_{S}=151\mathrm{GeV}$, produced from the decay of some heavier new scalar particle $H$ (with mass $m_{H}\geqslant 276\mathrm{GeV}$) into a lighter one $S$ and the SM Higgs $h$: $H\rightarrow Sh,SS$ according to $2\mathrm{HDM}+\mathrm{S}$ model \cite{mellado,budden2}. However, significance of this anomaly is debatable \cite{fowlie}. The CMS and ATLAS Collaborations reported about signal with the production cross section of a SM-like scalar $\phi$ with the mass $\sim 95\mathrm{GeV}$ which manifests itself as diphoton decay $pp\rightarrow\phi\rightarrow\gamma\gamma$ \cite{cms2,aguilar,biek} with a local significance of $1.7\sigma\ldots2.9\sigma$. During the search for additional Higgs bosons $\phi$ and vector leptoquarks in $\tau\tau$ final states the CMS found a $3.1\sigma$ excess of events for $pp\rightarrow\phi\rightarrow\tau\tau$ at an invariant mass $m_{\phi}\simeq 100\mathrm{GeV}$, and $2.6\sigma$ at an invariant mass $m_{\phi}\simeq 95\mathrm{GeV}$. Thus, low significance of these anomalies ($\lesssim 3\sigma$) does not make it possible to interpret them as unambiguous confirmation of multi-HDM models. It is possible, that these recorded signals correspond to some very heavy meson resonances or tetroquark resonances.

As we could see in Sects.\ref{spontanU1},\ref{spontanU1gauge},\ref{spontanSU2},\ref{spontanSU2U1}, due to the interband coupling the Goldstone modes from each band (oscillations of $\theta$ and $\vartheta$ phases) transform into following normal modes of the system. Twofold degenerated acoustic mode $q_{\mu}q^{\mu}=0$ which is common mode oscillations of the phases of the isospinor fields $\Psi_{1,2,3}$. Propagation of this mode is accompanied with the current $J^{\mu}\neq 0$, hence this mode is absorbed by the gauge fields $W_{\mu},W_{\mu}^{\ast},Z_{\mu}$. Other modes are Leggett modes, which are antiphase oscillations of the phases of the isospinor fields. Propagation of the L-modes are not accompanied with the current $J^{\mu}=0$, hence they do not interact with the gauge fields. Moreover, Leggett modes do not interact with Higgs modes in linear approximation if attractive interband coupling takes place $\epsilon<0$. Further, these modes do not interact with the Dirac fields $\psi_{1,2,3}$, unlike the Higgs modes. Thus, \emph{the Leggett modes do not interact with any particles, i.e. they are sterile. These modes can manifest itself through only gravity on astrophysical scales}. One of the L-modes is twofold degenerated acoustic mode $q_{\mu}q^{\mu}=0$, which we noted as $L_{3}\leftrightarrow L_{4}$ in Tab.\ref{tab2}. However, the massless bosons lose their energy in the process of space expansion, similar to relic photons. Moreover, ultrarelativistic particles cannot be assembled into self-gravitating halo. Therefore, such particles do not take part in the DM. However, other two modes $L_{1}$ and $L_{2}$ are massive with masses determined with the coefficient of interband coupling as $q_{\mu}q^{\mu}\sim\epsilon$. The masses of L-bosons can be calculated using Eqs.(\ref{6.13},\ref{6.14},\ref{8.6}):
\begin{equation}\label{8.18}
  m_{L1}=5.83\sqrt{|\epsilon|},\quad  m_{L2}=85.98\sqrt{|\epsilon|}.
\end{equation}
Since the L-bosons do not take part in the electro-weak interaction, hence they cannot decay, for example, into two photons, that is \emph{the L-bosons are stable particles}. Obviously, the massive L-bosons are able to form stable gravitationally bound structures (halo, clusters etc). Therefore, \emph{the massive L-bosons are suitable candidate for DM}. 

\section{The masses of Leggett bosons and the cuspy halo problem}\label{darkmatter}

In Sect.\ref{particle} we have found that the Leggett modes do not interact with any particles, i.e. these modes are sterile and they only can manifest itself through gravity on astrophysical scales. Hence, the massive L-bosons are particles of the so-called "dark matter" (massless, i.e ultrarelativistic, L-bosons cannot be accumulated in self-gravitating clusters). Masses of L-bosons are determined with the coefficient of interband coupling $\epsilon$ - Eq.(\ref{8.18}). This coefficient can be arbitrary small because effect of interband coupling is nonperturbative.

Observation of DM density distributions (halo around a galaxy) seems to prefer a central density as $\rho\sim r^{0}$. For example, the empirical core profiles can be described by the following function with two parameters: a scale radius $r_{0}$ and a scale density $\rho_{0}$ \cite{robles}:
\begin{equation}\label{10.2}
  \rho(r)=\frac{\rho_{0}}{1+(\frac{r}{r_{0}})^{2}}.
\end{equation}
However, in the large-scale simulations, which use the collisionless cold dark matter model, the inner region of the halo shows a density distribution described by a power law $\rho\sim r^{\alpha}$ with $\alpha=-1$. Such behaviour is what is now called as a "cusp". For example, Navarro–Frenk–White profile:
\begin{equation}\label{10.3}
  \rho(r)=\frac{\rho_{0}}{\frac{r}{r_{0}}\left(1+\frac{r}{r_{0}}\right)^{2}}.
\end{equation}
Since mass of L-bosons can be extremely small and, hence, critical temperature of BEC can be very high (because coefficient $|\epsilon|$ can be arbitrarily small), then the Bose-Einstein condensate dark matter (BEC DM or Scalar Field Dark Matter, Fuzzy, Wave, Ultra-light Dark Matter) can take place \cite{robles,lee,matos}. This means, that the halo is described by the macroscopic wave function:
\begin{equation}\label{10.4}
  \sqrt{M}\psi(r)=\sqrt{\rho(\mathbf{r},t)}e^{\mathrm{i}S(\mathbf{r},t)},
\end{equation}
where $M=mN$ is total mass of the DM halo, $m$ is the DM particle mass (mass of L-bosons - Eq.(\ref{8.18})), $N$ is number of the particles in the halo. Then an quantum Euler-Madelung equation for stationary case $\frac{D(\nabla S(\mathbf{r},t))}{Dt}=0$ is
\begin{equation}\label{10.5}
  \mathbf{g}-\frac{\nabla p}{\rho}+\sigma\nabla T+\frac{\nabla Q}{m}=0,
\end{equation}
where $\mathbf{g}$ is the gravitational field strength:
\begin{equation}\label{10.6}
  \mathbf{g}=-\frac{4\pi G\mathbf{r}}{r^{3}}\int_{0}^{r}\rho(r')r'^{2}dr'.
\end{equation}
L-bosons do not interact with anything except through gravity, then we can assume "dust" matter $p=0$. In BEC at $T\rightarrow 0$ the entropy can be suppose $\sigma=0$ or the profile can be suppose isothermal $\nabla T=0$. $Q$ is a quantum potential:
\begin{equation}\label{10.7}
  Q=\frac{\hbar^{2}}{2m}\frac{\Delta\sqrt{\rho}}{\sqrt{\rho}},
\end{equation}
i.e $\frac{1}{m}\nabla Q$ is the quantum pressure term.

Let us consider the central cusp of profile (\ref{10.3}) in a form $\rho(r)=\rho_{0}\frac{r_{0}}{r}$. Then the gravitational field strength is
\begin{equation}\label{10.8}
  \mathbf{g}=-4\pi G\rho_{0}r_{0}\frac{\mathbf{r}}{r},
\end{equation}
and the quantum pressure term takes the form:
\begin{equation}\label{10.9}
  \frac{\nabla Q}{m}=\frac{\hbar^{2}}{2m^{2}}\frac{1}{r^{3}}\frac{\mathbf{r}}{r}.
\end{equation}
We can see, that, while the field strength is finite, the quantum pressure is singular at $r=0$. Thus, the equilibrium cannot be achieved. Such a cusp should be blurred by itself.

Now, let us consider the observed profile (\ref{10.2}). At $r\rightarrow 0$ we obtain:
\begin{equation}\label{10.10}
  \mathbf{g}=-\frac{4\pi}{3} G\rho_{0}r\frac{\mathbf{r}}{r},\quad
  \frac{\nabla Q}{m}=\frac{\hbar^{2}}{m^{2}}\frac{6r}{r_{0}^{4}}\frac{\mathbf{r}}{r},
\end{equation}
from where we can see, that the Euler equation (\ref{10.5}) can be satisfied, when
\begin{equation}\label{10.11}
  \frac{4\pi}{3} G\rho_{0}=\frac{\hbar^{2}}{m^{2}}\frac{6}{r_{0}^{4}}.
\end{equation}
Obviously, that $\rho_{0}r_{0}^{3}\sim M$, then we have from Eq.(\ref{10.11}):
\begin{equation}\label{10.12}
  \rho_{0}\sim G^{3}\frac{m^{6}M^{4}}{\hbar^{6}},\quad r_{0}\sim \frac{\hbar^{2}}{Gm^{2}M}.
\end{equation}
Thus, due to the quantum pressure, the central density $\rho(r\rightarrow 0)$ is not singular. From Eq.(\ref{10.12}) we can see, that extremely small mass $m=m_{L}\sim 10^{-20}\mathrm{eV}$ can ensure small central density $\rho_{0}$ and large profile width $r_{0}$. At the same time, we can see, that the spatial distribution (\ref{10.2}) does not give the finite mass of a self-gravitating dark matter halo: $\int_{0}^{\infty}\frac{r^{2}dr}{1+\left(r/r_{0}\right)^{2}}=\infty$. A good approximation would be profile obtained in Ref.\cite{lev} for a self-gravitating system:
\begin{equation}\label{10.13}
  \rho(r)=\frac{\rho_{0}}{\cosh^{2}(\frac{r}{r_{0}})}.
\end{equation}
Such distribution becomes the profile (\ref{10.2}) at $r\ll r_{0}$, at the same time it gives finite mass of the halo: $M=\frac{\pi^{3}}{3}\rho_{0}r_{0}^{3}$. Unfortunately, we cannot verify (\ref{10.13}) by direct substitution in Eq.(\ref{10.5}) because the integral $\int_{0}^{r}\frac{r^{2}dr}{\cosh^{2}(r/r_{0})}$ cannot be calculated analytically. However we can verify it at another limit $r\gg r_{0}$. Then we have:
\begin{equation}\label{10.14}
  \mathbf{g}=-G\frac{M}{r^{2}}\frac{\mathbf{r}}{r},\quad
  \frac{\nabla Q}{m}=\frac{\hbar^{2}}{m^{2}}\frac{1}{r^{2}r_{0}}\frac{\mathbf{r}}{r}.
\end{equation}
Obviously, Eq.(\ref{10.5}) is satisfied at $r_{0}=\frac{\hbar^{2}}{Gm^{2}M}$, that corresponds to Eq.(\ref{10.12}). Thus, the spatial distribution (\ref{10.13}) can describe the DM halo.

Using Eq.(\ref{10.12}), mass and radius of the DM halo of Milky Way $M\sim 10^{12}M_{\odot}$ and $r_{0}\sim120\mathrm{kpc}$ \cite{batt} we can estimate mass of a L-boson $m=m_{L}\sim\sqrt{|\epsilon|}$ and, hence, magnitude of the parameter of interband coupling $|\epsilon|$ using Eq.(\ref{8.18}): $ m_{L}\sim 10^{-25}\mathrm{eV}\Rightarrow|\epsilon|\sim 10^{-54}\mathrm{eV}^{2}$. However, numerical simulations demonstrate, that the DM halo has some structure: a core from BEC of size $\sim 1\mathrm{kpc}$ and above-condensate Bose gas behaving as the cold DM, then mass $\sim10^{-22}\ldots10^{-20}\mathrm{eV}$ \cite{hu,lee,robles1,matos,ferr} is assumed. Indeed, observations of stellar kinematics of dwarf galaxies give the mass of just $\sim 10^{-22}\ldots 10^{-20}\mathrm{eV}$ \cite{broad,gold,pozo}. Then, we can suppose:
\begin{equation}\label{10.15}
  m_{L}\sim 10^{-20}\mathrm{eV}\Rightarrow|\epsilon|\sim 10^{-44}\mathrm{eV}^{2}.
\end{equation}
As mentioned above, the interband coupling is nonperturbative, hence even such a small magnitude of $\epsilon$ determines symmetry and spectrum of the system.

The L-bosons can appear due to the decaying of the vacuum after the inflation and precipitate into Bose-condensate. Temperature of Bose-condensation is $T_{\mathrm{\mathrm{BEC}}}\sim n_{\mathrm{cr}}^{2/3}\frac{h^{2}}{mk_{\mathrm{B}}}=\rho_{\mathrm{cr}}^{2/3}\frac{h^{2}}{m^{5/3}k_{\mathrm{B}}}\sim 10^{31}\mathrm{K}$, where $\rho_{\mathrm{cr}}$ is the critical density of universe. Thus, $T_{\mathrm{\mathrm{BEC}}}$ is commensurable with the Plank temperature $T_{\mathrm{plank}}\sim 10^{32}\mathrm{K}$. Thus, the L-bosons are so light, that $T_{\mathrm{BEC}}\sim T_{\mathrm{plank}}$, that is the L-bosons should be in BEC always. This indicates about a purely condensate nature of the DM halo, and not the two-component structure as condensate core of size $\lesssim 1\mathrm{kps}$ and cloud of above-condensate Bose-gas of size $\sim 120\mathrm{kps}$. L-bosons could condense in BEC at early eras of universe. Galaxies, galaxy clusters, and superclusters are immersed in Bose-condensate clouds from sterile massive L-bosons, that creates the effect of "dark matter".

\begin{figure}[ht]
\begin{center}
\includegraphics[width=9.0cm]{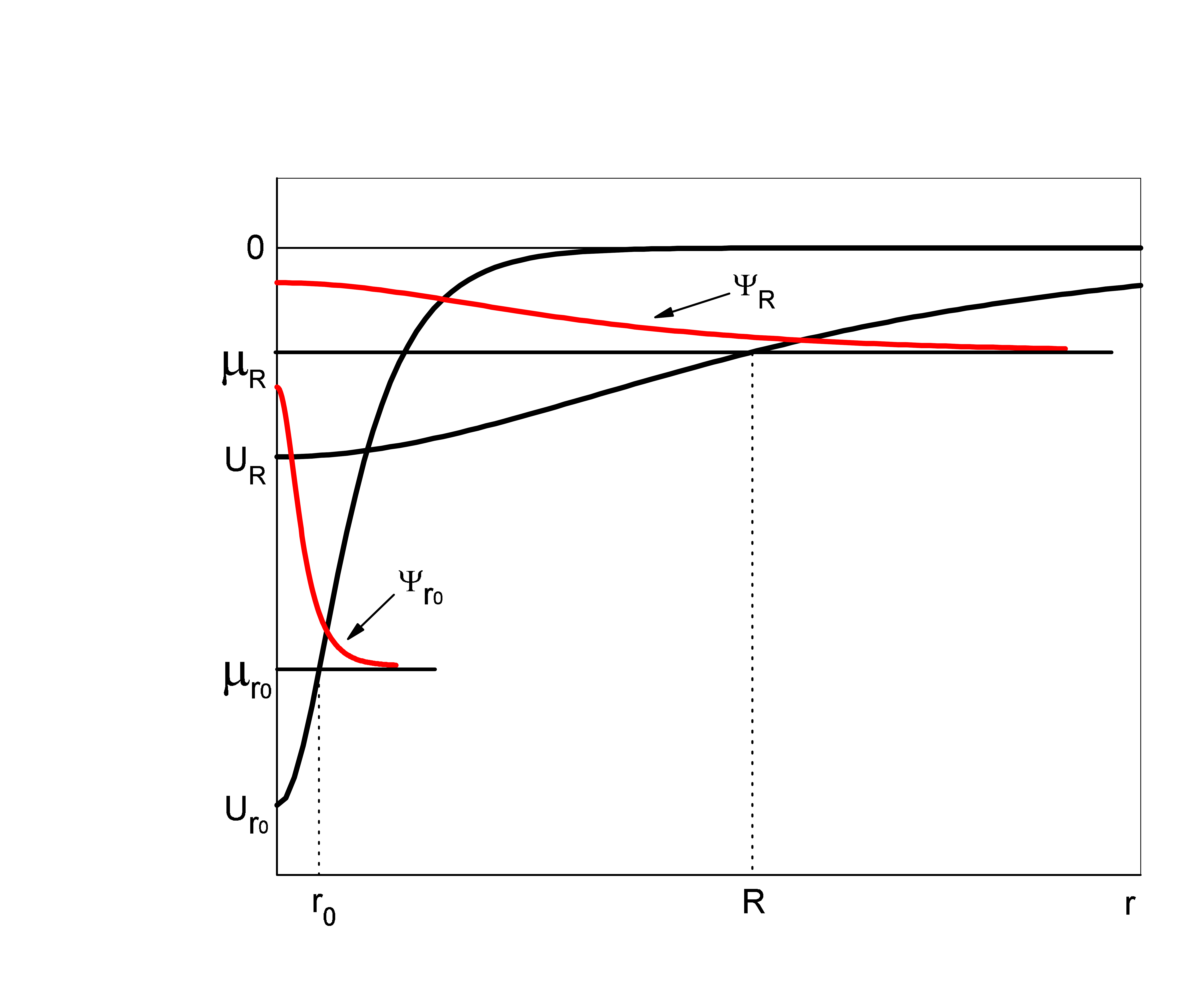}
\end{center}
\caption{Potentials $U_{r_{0}}$ and $U_{R}$ for the ground state $\Psi_{r_{0}}$ with energy $E_{r_{0}}=\mu_{r_{0}}\frac{M}{m}$ and some excited state $\Psi_{R}$ with energy $E_{R}=\mu_{R}\frac{M}{m}$ accordingly. Corresponding radii of the DM halos are $r_{0}$ and $R$.}
\label{Fig12}
\end{figure}

Let us estimate radius of the dark matter halo $r_{0}$ using Eq.(\ref{10.12}) and mass of Milky Way (mass of dark matter) $M\sim10^{12}M_{\odot}$, supposing $m_{L}\sim 10^{-20}\mathrm{eV}$. Then we obtain: $r_{0}\sim 10^{-5}\mathrm{pc}$, which is in no way comparable to radius of the DM halo being around $R\sim120\mathrm{kpc}$. In connection with this fact, a hypothesis about the formation of Bose stars, a large number of which can form the dark halo, has been proposed \cite{levkov1,levkov2}. However, we can propose another model. Let us compare energy of the halos with sizes $r_{0}$ and $R$ accordingly:
\begin{equation}\label{10.15a}
  E_{r_{0}}\sim-G\frac{M^{2}}{r_{0}}\sim -10^{63}\mathrm{J},\quad E_{R}\sim-G\frac{M^{2}}{R}\sim -10^{53}\mathrm{J}.
\end{equation}
These energies correspond to two different states of BEC - the ground state $\psi_{r_{0}}$ with energy $E_{r_{0}}$ and excited state $\psi_{R}$ with energy $E_{R}$, which are solutions of Gross-Pitaevskii equation:
\begin{equation}\label{10.16}
  -\frac{\hbar^{2}}{2m}\Delta\psi(r)-m\frac{4\pi GM}{r}\int_{0}^{r}|\psi(r')|^{2}r'^{2}dr'\psi(r)=\mu\psi(r),
\end{equation}
where $\mu=E/N=E\frac{m}{M}$. We can see, that Hamiltonian of the self-gravitating system is determined by its eigen state $\psi$. Thus, different states (ground and excited) correspond to different Hamiltonians of one and the same system. This means, that the states corresponding to different energies may not be orthogonal to each other, for example $\int\psi_{r_{0}}\psi_{R}^{+}d^{3}r\neq 0$. Let we excite the system from the ground state $\psi_{r_{0}}$ to some excited state $\psi_{R}$. Such transition stipulates the restructuring of the potential $U_{r_{0}}\rightarrow U_{R}$, so that our excited states $\psi_{R}$ becomes the ground state of new potential $U_{R}$ as demonstrated in Fig.\ref{Fig12}.

Let us enlarge radius of a system by $n>1$ times, i.e $r_{0}\rightarrow nr_{0}$, where $r_{0}$ is the "Bohr radius" for the self-gravitating system. Wave functions for the ground state and excited states have the corresponding forms:
\begin{eqnarray}
  \psi_{r_{0}}(r) &\sim& \frac{\sqrt{3/\pi^{3}}}{(r_{0})^{3/2}}\frac{1}{\cosh\left(\frac{r}{r_{0}}\right)}\label{10.17} \\
  \psi_{R}(r) &\sim& \frac{\sqrt{3/\pi^{3}}}{(nr_{0})^{3/2}}\frac{1}{\cosh\left(\frac{r}{nr_{0}}\right)},\quad n>1 \label{10.18}
\end{eqnarray}
Here the value $r_{0}$ plays role of Bohr radius, $R\equiv r_{0}n$. The average energy of self-gravitating system in the state $\psi$ is
\begin{equation}\label{10.19}
  E=\frac{M}{m}\frac{\hbar^{2}4\pi}{2m}\int_{0}^{\infty}(\nabla\psi)^{2}r^{2}dr-\frac{16\pi^{2}}{3}GM^{2}\int_{0}^{\infty}\psi^{4}r^{4}dr.
\end{equation}
Substitution the ground state wave function (\ref{10.17}) in Eq.(\ref{10.19}) we obtain:
\begin{equation}\label{10.20}
  E=\frac{6\hbar^{2}M}{\pi^{2}m^{2}}\frac{0.607}{r_{0}^{2}}-\frac{48}{\pi^{4}}GM^{2}\frac{0.249}{r_{0}}
\equiv\frac{A}{r_{0}^{2}}-\frac{B}{r_{0}}.
\end{equation}
Minimizing this energy by the radius $r_{0}$ we obtain:
\begin{equation}\label{10.21}
  r_{0}=\frac{2A}{B}=2.44\frac{\pi^{2}}{4}\frac{\hbar^{2}}{Gm^{2}M}
\Rightarrow E_{r_{0}}=-\frac{B^{2}}{4A}=
-\frac{B}{2r_{0}}=-\frac{5.98}{\pi^{4}}\frac{GM^{2}}{r_{0}}=-\frac{9.80}{\pi^{6}}\frac{G^{2}M^{3}m^{2}}{\hbar^{2}}.
\end{equation}
Substitution the excited state wave function (\ref{10.18}) in Eq.(\ref{10.19}) we obtain:
\begin{equation}\label{10.22}
  R=nr_{0}\Rightarrow E=\frac{A}{n^{2}r_{0}^{2}}-\frac{B}{nr_{0}}\Rightarrow E_{R}=E_{r_{0}}\left(\frac{2}{n}-\frac{1}{n^{2}}\right).
\end{equation}
For the highly excited states $n\gg 1$, we have $E_{R}=\frac{2E_{r_{0}}}{n}$, unlike the excited energies of a hydrogen atom: $E_{n}=\frac{E_{1}}{n^{2}}$.

Obviously, the self-gravitating system aspires to transit to any underlying state. For this the system must give somewhere the released energy $E(R_{1})-E(R_{2})>0$, where $R_{2}<R_{1}$. Let us consider transition of the system from a high "orbit" to a lower one. As a result, the cloud is collapsing heating up. However, as we could see above, the L-bosons are sterile particles, that is they do not scatter on each other or on barionic matter. Hence, the above-mentioned mechanism of collapse of the cloud does not work. Then there is only one way: in order to make transition from the state $\psi_{R}$ to the state $\psi_{r_{0}}$ it is necessary to radiate gravitation waves (as an atom making quantum transition from some exited state to some underlying state radiates photons). The energy loss rate and the halo compression rate due to gravitational radiation can be estimated from problem of two particles attracting according to Newton's law \cite{landau}:
\begin{equation}\label{10.23}
  \frac{dE}{dt}\sim\frac{G^{4}M^{5}}{c^{5}R^{5}}\sim  10^{21}\mathrm{J}/\mathrm{s}\quad
  \frac{dR}{dt}\sim\frac{G^{3}M^{3}}{c^{5}R^{3}}\sim  10^{-10}\mathrm{m}/\mathrm{s},
\end{equation}
from where we obtain the relaxation time to the ground state $\psi_{r_{0}}$:
\begin{equation}\label{10.24}
  \tau=\frac{(R^{4}-r_{0}^{4})c^{5}}{4G^{3}M^{3}}\sim 10^{32}\mathrm{s},
\end{equation}
which is incommensurably greater than the age of the universe $4\cdot 10^{17}\mathrm{s}$.  Thus, the DM halo of a galaxy is similar to Rydberg atoms (instead Coulomb interaction - self-gravity, and instead el.-mag. radiation - gravitational radiation, however instead electrons - $L$-bosons). A notable feature of the Rydberg atoms is a very long lifetime compared to lifetime of low-excited states, so, for hydrogen atom $\tau(n=2)\sim 10^{-8}\mathrm{s}$ against $\tau(n=1000)\sim 1\mathrm{s}$. As we could see above, analogous situation takes place for the DM halos. Thus, \emph{the DM halo is a Rydberg self-gravitating many-boson atom}. It should be noted that we have proposed the simplest model of the halo as an excited state of a self-gravitating many-boson system. However, the excited states can also be much more complex structures.

\section{Higgs modes at $T=T_{c}$}\label{higgsTc}

Let us consider a three-band system near the critical temperature $T_{c1},T_{c2},T_{c3}<T<T_{c}$. In this region $\varphi_{0i}^{2}\sim|\epsilon|/b_{i}$ \cite{grig2}. Then we have from Eq.(\ref{2.24}):
\begin{equation}\label{11.1}
  \alpha_{i}(T)=a_{i}(T)>0.
\end{equation}
Then coefficients $a,b,d$ (\ref{2.28}) in the dispersion equation (\ref{2.27}) take the form:
\begin{eqnarray}\label{11.2}
  b(T) &=& -a_{1}-a_{2}-a_{3}\nonumber\\
  c(T) &=& a_{1}a_{2}+a_{1}a_{3}+a_{2}a_{3}-3\epsilon^{2}\\
  d(T) &=& -a_{1}a_{2}a_{3}-2\epsilon^{3}+\epsilon^{2}(a_{1}+a_{2}+a_{3}),\nonumber
\end{eqnarray}
where we have accounted $\cos\theta_{ik}=1$. From equation for critical temperature (\ref{8.10}) we have $d(T_{c})=0$. Then, from the dispersion equation (\ref{2.27}) we obtain the corresponding dispersion relations at critical temperature:
\begin{eqnarray}
  q_{\mu}q^{\mu}(T_{c})&=&0\label{11.3a}\\
  q_{\mu}q^{\mu}(T_{c})&=&(-b\pm\sqrt{b^{2}-4c})/2>0.\label{11.3b}
\end{eqnarray}
For the first mode (\ref{11.3a}) (i.e for the common mode oscillations - Fig.\ref{Fig3}) the energy gap vanishes at the critical temperature, as it takes place in the single-band model. At the same time, the energy gaps of the second and third modes (\ref{11.3b}) (i.e for the antiphase oscillations - Fig.\ref{Fig3}) do not vanish at the critical temperature. So, for symmetrical bands $\alpha_{1}=\alpha_{2}=\alpha_{3}\equiv\alpha$ the massive modes have the same spectrum ($b^{2}-4c=0$ taking into account the condition $d(T_{c})=0\Rightarrow a(T_{c})=2|\epsilon|$):
\begin{equation}\label{11.4}
    q_{\mu}q^{\mu}(T_{c})=3|\epsilon|\Rightarrow m_{H1,2}(T_{c})=\sqrt{3|\epsilon|}.
\end{equation}
Thus, the energy gaps of the second and third Higgs modes are determined with the interband coupling $\epsilon$. At the same time, at $T=T_{c}$ the second order phase transition takes place: all equilibrium scalar fields become zero $\varphi_{01}(T_{c})=\varphi_{02}(T_{c})=\varphi_{03}(T_{c})=0$ -  Fig.\ref{Fig5}. Higgs bosons are oscillations of the modules $|\varphi_{1}|,|\varphi_{2}|,|\varphi_{3}|$ of the condensates. \emph{Since all $\varphi_{0i}(T_{c})=0$, then the nonzero energy gap $q_{\mu}q^{\mu}(T_{c})\neq 0$ of the Higgs modes at $T=T_{c}$ is a non-physical property}. In other words, at $T\geq T_{c}$ it is nothing to oscillate, there are fluctuations only, where $\langle\varphi_{i}\rangle=0$, $\langle\varphi_{i}^{2}\rangle\neq 0$ \cite{tinh}. Thus, nonzero masses of Higgs bosons at $T=T_{c}$ is incompatible with the second order phase transition.

In order to solve this problem in Ref.\cite{grig2,grig3,grig0} it has been proposed the intergradient interaction in the form as $\eta_{ik}\left(\partial_{\mu}\Psi_{i}\partial^{\mu}\Psi_{k}^{+}+\partial^{\mu}\Psi_{i}^{+}\partial_{\mu}\Psi_{k}\right)$. At special choice of the coefficients $\eta_{ik}$ we obtain single Higgs mode with correct dispersion law $q_{\mu}q^{\mu}(T_{c})=0$ (but $q_{\mu}q^{\mu}(T<T_{c})>0$) and single Goldstone modes $q_{\mu}q^{\mu}=0$ (i.e. the Leggett modes are absent). However, unlike superconductivity, in the field theory there is no restriction on the type of phase transition. So, the second order phase transition can be turned into the first-order phase transition by, for example, quantum corrections to Lagrangian of the scalar field which interacts with the gauge fields \cite{linde}. Or, we can use the effective potential \cite{rubakov2} in the following form:
\begin{equation}\label{11.5}
  U(\varphi,T)=\frac{1}{2}\mathcal{N}\left(\frac{T^{2}}{T_{-}^{2}}-1\right)\varphi^{2}-\frac{1}{3}cT\varphi^{3}
  +\frac{1}{4}bT\varphi^{4},
\end{equation}
where $T_{-}$ is the lower spinodal temperature. Due to presence of the cubic term $cT\varphi^{3}$, the potential describes the first order phase transition at critical temperature $T_{c}=\frac{T_{-}}{\sqrt{1-2c^{2}T_{-}^{2}/9b\mathcal{N}}}$ and the jump of the density of condensate $\frac{\Delta\varphi_{0}(T_{c})}{T_{c}}=\frac{3}{2}\frac{c}{b}$. Thus, the presence of the jump, i.e $\varphi_{0}(T_{c})\neq 0$, allows existence of the nonzero energy gap $q_{\mu}q^{\mu}(T_{c})\neq 0$ of the Higgs modes at $T=T_{c}$. Any other options can be considered else.

\section{Results}\label{results}

In this work we proposed an extension of the Glashow-Weinberg-Salam model of electro-weak interaction using analogy with three-band superconductors with interband Josephson couplings. The proposed model describes important phenomena formulated in Sect.\ref{intro}:
\begin{itemize}
  \item There are two ultra-light sterile bosons - the Leggett bosons, the Bose-Einstein condensate of which plays role of the dark matter halo. The halo is in an excited but stable quantum state, where there is no a central cusp, due to quantum pressure counteracting to gravitational compression. In order to obtain the L-boson at least two bands are required. In the case of multi-band system the attractive interband coupling $\epsilon<0$ should take place in order for fermions to acquire Dirac masses.
  \item Dirac neutrinos receive effective masses which manifest themselves in the neutrino oscillations and $\beta$-decays. In order to obtain the neutrino oscillations and violation of CP-invariance at least three bands are required. Mixing angles for charged leptons are negligibly small, so that the flavour oscillations electron-muon-tauon cannot be observable.
  \item There are neutral Higgs bosons of three flavours: $H_{e},H_{\mu},H_{\tau}$, which each interacts with corresponding generation of fermions only, where the heaviest boson $H_{\tau}$ is associated with the observed H-boson. Therefore decays of this H-boson into fermions of the second  and first generations through Yukawa interaction are prohibited. Another more light flavours $H_{e}$ and $H_{\mu}$ require detection as experimental test of the proposed model, at the same time, these two additional H-bosons very weakly interact with gauge and Dirac fields, which makes them difficult to detect.
  \item Masses of each generation of fermions are determined by the Yukawa coupling with amplitudes of corresponding condensates $\varphi_{0e},\varphi_{0\mu},\varphi_{0\tau}$ of the scalar fields. The slight mass asymmetry $m_{He}<m_{H\mu}<m_{H\tau}$ leads to the strong band asymmetry $\varphi_{0e}\ll\varphi_{0\mu}\ll\varphi_{0\tau}$. Hence, the fermion masses differ by orders of magnitude $m_{e}\ll m_{\mu}\ll m_{\tau}$.
\end{itemize}
It should be noted, that, unlike such extension of SM as nHDM or nHDM+S, the proposed model does not generate a large number of other particles (for example, charged Higgs bosons), which can essentially interact with ordinary matter. In addition, the proposed particles-candidates for DM - Leggett bosons are absolutely sterile, that is, they cannot weakly interacts with matter (as neutrino) even.

\section*{Acknowledgments}
This research was supported by theme grants of Department of physics and astronomy of NAS of Ukraine: "Noise-induced dynamics and correlations in nonequilibrium systems" 0120U101347, "Stochastic processes in condensed media, biological systems and radiation fields" 0125U000031 and by grant of Simons Foundation.

\appendix
\section{Some symmetric 3HDM potentials}\label{symm}

Following to \cite{keus} a scalar 3HDM potential symmetric under a group $G$ can be written as
\begin{equation}\label{A1}
  V=V_{0}+V_{G},
\end{equation}
where
\begin{eqnarray}\label{A2}
    V_{0}&=&\sum_{i=1}^{3}a_{i}\left|\Psi_{i}\right|^{2}+\frac{b_{i}}{2}\left|\Psi_{i}\right|^{4}
    +b_{12}|\Psi_{1}|^{2}|\Psi_{2}|^{2}+b_{13}|\Psi_{1}|^{2}|\Psi_{3}|^{2}+b_{23}|\Psi_{2}|^{2}|\Psi_{3}|^{2}\nonumber\\
    &+&b_{12}'(\Psi_{1}^{+}\Psi_{2})(\Psi_{2}^{+}\Psi_{1})+b_{13}'(\Psi_{1}^{+}\Psi_{3})(\Psi_{3}^{+}\Psi_{1})+
    b_{23}'(\Psi_{2}^{+}\Psi_{3})(\Psi_{3}^{+}\Psi_{2})
\end{eqnarray}
is invariant under the most general $U(1)\otimes U(1)$ gauge transformation and $V_{G}$ is a collection of extra terms
ensuring the symmetry group $G$. The $U(1)\otimes U(1)$ group is generated by
\begin{equation}\label{A3}
  \left(
     \begin{array}{ccc}
       e^{-\mathrm{i}\alpha} & 0 & 0 \\
       0 & e^{\mathrm{i}\alpha} & 0 \\
       0 & 0 & 1 \\
     \end{array}
   \right)
   \left(
     \begin{array}{ccc}
       e^{-2\mathrm{i}\beta/3} & 0 & 0 \\
       0 & e^{\mathrm{i}\beta/3} & 0 \\
       0 & 0 & e^{\mathrm{i}\beta/3} \\
     \end{array}
   \right).
\end{equation}
However, in the present work we use the minimal model, where
$b_{ik}=b_{ik}'=0$. A potential symmetric under the $U(1)$ group is
\begin{equation}\label{A4}
  V_{U(1)}=V_{0}+\lambda_{123}\left[(\Psi_{1}^{+}\Psi_{3})(\Psi_{2}^{+}\Psi_{3})+(\Psi_{1}\Psi_{3}^{+})(\Psi_{2}\Psi_{3}^{+})\right].
\end{equation}
The $U(1)$ group is generated by
\begin{equation}\label{A5}
  \left(
     \begin{array}{ccc}
       e^{-\mathrm{i}\alpha} & 0 & 0 \\
       0 & e^{\mathrm{i}\alpha} & 0 \\
       0 & 0 & 1 \\
     \end{array}
   \right).
\end{equation}
A potential symmetric under the $U(1)\otimes Z_{2}$ group is
\begin{equation}\label{A6}
  V_{U(1)\otimes Z_{2}}=V_{0}+\lambda_{23}\left[(\Psi_{2}^{+}\Psi_{3})^{2}+(\Psi_{2}\Psi_{3}^{+})^{2}\right].
\end{equation}
The $U(1)\otimes Z_{2}$ group is generated by
\begin{equation}\label{A7}
  \left(
     \begin{array}{ccc}
       e^{-2\mathrm{i}\beta/3} & 0 & 0 \\
       0 & e^{\mathrm{i}\beta/3} & 0 \\
       0 & 0 & e^{\mathrm{i}\beta/3} \\
     \end{array}
   \right)
   \left(
     \begin{array}{ccc}
       -1 & 0 & 0 \\
       0 & -1 & 0 \\
       0 & 0 & 1 \\
     \end{array}
   \right).
\end{equation}
A potential symmetric under the $Z_{2}$ group is
\begin{eqnarray}\label{A8}
  V_{Z_{2}}=V_{0}&+&\epsilon_{12}\left[\Psi_{1}^{+}\Psi_{2}+\Psi_{1}\Psi_{2}^{+}\right]\nonumber\\
  &+&\lambda_{12}\left[(\Psi_{1}^{+}\Psi_{2})^{2}+(\Psi_{1}\Psi_{2}^{+})^{2}\right]
  +\lambda_{13}\left[(\Psi_{1}^{+}\Psi_{3})^{2}+(\Psi_{1}\Psi_{3}^{+})^{2}\right]
  +\lambda_{23}\left[(\Psi_{2}^{+}\Psi_{3})^{2}+(\Psi_{2}\Psi_{3}^{+})^{2}\right].
\end{eqnarray}
The $Z_{2}$ group is generated by
\begin{equation}\label{A9}
   \left(
     \begin{array}{ccc}
       -1 & 0 & 0 \\
       0 & -1 & 0 \\
       0 & 0 & 1 \\
     \end{array}
   \right).
\end{equation}


\end{document}